\documentclass[a4paper,useAMS]{mn2e}
\input{psfig}
\usepackage{times}

\usepackage[totalwidth=504pt, totalheight=686pt, letterpaper]{geometry}

\newif\ifAMStwofonts

\newcommand{\kms}{\ensuremath{\hbox{km}\cdot \hbox{s}^{-1}}}
\newcommand{\hmpc}{\ensuremath{h^{-1}\,\hbox{Mpc}}}

\newcommand{\de}{\delta}

\newcommand{\f}{\frac}

\newcommand{\bfr}{{\bf r}}
\newcommand{\bfs}{{\bf s}}

\newcommand{\bfk}{{\bf k}}
\newcommand{\bfv}{{\bf v}}

\newcommand{\bfg}{{\bf g}}

\newcommand{\calO}{{\cal O}}

\newcommand{\bc}{\begin{center}}
\newcommand{\be}{\begin{equation}}
\newcommand{\ee}{\end{equation}}
\newcommand{\ec}{\end{center}}
\newcommand{\lan}{\langle}
\newcommand{\ran}{\rangle}
\newcommand{\spose}[1]{\hbox to 0pt{#1\hss}}
\newcommand{\lta}{\mathrel{\spose{\lower 3pt\hbox{$\mathchar"218$}}
 \raise 2.0pt\hbox{$\mathchar"13C$}}}
\newcommand{\gta}{\mathrel{\spose{\lower 3pt\hbox{$\mathchar"218$}}
 \raise 2.0pt\hbox{$\mathchar"13E$}}}

\newcommand{\etal}{{et al.}~}

\begin{document}

\title[]{Cosmic velocity--gravity relation in redshift space}

\author[Colombi, Chodorowski \& Teyssier]
{St\'ephane Colombi,$^1$\thanks{E-mail: colombi@iap.fr} 
Micha{\l} J.\ Chodorowski$^2$\thanks{E-mail: michal@camk.edu.pl}
and Romain Teyssier$^3$\thanks{E-mail: romain.teyssier@cea.fr} \\
$^1$Institut d'Astrophysique de Paris, CNRS, 98 bis Boulevard Arago, 
			75014 Paris, France\\
$^2$Copernicus Astronomical Center, Bartycka 18, 00--716 Warsaw, 
Poland \\
$^3$Commissariat \`a l'Energie Atomique, Direction des Sciences de la 
Mati\`ere, Service d'Astrophysique, \\
\ \ Centre d'\'Etudes de Saclay, L'orme des Merisiers,
91191 Gif-sur-Yvette Cedex, France
}

\maketitle
\begin{abstract}
We propose a simple way to estimate the parameter $\beta\simeq
\Omega^{0.6}/b$ from three-dimensional galaxy surveys, where $\Omega$
is the non-relativistic matter density parameter of the Universe and
$b$ is the bias between the galaxy distribution and the total matter
distribution. Our method consists in measuring the relation between
the cosmological velocity and gravity fields, and thus requires
peculiar velocity measurements. The relation is measured {\em directly
in redshift space}, so there is no need to reconstruct the density
field in real space. In linear theory, the radial components of the
gravity and velocity fields in redshift space are expected to be
tightly correlated, with a slope given, in the distant observer
approximation, by
\begin{displaymath}
\frac{g}{v} =  \frac{\sqrt{1+6\beta/5+3\beta^2/7}}{\beta}.
\end{displaymath}
We test extensively this relation using controlled numerical
experiments based on a cosmological $N$-body simulation. To perform the
measurements, we propose a new and rather simple adaptive
interpolation scheme to estimate the velocity and the gravity field on
a grid.

One of the most striking results is that nonlinear effects, including
`fingers of God', affect mainly the tails of the joint probability
distribution function (PDF) of the velocity and gravity field: the
$1$--$1.5 \sigma$ region around the maximum of the PDF is {\em dominated
by the linear theory regime}, both in real and redshift space. This is
understood explicitly by using the spherical collapse model as a proxy
of nonlinear dynamics. 

Applications of the method to real galaxy catalogs are discussed,
including a preliminary investigation on homogeneous (volume limited)
``galaxy'' samples extracted from the simulation with simple
prescriptions based on halo and sub-structure identification, to
quantify the effects of the bias between the galaxy distribution and
the total matter distribution, as well as the effects of shot noise.
\end{abstract}


\section{Introduction}
\label{sec:intro}

Analyzes of large-scale structure of the Universe provide estimates of
cosmological parameters that are complementary to those from the
cosmic microwave background measurements. In particular,
comparing the large-scale distribution of galaxies to their peculiar
velocities enables one to constrain the quantity $\beta \equiv
\Omega^{0.6}/b$. Here, $\Omega$ is the cosmological
non-relativistic matter density parameter and $b$ is the linear bias of
galaxies that are used to trace the underlying mass distribution. This
is so because the peculiar velocity field, $\bfv$, is induced
gravitationally and is therefore tightly coupled to the matter
distribution. In the linear regime, this relationship takes the form
(Peebles 1980)

\be
\bfv(\bfr) = \Omega^{0.6} \int \frac{{\rm d}^3\bfr'}{4 \pi} 
\delta(\bfr') \frac{\bfr'-\bfr}{|\bfr'-\bfr|^3} \,,
\label{eq:lin_vel}
\ee
where $\de$ denotes the mass density contrast and distances have been
expressed in \kms. Under the assumption of linear bias $\de = b^{-1}
\de^{(g)}$, where $\de^{(g)}$ denotes the density contrast of 
galaxies, the amplitude of peculiar velocities depends linearly on
$\beta$.

Density--velocity comparisons are done by extracting the density field
from full-sky redshift surveys (such as the PSCz, Saunders \etal
2000; or the 2MRS, Erdo$\breve{\rm g}$du \etal 2006), 
and comparing it to the observed velocity field from peculiar
velocity surveys (such as the Mark III catalog, Willick \etal 1997;
ENEAR, da Costa \etal 2000; and more recently SFI++, Masters \etal
2005).  The methods for doing this fall into two broad categories. One
can use equation~(\ref{eq:lin_vel}) to calculate the predicted
velocity field from a redshift survey, and compare the result with the
measured peculiar velocity field; this is referred to as a
velocity--velocity comparison (e.g., Kaiser et al. 1991; Willick \&
Strauss 1998).  Alternatively, one can use the differential form of
this equation, and calculate the divergence of the observed velocity
field to compare directly with the density field from a redshift
survey; this is called a density--density comparison (e.g., Dekel et
al. 1993; Sigad \etal 1998). The advantage of density--density
comparisons is that they are purely local, but they are significantly
sensitive to shot noise because the divergence of the observed
velocity field is estimated from the sparse velocity sample. The
integral form of velocity--velocity comparisons make them much less
sensitive to such a noise, but their non-local nature make them
affected by the tides due to the presence of fluctuations outside the
survey volume (see, e.g., Kaiser \& Stebbins 1991). 

Still, the common problem with both types of these
comparisons is that while equation~(\ref{eq:lin_vel}) involves the
density field in real space, the observed density field is given
solely in redshift space. One approach to tackle with this problem is
to reconstruct the real-space density field from the redshift-space
one. The transformation from the real space coordinate, $\bfr$, to the
redshift space coordinate, $\bfs$, is

\be 
\bfs = \bfr + v(\bfr)\, \hat{\bfr} \,,
\label{eq:rtos}
\ee
where $v(\bfr) = \bfv \cdot \hat{\bfr}$; $\bfv(\bfr)$ is the
real-space velocity field, and velocities are measured
relative to the rest frame of CMB.   
Therefore, one has to correct galaxy positions for
their peculiar velocities. To do so, equation~(\ref{eq:lin_vel}) is
used and a self-consistent solution for the density field in real
space is usually obtained iteratively (Yahil \etal 1991). In the first
iteration, to predict peculiar velocities according to
equation~(\ref{eq:lin_vel}), the real-space density field appearing on
the r.h.s. of this equation is approximated by the redshift
space-density field, and so on, until convergence. However, from
equation~(\ref{eq:lin_vel}) it is obvious that the amplitude of
peculiar velocities depends on $\beta$, the parameter to be
subsequently estimated. Therefore, to perform a density--velocity
comparison self-consistently, one has to reconstruct the real space
density field for a range of different values of $\beta$. For example,
Branchini \etal (1999) performed such reconstructions for 10 different
values of $\beta$ in the range $0.1$--$1.0$.
Note that instead of this traditional algorithm,
more sophisticated methods now rely on Euler-Lagrange action minimization (e.g., 
Peebles 1989; Shaya, Peebles \& Tully 1995; Nusser \& Branchini 2000; Phelps 2002; 
Phelps et al. 2006) or resolution
of optimal assignement problems (e.g., Croft \& Gazta\~naga 1997; Frisch
et al. 2002; Mohayaee et a. 2003; Mohayaee \& Tully 2005). 

Another approach, proposed by Nusser \& Davis (1994; hereafter ND), is
to perform the comparison directly in redshift space. ND derived the
density--velocity relation in redshift space in the linear
regime. Because velocity--velocity comparisons seem in practice
more robust than
density--density ones, they aimed at transforming this relation to an
integral form, i.e. at solving for the velocity as a functional of the
redshift-space density. Due to the radial character of redshift-space
distortions, it turned out to be possible only via a modal expansion of
the density and velocity fields in spherical harmonics. Using this
expansion, ND were able to put constraints on $\beta$ (ND; 
Davis, Nusser \& Willick 1996). However, the approach with the
reconstruction of the real-space density field remained
popular. Apparently, equation~(\ref{eq:lin_vel}) is appealing by its
simplicity, both in terms of reconstruction of the velocity field, and
of estimation of the parameter $\beta$.

This paper is devoted to finding an equivalent of
equation~(\ref{eq:lin_vel}) that would hold for redshift-space
quantities, but would share its simplicity. Specifically, let us
define the {\em scaled} gravity:

\be
\bfg(\bfr) \equiv \int \frac{{\rm d}^3\bfr'}{4 \pi} 
\delta^{(g)}(\bfr') \frac{\bfr'-\bfr}{|\bfr'-\bfr|^3} \,.
\label{eq:lin_grav}
\ee
Under the assumption of linear bias, adopted here, $\bfg$ is
proportional to gravitational acceleration, and can be directly
measured from a 3-D galaxy survey. Equation~(\ref{eq:lin_vel}) then
implies

\be
\bfg = \beta^{-1}\, \bfv \,.
\label{eq:g-v_real}
\ee
Now, let us assume that we measure the gravitational acceleration
directly in redshift space:
\be
\bfg_\bfs(\bfs) = \int \frac{{\rm d}^3\bfs'}{4 \pi} 
\delta^{(g)}_\bfs(\bfs') \frac{\bfs'-\bfs}{|\bfs'-\bfs|^3} \,,
\label{eq:grav_red}
\ee
where $\de^{(g)}_\bfs$ denotes the density contrast of galaxies in
redshift space. We will follow ND in a natural definition of the
redshift-space velocity field:

\be
\bfv_\bfs(\bfs) \equiv \bfv(\bfr) \,.
\label{eq:vel_red}
\ee
What is the relation between $\bfg_\bfs$ and $\bfv_\bfs$?

Equation~(\ref{eq:g-v_real}) holds strictly in the linear
regime. Nevertheless, numerical simulations (Cieciel\c ag \etal 2003)
have shown that it remains accurate to a few percent for fully
nonlinear gravity and velocity fields. These results will be fully
confirmed by the present work and can be explained by the fact that
velocity and gravity fields are dominated by long-wavelength, linear
modes. Therefore, in deriving the redshift-space counterpart of
Equation~(\ref{eq:g-v_real}), we will apply linear
theory. Unfortunately, there is no local deterministic relation
between $\bfg_\bfs$ and $\bfv_\bfs$. However, as shown below,
$\bfg_\bfs$ and $\bfv_\bfs$ are strongly correlated, so the mean
relation will be a useful quantity. Only radial components of
velocities are directly measurable, so we will be interested in the
relation between $g_\bfs = \bfg_\bfs \cdot \hat{\bfs}$ and $v_\bfs =
\bfv_\bfs \cdot \hat{\bfs}$.

This paper is thus organized as follows. In \S~\ref{sec:calcul}, we
compute the properties of the joint probability distribution function
(PDF) of the fields $g_\bfs$ and $v_\bfs$ in the framework of linear
theory and distant observer limit.  In the linear regime, this PDF is
expected to be a Gaussian, entirely determined by its second order
moments. The quantities of interest are ${\lan v_\bfs g_\bfs
\ran}/{\lan v_\bfs^2 \ran}$, ${\lan g_\bfs^2 \ran}/{\lan v_\bfs g_\bfs
\ran}$ and $\sqrt{{\lan g_\bfs^2 \ran}/{\lan v_\bfs^2 \ran}}$: they
all give an estimate of the expected ratio between $g_\bfs$ and
$v_\bfs$ and the difference between them can be used to compute the
scatter on the relation, that will be shown to be small. The validity
of our assumptions, in particular the distant observer approximation,
 is examined {\em a posteriori} in \S~\ref{sec:valdol}. We also justify
our choice of  CMB rest frame redshifts, needed to 
avoid the so-called rocket effect (Kaiser 1987; Kaiser \& Lahav 1988).
In \S~\ref{sec:simus}, linear theory is tested against numerical
experiments, both in real and redshift space. To do that we use a dark
matter cosmological $N$-body simulation with high resolution,
allowing us to probe the highly nonlinear regime. We propose a new and
simple algorithm to interpolate the velocity and gravity field on a
grid from a distribution of particles. We address extensively a number of
issues, such as the validity of the distant observer limit, edge
effects, cosmic variance effects, effects arising from non-linear
dynamics, in particular so called `fingers of God' (FOG) and effects
of dilution (numbers of tracers used to probe the velocity and gravity
fields).  We use
spherical top-hat model to support our interpretations of the
measurements (technical details are given in Appendix A).
The effect of the bias is also examined briefly
by extracting from the dark matter distribution two kinds of subsamples,
one where each ``galaxy'' is identified with a dark matter halo, and the
other one where each ``galaxy'' is identified with a dark matter substructure.
Finally, the
main results of this work are summarized in \S~\ref{sec:concl}.  In
this last section, we discuss observational issues such as
discreteness effects, appropriate treatment of biasing, incompleteness, errors on peculiar
velocity estimates, etc. These issues will be addressed in detail in a
forthcoming work, where the method will be applied to real data.

\section{Linear theory predictions}
\label{sec:calcul}

Linear regression of $v_\bfs$ on $g_\bfs$ yields
\be
g_\bfs = \f{\lan v_\bfs g_\bfs \ran}{\lan v_\bfs^2 \ran} v_\bfs \,.
\label{eq:regress}
\ee 
The symbols $\lan \cdot \ran$ denote ensemble averaging. We can thus
characterize the linear relation between gravity and velocity (gravity
in terms of velocity) by its slope, 

\be
a_F \equiv \f{\lan v_\bfs g_\bfs \ran}{\lan v_\bfs^2 \ran} \,.
\label{eq:a_F}
\ee 
Alternatively, one can study the inverse relation (velocity in terms
of gravity). Linear regression then yields $v_\bfs = (\lan v_\bfs
g_\bfs \ran/\lan g_\bfs^2 \ran) g_\bfs$, so the inverse slope is $a_I
\equiv \lan v_\bfs g_\bfs \ran/ \lan g_\bfs^2 \ran$. To describe the
linear velocity--gravity relation, we have thus at our disposal two
estimators of its slope, the forward slope, $a_F$, and the reciprocal
of the slope of the inverse relation, 
\be
a_I^{-1} = \frac{\lan g_\bfs^2 \ran}{\lan v_\bfs g_\bfs \ran} \,. 
\label{eq:a_I}
\ee 
Due to the scatter, these two estimators are not equal:

\be
a_I^{-1} - a_F = a_I^{-1} (1 - r^2) > 0 \,,
\label{eq:two_slopes}
\ee
where 

\be
r = \frac{\lan v_\bfs g_\bfs \ran}{\sigma_v \sigma_g} 
\label{eq:cross}
\ee 
is the cross-correlation coefficient of the velocity and gravity
fields; $\sigma_v^2 = \lan v_\bfs^2 \ran$ and $\sigma_g^2 = \lan
g_\bfs^2 \ran$. The linear regression, although the best among all
linear fits to a cloud of points, visually looks biased. For two
correlated Gaussian variables, an unbiased slope of the isocontours
(ellipsoids) of their joint PDF is given by the square root of the
ratio of the variances of the two variables. Thus we have third
estimator of the slope,

\be 
a_U \equiv \sqrt{\frac{\lan g_\bfs^2 \ran}{\lan v_\bfs^2 \ran}} \,.
\label{eq:a_U}
\ee 
This is easy to show that it predicts intermediate values between
$a_I^{-1}$ and $a_F$.

Second-order moments of the joint PDF for gravity and velocity, which
appear in the above three estimators of its slope, are much easier to
perform in Fourier space. We will adopt the distant observer limit
(DOL), in which the Jacobian of the transformation from real to
redshift space, equation~(\ref{eq:rtos}), simplifies to

\be
J(\bfr) = 1 + \f{\partial v}{\partial r} \,.
\label{eq:jac}
\ee
The Fourier transform of the redshift-space velocity field, using the
mapping~(\ref{eq:rtos}), is
\begin{eqnarray}
\tilde \bfv_\bfs(\bfk) 
\!\!\!\! &\equiv& \!\!\!\!\! 
\int {\rm e}^{i \bfk \cdot \bfs} \bfv_\bfs(\bfs)\, {\rm d}^3 s\, = 
\!\! \int {\rm e}^{i \bfk \cdot \bfr} {\rm e}^{i k v} \bfv(\bfr) 
J(\bfr)\, {\rm d}^3 r \nonumber \\ 
&=& \!\!\!\!\! \int {\rm e}^{i \bfk \cdot \bfr} \bfv(\bfr)\, 
{\rm d}^3 r + \calO(v \de) \simeq \tilde\bfv(\bfk) \,,
\label{eq:vel(k)}
\end{eqnarray}
where $\tilde\bfv(\bfk)$ is the Fourier transform of the real-space
velocity field. The linearized continuity equation yields
$\tilde\bfv(\bfk) = - f (i \bfk/k^2) \tilde\de(\bfk)$, where $f \simeq
\Omega^{0.6}$ and $\tilde\de(\bfk)$ is the Fourier transform of the
real-space density field, hence

\be 
\tilde \bfv_\bfs(\bfk) = - f \f{i \bfk}{k^2} \tilde\de(\bfk) \,.
\label{eq:vel(k)_fin}
\ee
From equation~(\ref{eq:grav_red}) we have

\be 
\tilde \bfg_\bfs(\bfk) = - \f{i \bfk}{k^2} 
\tilde\de^{(g)}_\bfs(\bfk) \,.
\label{eq:grav(k)}
\ee 
This equation looks similar to the preceding one, but here appears the
Fourier transform of the {\em redshift-space} galaxy density
field. Moreover, unlike the preceding equation, the above equation is
exact.
  
Radial components of the redshift-space velocity and gravity fields
are $\tilde v_\bfs(\bfk) = - f\, (i\mu/k)\, \tilde\de(\bfk)$ and
$\tilde g_\bfs(\bfk) = - (i\mu/k)\, \tilde\de^{(g)}_\bfs(\bfk)$, where
$\mu = \bfk\cdot\bfs/(k s)$.  From conservation of numbers of galaxies
in real and redshift space, it is straightforward to write down an
equation for the Fourier transform of the redshift-space galaxy
density contrast. In the linear regime it reduces to
$\tilde\de^{(g)}_\bfs(\bfk) = \tilde\de^{(g)}(\bfk) + f \mu^2
\tilde\de(\bfk)$ (Kaiser 1987), or 

\be
\tilde\de^{(g)}_\bfs(\bfk) = b (1+ \beta \mu^2) \tilde\de(\bfk) \,.
\label{eq:dg_s}
\ee
Therefore, we obtain

\be
\tilde v_\bfs(\bfk) = - f \f{i \mu}{k} \tilde\de(\bfk) \,,
\label{eq:v(k)}
\ee
and

\be
\tilde g_\bfs(\bfk) = - b (1+ \beta \mu^2) \f{i \mu}{k} 
\tilde\de(\bfk) \,. 
\label{eq:g(k)}
\ee

The above pair of equations enables us to calculate the averages
appearing in equations~(\ref{eq:a_F}), (\ref{eq:a_I}),
and~(\ref{eq:a_U}). Specifically,

\be
\lan v_\bfs^2 \ran = (2\pi)^{-6} \!\!\! \int {\rm e}^{-i (\bfk+\bfk') 
\cdot \bfs} \, f \f{i \mu}{k} f \f{i \mu'}{k'}\, \lan\tilde\de(\bfk)
\tilde\de(\bfk')\ran\, {\rm d}^3 k\, {\rm d}^3 k'.
\label{eq:int_F}
\ee
For a homogeneous and isotropic random process, $\lan\tilde\de(\bfk)
\tilde\de(\bfk')\ran = (2\pi)^3 \de_D(\bfk + \bfk') P(k)$,
where $\de_D$ is Dirac's delta and $P(k)$ is the power spectrum of the
real-space density field. Performing the integral over $\bfk'$ yields

\be
\lan v_\bfs^2 \ran = \f{f^2}{(2\pi)^{3}} \!\!\! 
\int \f{\mu^2}{k^2} P(k)\, {\rm d}^3 k\ = \f{f^2}{6\pi^2} 
\int_0^\infty \! P(k)\, {\rm d} k \,.
\label{eq:int_Fvv}
\ee
Similarly, 
\begin{eqnarray}
\lan v_\bfs g_\bfs \ran 
\!\!\! &=&\!\!\! \f{b f}{(2\pi)^{3}} \!\!\! \int (1 + \beta
\mu^2) \f{\mu^2}{k^2} P(k)\, {\rm d}^3 k \nonumber \\
\!\!\! &=&\!\!\!  
\f{b f[1+ (3/5)\beta]}{6\pi^2} \int_0^\infty \! P(k)\, {\rm d} k \,.
\label{eq:int_Fvg}
\end{eqnarray}
Finally,
\begin{eqnarray}
\lan g_\bfs^2 \ran
\!\!\! &=&\!\!\! \f{b^2}{(2\pi)^{3}} \!\!\! \int (1 + \beta
\mu^2)^2 \f{\mu^2}{k^2} P(k)\, {\rm d}^3 k \nonumber \\
\!\!\! &=&\!\!\!  
\f{b^2}{6\pi^2} \left[1 + \f{6\beta}{5} + \f{3\beta^2}{7} \right] 
\int_0^\infty \! P(k)\, {\rm d} k \,.
\label{eq:int_Fgg}
\end{eqnarray}
This yields
\begin{eqnarray}
\displaystyle
a_F & \equiv & \f{\lan v_\bfs g_\bfs \ran}{\lan v_\bfs^2 \ran} = \frac{1 + 3 \beta/5}{\beta} \,, \label{eq:F} \\
a_I^{-1} & \equiv & \frac{ \lan g_\bfs^2 \ran}{\lan v_\bfs g_\bfs \ran} = \frac{1+6\beta/5+3\beta^2/7}{\beta(1+3\beta/5)} \,, 
\label{eq:I} \\
a_U & \equiv  & \sqrt{\frac{\lan g_\bfs^2 \ran}{\lan v_\bfs^2 \ran}} =\frac{\sqrt{1+6\beta/5+3\beta^2/7}}{\beta} \,. \label{eq:U}
\end{eqnarray}

Equations~(\ref{eq:F})--(\ref{eq:U}) provide a way to estimate $\beta$
comparing gravitational accelerations to peculiar velocities of
galaxies directly in redshift space.  All three estimators predict the
slope that is greater than the corresponding one for real-space
quantities ($\beta^{-1}$), the more the bigger $\beta$. This is
expected since redshift distortions enhance linear density contrasts
(eq.~\ref{eq:dg_s}), and the amplitude of distortions scales linearly
with $\beta$. We can therefore consider the factor $\beta^{-1}$ as the
one coming from real-space dynamics, and the additional factors
involving $\beta$ as redshift-space corrections.

As stated above, in redshift space, unlike in real space, the
velocity--gravity relation is not deterministic even in the linear
theory, so $a_I^{-1} > a_U > a_F$. However, the linear gravity and
velocity fields are tightly correlated, so
Equations~(\ref{eq:F})--(\ref{eq:U}) yield values of the ratio between
$g_\bfs$ and $v_\bfs$ very close to each other: e.g., for
$\beta=0.512$ (the value used in our numerical experiments,
\S~\ref{sec:simus}) $a_I^{-1}=2.579$, $a_U=2.565$, and $a_F=2.552$.

To illustrate this point furthermore, one can examine the scatter on
the conditional average  $\lan g \ran_{|v}$ (mean $g$ given $v$), where
we have dropped out the subscripts
`$\bfs$' for simplicity of notation.
For $g$ and $v$ being Gaussian-distributed, $\lan g
\ran_{|v} = \lan v g \ran v/\lan v^2\ran$, in agreement with linear
regression, equation~(\ref{eq:regress}). 
The scatter in this relation, $\sigma_{g|v}$, 
 is the square root of the
conditional variance, $\lan (g - \lan g \ran_{|v})^2 \ran_{|v}$.  
The conditional variance is
then equal to $(1 - r^2) \sigma_g^2$ (see, for example, Appendix B of
Cieciel\c ag \etal 2003). Thus, for Gaussian velocity and gravity
fields in redshift space we have
\be
\sigma_{g|v}^2 = (1 - r^2) \sigma_g^2 \,,
\label{eq:scatt}
\ee 
where $r$ is the cross-correlation coefficient defined in eq.~(\ref{eq:cross}).
Note that the scatter is just one number, i.e., it is independent of
the value of $v$. Equations~(\ref{eq:int_Fvv})--(\ref{eq:int_Fgg})
yield
\be
r^2 = 1 - \frac{12}{175}\, \frac{\beta^2}{1 + 6\beta/5 +
3 \mbox{$\beta^2$}/7} \,.
\label{eq:err^2_other}
\ee
In particular, for $\beta = 1$, $r \simeq 0.987$, and for $\beta =
0.5$, $r \simeq 0.995$: again, redshift-space velocity and gravity in
the linear regime are (though not simply mutually proportional, like
in real space) very tightly correlated. Inserting
equation~(\ref{eq:err^2_other}) into~(\ref{eq:scatt}) yields
\be
\sigma_{g|v}^2 = \frac{12}{175}\, \frac{\beta^2}{1 + 6\beta/5 +
3 \mbox{$\beta^2$}/7} \sigma_g^2 \,.
\label{eq:scatt_other}
\ee
Hence, for $\beta = 1$, $\sigma_{g|v} \simeq 0.16 \sigma_g$, and for
$\beta = 0.5$, $\sigma_{g|v} \simeq 0.10 \sigma_g$. This implies that
the `signal to noise ratio', $S/N$, of the estimate of a single
galaxy's peculiar velocity from its gravitational acceleration can be
as high as $10$ ($S/N \sim g/\sigma_{g|v} \sim \sigma_g/\sigma_{g|v}
\simeq 10$).  This is to be contrasted with the signal to noise ratio
of the estimate of a galaxy's peculiar velocity from its distance and
redshift, which is typically below unity: an expected typical $20$\% relative
error in distance,\footnote{See, for instance, Strauss \& Willick (1995) 
for a review on issues related to
peculiar velocity estimates.} 
for a galaxy at a distance of $40$ \hmpc,\footnote{Where $h=H_0/100$ and $H_0$ is the
Hubble constant expressed in km/s/Mpc.} 
translates to the velocity error of $800$ \kms, greater than typical
peculiar velocities of galaxies. As a corollary, the intrinsic
scatter in the redshift-space {\em linear\/} velocity--gravity
relation is negligible compared to that introduced by the errors of
measurements of peculiar velocities.

%
\section{Validity of the distant observer limit in the rest frame of CMB}
\label{sec:valdol}
%
While equation (\ref{eq:rtos}) was written in the CMB rest frame, the
choice of the reference frame for computing redshifts of galaxies can be
more general,
\be 
\bfs = \bfr + \left[ v(\bfr)-v_{\rm orig}\right]\, \hat{\bfr} \,
\label{eq:rtosgen}
\ee
where $v_{\rm orig}$ is the (angle dependent) 
radial velocity of the origin of the system
of coordinates (in the CMB rest frame). 

The Jacobian of eq.~(\ref{eq:rtosgen}) yields (Kaiser 1987)
\begin{equation}
1+\delta^{(g)}_\bfs(\bfs)=\left[1+\frac{v(\bfr)-v_{\rm orig}}{r} \right]^{-2} 
\left( 1 + \frac{\partial v}{\partial r} \right)^{-1} \left[ 1+\delta^{(g)} \right].
\label{eq:mapping}
\end{equation}
This equation is valid as long as the mapping (\ref{eq:rtosgen}) does
not induce any shell crossing. In general,
non trivial singularities can appear in $\delta^{(g)}_\bfs(\bfs)$, even if
$\delta^{(g)}$ is finite. However, in practice, since one always performs
some additional smoothing to the data,  $\delta^{(g)}_\bfs(\bfs)$ remains
finite. The small $r$ limit can be still problematic as it corresponds
to a singularity in the system of coordinates. If the field $v(\bfr)$
is smooth enough, then in the neighborhood of the origin, 
one can write 
\begin{equation}
1+\frac{v(\bfr)-v_{\rm orig}}{r} \simeq 1+\frac{v({\bf 0})-v_{\rm orig}}{r}.
\end{equation}
Choosing a spherical coordinate system such that the $z$ axis is parallel
to ${\bf v}({\bf 0})-{\bf v}_{\rm orig}$, 
the term $\{ 1+[v(\bfr)-v_{\rm orig}]/r\}^{-2}$ in eq.~(\ref{eq:mapping})
will create a singular surface of equation
\begin{equation}
r=-|{\bf v}({\bf 0})-{\bf v}_{\rm orig}|\cos(\theta), \quad \theta \geq \pi/2
\label{eq:sing1}
\end{equation}
that will concentrate to the origin, ${\bfs}={\bf 0}$, in redshift space. 
Similarly the term $(1+\partial v/\partial r)^{-1}$ might become singular,
but we see here that the situation nearby the origin is not different
from what happens far away from it.

The singular behavior of the form (\ref{eq:sing1}) is expected to occur
only in the neighborhood of the observer, but might be problematic when
estimating the gravitational acceleration. While inserting this singularity
in eq.~(\ref{eq:grav_red}), we notice that it should be an issue only for 
${\bfs}={\bf 0}$, where it coincides with the
Green function singularity (this is due to the finite mass of the singular
surface). In other words, even though its small $s$ behavior is difficult
to predict, the redshift space gravity should not be significantly affected
by such a singular behavior at distances sufficiently large
from the observer. In the CMB rest frame, $v_{\rm orig}=0$ and $v({\bf 0})$
would correspond to the Local Group velocity, say, $|{\bf v}({\bf 0})|\simeq 630$ km/s
(see e.g., Erdo$\breve{\rm g}$du et al. 2006), so we would
need $r$ and therefore $s$ 
large enough compared to 6 $h^{-1}$ Mpc.
More specifically, let us consider the real space sphere of radius
$r=|{\bf v}({\bf 0})|$. Its content is embedded in a sphere of radius $s=2 |{\bf v}({\bf 0})|$
in redshift space. The mass inside such a volume remains finite, but its internal
distribution affects the gravity field at larger $s$ in a non trivial way. What matters
is that the multipole contributions of higher order than the monopole (the substructures
within this volume) have negligible contribution on the gravity field. Using
the wisdom from treecode simulations techniques (e.g. Barnes \& Hut 1986;
Barnes \& Hut 1989), this amounts to
\begin{equation}
\frac{2 |{\bf v}({\bf 0})|}{s} \la 1.
\end{equation}
According to that criterion,
the effect of the singular behavior near the origin should be
of little consequence if $s \ga 2 |{\bf v}({\bf 0})|=13 h^{-1}$ Mpc. 

If it is supposed now that $s$ is indeed large enough, one can linearly expand
eq.~(\ref{eq:mapping}) to obtain (Kaiser 1987)
\begin{equation}
\delta^{(g)}_\bfs(\bfs)=\delta^{(g)} -2 \frac{v(\bfr)-v_{\rm orig}}{r} 
- \frac{\partial v}{\partial r}.
\label{eq:toto}
\end{equation}
(Note that, at linear order, one can
assume ${\bfs} \simeq \bfr$.) 
The distant observer limit would consist in
dropping the term $2 [{v(\bfr)-v_{\rm orig}}]/{r}$ from this equation.
However, as extensively discussed in Kaiser (1987) and Kaiser \& Lahav (1998),
this term is in fact non negligible in the redshift space gravitational 
acceleration as it induces the so-called rocket effect, resulting in 
a large $r$ logarithmic divergence if $v_{\rm orig} \neq 0$. 
This justifies the choice of CMB rest frame coordinate system, 
$v_{\rm orig} \equiv 0$. Still the remaining contribution of $v(\bfr)/r$,
although zero in average, might introduce some significant fluctuations
on the large scale redshift space gravity field. 
In the linear perturbation theory framework, $v(\bfr)$ does
not correlate with either $\delta^{(g)}$ or $\partial v/\partial r$.
As a result, while computing the sum of the fluctuations of $\delta^{(g)}_\bfs$
in a sphere of radius $r=s'$ in eq.~(\ref{eq:grav_red}), what matters is to see whether the
fluctuations added by $2 v(\bfr)/r$ are small, or not, compared
to the fluctuations added by $\delta^{(g)}-\partial v/\partial r$.

So let us estimate the ratio
\begin{equation}
{\cal R}^2 \equiv \frac{ \left\langle \left[2 v(\bfr)/r \right]^2 \right\rangle }
    { \left\langle \left[ \delta^{(g)}-\partial v/\partial r \right]^2 \right\rangle}.
\end{equation}
Calculations are similar to \S~\ref{sec:calcul}, and one simply finds, in
the linear regime,
\begin{equation}
{\cal R}^2=\frac{4 \beta^2}{3\left( 1 + 2\beta/3 + \beta^2/5 \right)}
\frac{\int_0^{\infty} P(k)dk}{r^2 \int_0^{\infty} P(k) k^2 dk}.
\label{eq:rexp}
\end{equation}
One can furthermore assume that the linear fields
are smoothed, e.g. with a Gaussian window of size $\ell$. In that
case one has to replace $P(k)$ with $P(k) \exp[-(k\ell)^2]$ in 
eq.~(\ref{eq:rexp}).\footnote{We suppose here that smoothing on $v(\bfr)/r$ is
approximately equivalent to smoothing on $v(\bfr)$, prior to dividing by $r$,
which should be reasonable.}
For scale-free power spectra, $P(k) \propto k^n$, one finds using eq.~(4.10b)
of Bardeen et al. (1986)
\begin{equation}
{\cal R}^2=\frac{8 \beta^2}{3 (n+3) \left( 1 + 2\beta/3 + \beta^2/5 \right) } 
\left( \frac{\ell}{r} \right)^2,
\end{equation}
which gives 
\begin{equation}
{\cal R}^2\simeq \frac{0.48}{n+3}  \left( \frac{\ell}{r} \right)^2, \quad {\rm for} \ \ \beta=0.5.
\end{equation}
This means that smoothing {\em increases} the relative contribution of the
term $v(\bfr)/r$! In other words, {\em smoothing makes the distant observer
approximation worse}.

The standard cold dark matter (CDM) cosmology considered 
in the forthcoming numerical analyses assumes the non-relativistic matter density
parameter $\Omega=0.3$, the cosmological constant
$\Lambda=0.7$ and the Hubble constant $H_0=70 \equiv 100\, h$
$\rm km\cdot s^{-1}\cdot Mpc^{-1}$. Using for instance the package of Eisenstein \& Hu (1998)
to compute $P(k)$ with these cosmological parameters (and $\Omega_{\rm baryons}=0.04$), 
one obtains numerically, in the absence of bias, $b=1$,
\begin{eqnarray}
{\cal R} & \simeq & \displaystyle \frac{0.17\,h^{-1}\,{\rm Mpc}}{r}, \quad {\rm no\ smoothing},\\
{\cal R} & \simeq & \displaystyle \frac{2.7\,h^{-1}\,{\rm Mpc}}{r}, \quad \ell=1\,h^{-1}\ {\rm Mpc}\\
{\cal R} & \simeq & \displaystyle \frac{10\,h^{-1}\,{\rm Mpc}}{r}, \quad \ell=10\,h^{-1}\ {\rm Mpc}.
\end{eqnarray}
Clearly, for a catalog depth of the order of
$70\,h^{-1}$ Mpc as considered below, we expect significant deviations
from our theoretical predictions if the smoothing scale is as large as
$10\,h^{-1}\ {\rm Mpc}$.
%
%
\section{Numerical experiments} 
\label{sec:simus}
%
%
In this section we perform controlled numerical experiments to test the
velocity--gravity relation, both in real and redshift space, on the
dark matter distribution. These analyzes extend the work of
Cieciel\c{a}g et al. (2003) who performed similar work but only in
real space and on simulations using a pure hydrodynamic code
approximating the dynamics of dark matter, namely the Cosmological
Pressureless Parabolic Advection code of Kudlicki, Plewa \&
R\'o\.zyczka (1996, see also Kudlicki et al. 2000).

This section is organized as follows. In \S~\ref{sec:algomes}, we
present a new algorithm to interpolate the velocity and the gravity
fields on a grid using an adaptive interpolating procedure inspired
closely from smooth particle hydrodynamics (SPH).  In
\S~\ref{sec:nbodysamples}, we describe the $N$-body simulation set
used in this work. In \S~\ref{sec:dmsamples}, we discuss measurements
on our dark matter samples and diluted counterparts, while mock galaxy
catalogs are addressed in \S~\ref{sec:mockmes}.

Since we assume flat cosmology, $\Omega+\Lambda=1$, we
adopt the approximation 
\begin{equation}
f(\Omega)\simeq \Omega^{5/9}
\end{equation}
for function $f(\Omega)$, as it is known to be slightly more accurate
in that case than the traditional $\Omega^{0.6}$ fit (Bouchet et
al. 1995). Finally, let us recall that `DOL' means ``distant observer
limit''.
%
%
%
\subsection{Algorithm used to estimate velocity and gravity fields}
\label{sec:algomes}
%
In real observations, the catalogs used to estimate cosmic velocity
and gravity are in general different. In particular, the velocity
field tracers are not necessarily representative of the underlying
density field.  Here we assume to simplify that the same catalog is
used to probe velocity and gravity fields and that all the objects in
the catalog have the same weight, or from the dynamical point of view,
the same mass. 
%
Also, since gravitational force is of long range, it is necessary to
estimate it in a domain $V_{\rm grav}$ large enough compared to the
effective volume $V$ where the velocity--gravity comparison is
actually performed. We assume here that it is indeed the case. In this
paper, $V_{\rm grav}$ will simply be the simulation cube, while $V$
will be a sub-volume included in this cube.

The basics of our method is to perform adaptive smoothing on the
particle distribution to obtain a smooth velocity and a smooth density
field on a regular grid encompassing $V_{\rm grav}$ and of resolution
$N_{\rm g}$, while preserving as much as possible all the
information. Then additional smoothing with a fixed window,
preferably Gaussian, can be performed {\em a posteriori} in order,
e.g., to be in the linear regime, since our analytic predictions in
principle apply to that regime. We shall see however below that this
additional step is not necessarily needed and can in fact complicate
the analyzes (see also \S~\ref{sec:valdol}). 

Whether we perform the measurements in real or redshift space does not
change our approach: we assume that we have at our disposal a set of
$N$ points with three-dimensional coordinates and a scalar velocity.
This latter is either the radial velocity when we consider
redshift space measurements, or the $z$ coordinate of the velocity
vector when we consider real space measurements or redshift space
measurements in the DOL approximation.

The main difficulty is to reconstruct a smooth velocity field on the
sampling grid. Let us remind as well that the velocity field we aim to
estimate is a purely Eulerian quantity, in other words, a mean flow
velocity.  In particular, what we aim to measure, in terms of
dynamics, is something as close as possible to a moment of the density
distribution function $f({\bf r}_{\rm L},{\bf v}_{\rm L})$ in
phase-space: 
\begin{equation}
{\bf v}({\bf r})\equiv \frac{1}{\rho} \int d^3 {\bf v}_{\rm L} {\bf
v}_{\rm L} f({\bf r},{\bf v}_{\rm L})
\label{eq:momvla}
\end{equation}
where $\rho$ is density, the source term of the Poisson equation to
estimate the gravitational potential, given by
\begin{equation}
\rho({\bf r})=\int d^3 {\bf v}_{\rm L}
f({\bf r},{\bf v}_{\rm L}).
\label{eq:momvla2}
\end{equation}
Note that equation (\ref{eq:momvla}) applies only to real space, but we shall
come back to that below. 

Since we aim to estimate velocity and density on a finite resolution
grid, a more sensible way of performing the calculations is to
integrate equations (\ref{eq:momvla}) and (\ref{eq:momvla2}) on a
small cubic patch $\Delta {\bf r}^3$ corresponding to a grid element
to reduce noise as much as possible and to make the calculation
conservative (i.e., the total mass and momentum is conserved). In
practice, we do not perform this calculation exactly, but only
approximately using an approach inspired from smooth particle
hydrodynamics (SPH, see, e.g. Monaghan 1992), as we now describe in
details.

In the SPH approach, each particle is represented as a smooth cloud of
finite varying size $2R_{\rm SPH}$ depending on local density,
i.e. on the typical distance between the particle and its $N_{\rm
SPH}$ closest neighbors (which can be found quickly with e.g.
standard KD-tree algorithm), where $N_{\rm SPH}$ is usually of the
order of a few tens. As a result, smooth representation of the density
and velocity fields can be obtained at any point of space by summing
up locally the contribution of all the clouds associated to each
particle. With appropriate choice of the SPH kernel, these functions
can be easily integrated over each cell of the grid.

The problem with this approach is that it does not guarantee that local
reconstructed density is strictly positive everywhere, which can leave
regions where the velocity field is undefined.  To solve that problem,
we start from the grid points, which we assume to be virtual particles
for which we find the $N_{\rm SPH}$ closest neighbors to define the
SPH kernel associated to this grid site. However, one has to take into
account the fact that the sought estimates should roughly correspond
to an integral in a small cubic patch $\Delta {\bf r}^3$.  In
particular, all particles belonging to a grid site should participate
to such an integral.  This issue can be addressed in an approximate
way as follows:
\begin{enumerate}
\item Count and store for each grid site, the number $N_{\rm p}$ of
particles it contains;
\item If $N_{\rm p} \geq N_{\rm SPH}$, then perform SPH interpolation
at the grid site as explained below using the $N_{\rm p}$ particles
contributing to it instead of the $N_{\rm SPH}$ closest neighbors;
\item If $N_{\rm p} < N_{\rm SPH}$, then find for the grid site
the $N_{\rm SPH}$ closest particles and perform SPH interpolation
at the grid site as explained below.
\end{enumerate}

Given a choice of the SPH kernel, ${\cal S}(x)$ [which should be a
monotonic function verifying ${\cal S}(0)=1$ and ${\cal S}(x \ge
2)=0$], and a number $N_{\rm X}$ of neighbors, with $N_{\rm X}=N_{\rm p}$
or $N_{\rm SPH}$ according to the procedure described above, the
interpolation of a quantity $A$ on the grid site is given by
\begin{equation}
{\tilde A}(i,j,k)=\frac{1}{[R_{\rm SPH}(i,j,k)]^3} \left[
\sum_{l=1}^{N_{\rm X}-1} A_l W_l {\cal S}\left(\frac{d_l}{R_{\rm 
SPH}(i,j,k)}\right)\right] \! .
\label{eq:interp}
\end{equation}
In this equation, $A_l$ is the value of $A$ associated to each
particle $l$, $R_{\rm SPH}(i,j,k)$ is half the distance to the
furthest neighbor of the grid site $(i,j,k)$ among the $N_{\rm X}$,
$d_l$ is the distance of the $l$-th particle to the grid site, and
$W_l$ is a weight given to each particle such that the total
contribution of every particle to all the grid sites is exactly unity:
\begin{equation}
W_l=1/S_l=\left[ \sum_{i,j,k} \frac{1}{[R_{\rm SPH}(i,j,k)]^3} 
{\cal S}\left(\frac{d_l}{R_{\rm SPH}(i,j,k)}\right) \right]^{-1}\! .
\label{eq:poids}
\end{equation}
In practice, the interpolated density thus reads
\begin{equation}
{\tilde \rho}(i,j,k)={\tilde m}(i,j,k)/\Delta {\bf r}^3,
\end{equation}
by taking $A_l=m_l$ in equation (\ref{eq:interp}), where $m_l$ is the
mass of each particle. An interpolated velocity reads
\begin{equation}
{\tilde v}(i,j,k)={\tilde p}(i,j,k)/{\tilde m}(i,j,k),
\end{equation}
where ${\tilde p}(i,j,k)$ is the interpolated momentum in cell
$(i,j,k)$. It is obtained by taking $A_l=p_l\equiv m_l v_l$ in
equation (\ref{eq:interp}), where $v_l$ is the velocity (the radial
component, or a coordinate) of particle $l$.

Note that there can be some particles for which $S_l=0$ in equation
(\ref{eq:poids}) (in that case $W_l=0$ by definition), i.e. which do
not contribute at all to the interpolation. 
In the practical measurements described later, this
can happen at most only for a very small fraction of the particles,
typically of the order of 0.1 percent, and affects the results
insignificantly.  There is, however, another noticeable defect in our
method, due to the unsmooth transition between the two interpolation
schemes when the number of particles $N_{\rm p}$ per grid site becomes
larger than $N_{\rm SPH}$, that affects in a non trivial way the
interpolated density, but again, it does not have any significant
consequences for the present work.

In practice, we take $N_{\rm SPH}=32$ for all the measurements
described below and the following spline for the SPH kernel (Monaghan
1992)
\begin{eqnarray}
{\cal S}(x) \!\!\! & = & \!\!\! 1-1.5\ x + 0.75\ x^3, \quad 0 \leq x
\leq 1 \nonumber \\ 
{\cal S}(x) \!\!\! & = & \!\!\! 0.25\ (2-x)^3, \quad 1 \leq x \leq 2
\nonumber \\ 
{\cal S}(x) \!\!\! & = & \!\!\! 0, \quad x \geq 2.
\end{eqnarray}

Once the density field is interpolated on the grid, it is easy to
estimate the gravitational potential from it by solving the Poisson
equation in Fourier space, keeping in mind that the edge of the volume
$V$, where the velocity--gravity relation is tested, should be
sufficiently far away from the edges of the volume $V_{\rm grav}$,
which is itself included in the grid.  It would be beyond the scope of
this paper to discuss other problems related to incompleteness or edge
effects.  The main one is related to the uncertainties on the
gravitational potential induced by the obscuring due to our own
Galaxy. While this can be certainly an issue, other incompleteness
problems such as segregation in luminosity can be addressed by giving
the proper weight (or mass) to the galaxies in the catalog. This of
course needs strong assumption on the bias, and can work only for
populations of not too bright galaxies.

Note finally that, as explained before, we perform ``naively'' our SPH
interpolation whether we are working in real space or in redshift
space. The fact that we use SPH interpolation in redshift space is
sufficient for our purpose as long as we are in the DOL limit,
although it does not correspond anymore exactly to a simple moment of
the Vlasov equation. However, the interpolation becomes somewhat
questionable when the assumption of DOL is dropped: the nature of the
interpolation changes with distance from the observer. In particular,
if we assume that the smoothing kernel was fixed, projections,
e.g. passing from the velocity vector to a one dimensional quantity
such as the radial velocity, and smoothing do not commute anymore.
The same problem arises for commutation between the calculation of the
radial part of the gravitational force and smoothing.  It is therefore
necessary to carefully check that the simplistic nature of our
interpolation does not introduce any systematic bias in the
measurements, when performing them in redshift space relaxing the
approximation of a distant observer.

%
%
\subsection{The $N$-body simulation set}
%
%
\label{sec:nbodysamples}
We performed a high resolution simulation using the adaptive mesh
refinement (AMR) code RAMSES (Teyssier 2002).
As already mentioned earlier, the
cosmology considered here assumes  $\Omega=0.3$, $\Lambda=0.7$
 and $H_0=70 \equiv 100\, h$
$\rm km\cdot s^{-1}\cdot Mpc^{-1}$. Initial conditions were set up
using Zel'dovich approximation (Zel'dovich 1970) to perturb a set of
particles disposed on a regular grid pattern to generate initial
Gaussian fluctuations with a standard $\Lambda$CDM power spectrum.
To do that, we used
the COSMICS package of Berstchinger (1995). 
The simulation involves $512^3$ dark matter particles
on the AMR grid initially regular of size $512^3$, in a periodic cube
of size $200\ h^{-1}$ Mpc. Then, additional refinement is allowed
during runtime: if cells contain more than 40 particles they are
divided using standard AMR technique
(with a maximum of 7 levels of refinement). 
Note finally for completeness that the
normalization of the amplitude of initial fluctuations was chosen 
such that the
variance of the density fluctuations in a sphere of radius $8\ h^{-1}$
Mpc extrapolated linearly to the present time was given by
$\sigma_8=0.93$.

%
%
\subsection{Measurements on the dark matter distribution}
%
\label{sec:dmsamples}

From our RAMSES simulation, we extracted a number of dark-matter
samples, as described in details in Table~\ref{tab:table1}, namely:
\begin{description}
\item[(i)] A high resolution grid sample of size $N_{\rm g}=512$
for testing the velocity--gravity relation in real space;
\item[(ii)] The analog of (i) for testing the velocity--gravity
relation in redshift space, using the DOL approximation;
\item[(iii)] The analog of (ii), but without using the DOL, for
testing the validity of this approximation. In that case the radial
coordinate of the velocity and gravity field was estimated using an
observer at rest at the center of the simulation box. The comparison
of velocity to gravity was performed in a sphere of radius 70 $h^{-1}$ Mpc
to avoid edge effects due to
the loss of periodicity while projecting in redshift space;
\item[(iv)] A set of 125 low resolution grids ($N_{\rm g}=128$) in
redshift space, for estimating (at least partly) cosmic variance
effects and effects of structures nearby the observer.  These samples
were generated by locating the observer in the simulation box on a
regular grid of size $5 \times 5 \times 5$.  Again, in these samples,
the velocity--gravity relation was tested in a volume of radius 70 $h^{-1}$ Mpc
centered on the observer (exploiting the periodic nature of the
simulation box).
\end{description}
Since the catalogs we consider in point (iv) represent a significant
fraction of the simulation volume, we know however that effects of cosmic
variance are likely to be underestimated. 
\begin{table*}
\caption[]{Summary of the characteristics of the dark matter samples
used in this paper and the corresponding main quantitative
results. The first column gives the name of the sample as used in the
text. The second one  mentions whether the measurements
are performed in real or redshift space, the acronym DOL meaning that
the approximation of an infinitely remote observer was used. The third
column gives the number $N_{\rm rea}$ of realizations considered. The
fourth column gives the size of the sample, which is the full periodic
simulation cube of size $L$ when the letter $L$ is used and a sphere
of radius $R$ when the letter $R$ is used. Scales are expressed in
units of $h^{-1}$ Mpc. The fifth column gives the resolution $N_{\rm
g}$ of the rectangular grid (matching the simulation box) used to
sample various fields from the particle distribution.  The sixth
column gives the number $N_{\rm obj}$ of objects that are actually
contained in the sample. A star means that this number is only
approximate, corresponding to a simple rescaling of the total number
of objects in the periodic cube by taking into account the size of the
sampled volume, a sphere of radius $R$. The 7th, 8th and 9th columns
give the measured $\Omega$ using $\langle g_i^2 \rangle/\langle v_i
g_i \rangle$ (1st number), $\sqrt{\langle g_i^2 \rangle/\langle v_i^2
\rangle}$ (2nd number) and $\langle v_i g_i \rangle / \langle v_i^2
\rangle$ (3rd number) as estimators of the velocity--gravity relation,
where $i=s$ if the measurements are performed in redshift space and
$i=z$ if the measurements are performed in real space or distant
observer redshift space. In principle, all these estimators should
give the same answer (nearly the same answer in redshift space) in the
ideal case where the linear theory applies, and various sources of
noise (such as shot noise, cosmic variance, systematics, etc.) are
negligible. The disagreement between these various estimates can be
used as a proxy to estimate error bars on actual measurements. Where
possible, an error bar obtained from the dispersion among various
realizations is quoted as well.  The last column explains what exactly
the quoted numbers correspond to, in particular if the fields were
smoothed with a Gaussian window of size $\ell=10\ h^{-1}$ Mpc
(``smoothed'') or not (``no smoothing''), if all the PDF was used to
perform the measurements (``all'') or only the most likely region
defined by Equation~(\ref{eq:con68}) (``1.5$\sigma$ isocontour'').  The
symbol $\langle\cdots\rangle$ means that an average was performed over
the number of available realizations, when relevant.}
\begin{tabular}{llclcccccl}
\hline
Samp. & Content & $N_{\rm rea}$      & Size & $N_{\rm g}$ & $N_{\rm obj}$ &  \multicolumn{3}{c}{Measured $\Omega$ from:} & Comment \\
       &         &                    &      &             &               &  $\frac{\langle g^2 \rangle}{\langle v g \rangle}$ &
                                                                              $\sqrt{\frac{\langle g^2 \rangle}{\langle v^2 \rangle}}$ &
                                                                              $\frac{\langle v g \rangle}{\langle v^2 \rangle}$ & \\ 
\hline \hline
(i)    & real space                                      & 1        & $L=200$ & 512 & $512^3$                  & 0.103 & 0.157 & 0.240 & all, no smoothing \\
       &                                                         &          &         &     &                          & 0.251 & 0.257 & 0.264 & all, smoothed \\
       &                                                         &          &         &     &                          & 0.299 & 0.303 & 0.307 & 1.5$\sigma$ 
                                                                                                                                                 isocontour,
                                                                                                                                                 no smoothing \\
       &                                                         &          &         &     &                          & 0.295 & 0.297 & 0.299 & 1.5$\sigma$ 
                                                                                                                                                 isocontour,
                                                                                                                                                 smoothed\\
(ii)   & DOL redshift space                              & 1        & $L=200$ & 512 & $512^3$                  & 0.143 & 0.280 & 0.578 & all, no smoothing \\
       &                                                         &          &         &     &                          & 0.206 & 0.223 & 0.242 & all, smoothed \\ 
       &                                                         &          &         &     &                          & 0.296 & 0.303 & 0.311 & 1.5$\sigma$ 
                                                                                                                                                 isocontour,
                                                                                                                                                 no smoothing \\ 
       &                                                         &          &         &     &                          & 0.274 & 0.279 & 0.284 & 1.5$\sigma$ 
                                                                                                                                                 isocontour,
                                                                                                                                                 smoothed\\
(iii)  & redshift space                                  & 1        & $R=70$  & 512 & ${2.41{\rm e}7}^{*}$ & 0.098 & 0.287 & 0.987 & all, no smoothing \\ 
       &                                                         &          &         &     &                          & 0.167 & 0.197 & 0.231 & all, smoothed \\
       &                                                         &          &         &     &                          & 0.316 & 0.332 & 0.349 & 1.5$\sigma$ 
                                                                                                                                                 isocontour,
                                                                                                                                                 no smoothing \\
       &                                                         &          &         &     &                          & 0.269 & 0.284 & 0.300 & 1.5$\sigma$ 
                                                                                                                                                 isocontour,
                                                                                                                                                 smoothed\\
(iv)   & redshift space                                  & 125      & $R=70$  & 128 & ${2.41{\rm e}7}^{*}$ & 0.115 & 0.280 & 0.817 & $\langle$ all $\rangle$, no smoothing \\
       &                                                         &          &         &     &                          & 0.023 & 0.030 & 0.269 & error from dispersion \\
       &                                                         &          &         &     &                          & 0.177 & 0.200 & 0.226 & $\langle$ all $\rangle$, smoothed \\
       &                                                         &          &         &     &                          & 0.022 & 0.021 & 0.023 & error from dispersion \\ 
       &                                                         &          &         &     &                          & 0.283 & 0.294 & 0.304 & $\langle$ 1.5$\sigma$ 
                                                                                                                                                 isocontours $\rangle$,
                                                                                                                                                 no smoothing\\
       &                                                         &          &         &     &                          & 0.039 & 0.040 & 0.041 & error from dispersion \\ 
       &                                                         &          &         &     &                          & 0.255 & 0.262 & 0.270 & $\langle$ 1.5$\sigma$ 
                                                                                                                                                 isocontours $\rangle$,
                                                                                                                                                 smoothed\\
       &                                                         &          &         &     &                          & 0.031 & 0.032 & 0.034 & error from dispersion \\
\hline
\end{tabular}
\label{tab:table1}
\end{table*}

We now perform a visual inspection of the fields, followed by
measurements of the joint PDF of velocity and gravity, first in real
space, then in redshift space.

\subsubsection{Visual inspection}
\label{sec:visu}
\begin{figure*}
\centerline{\hbox{
\psfig{file=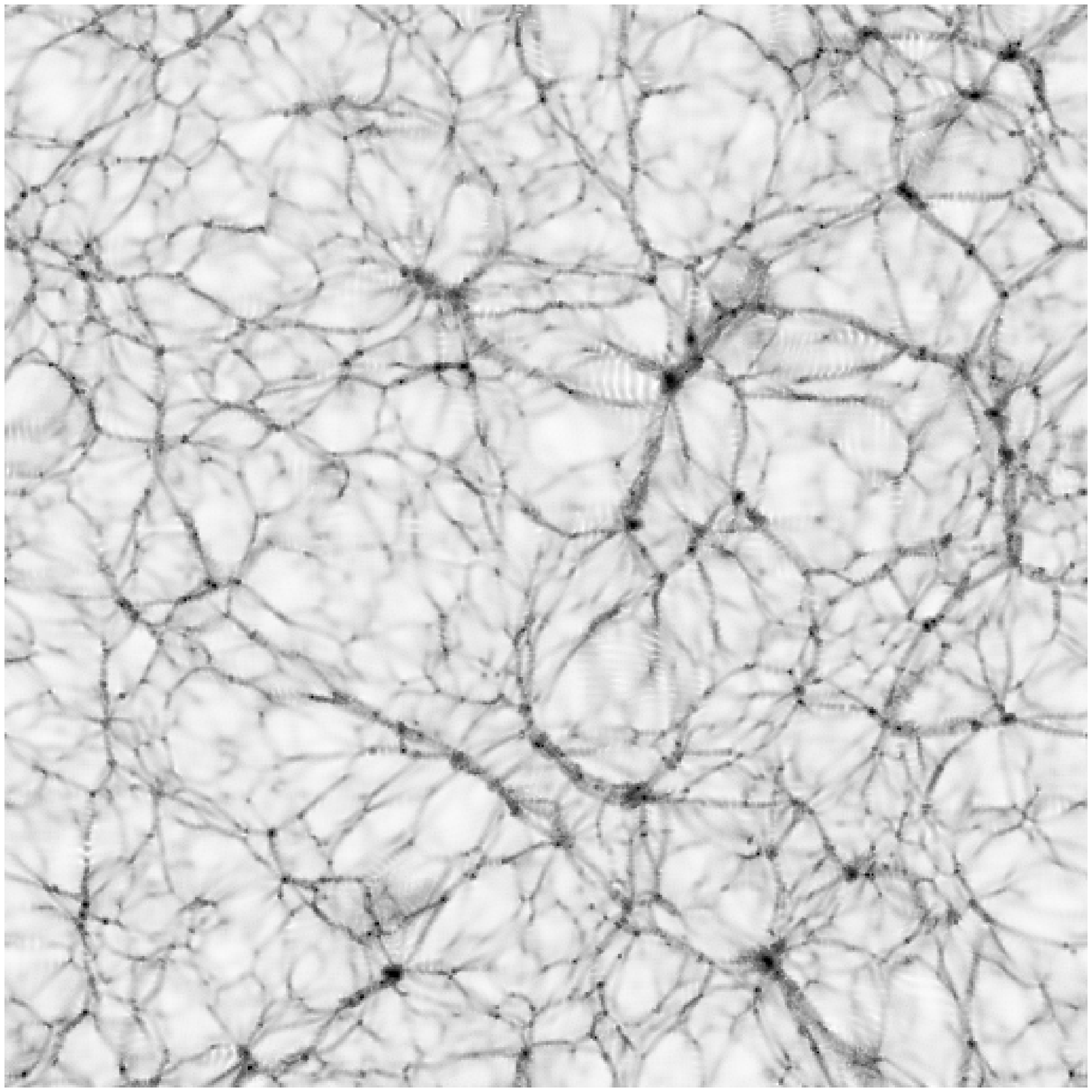,width=7cm}
\hskip 0.2cm
\psfig{file=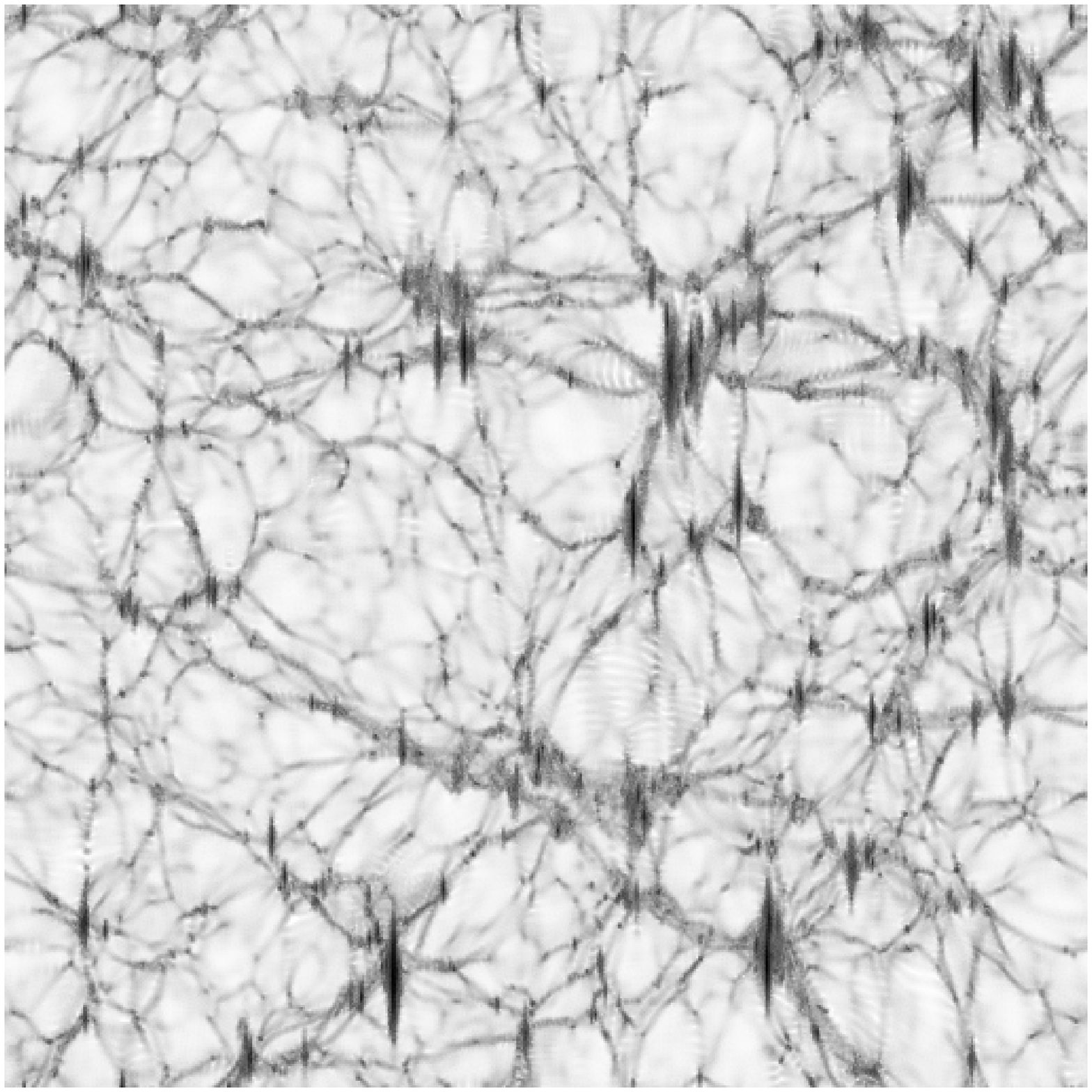,width=7cm}
}}
\vskip 0.2cm
\centerline{\hbox{
\psfig{file=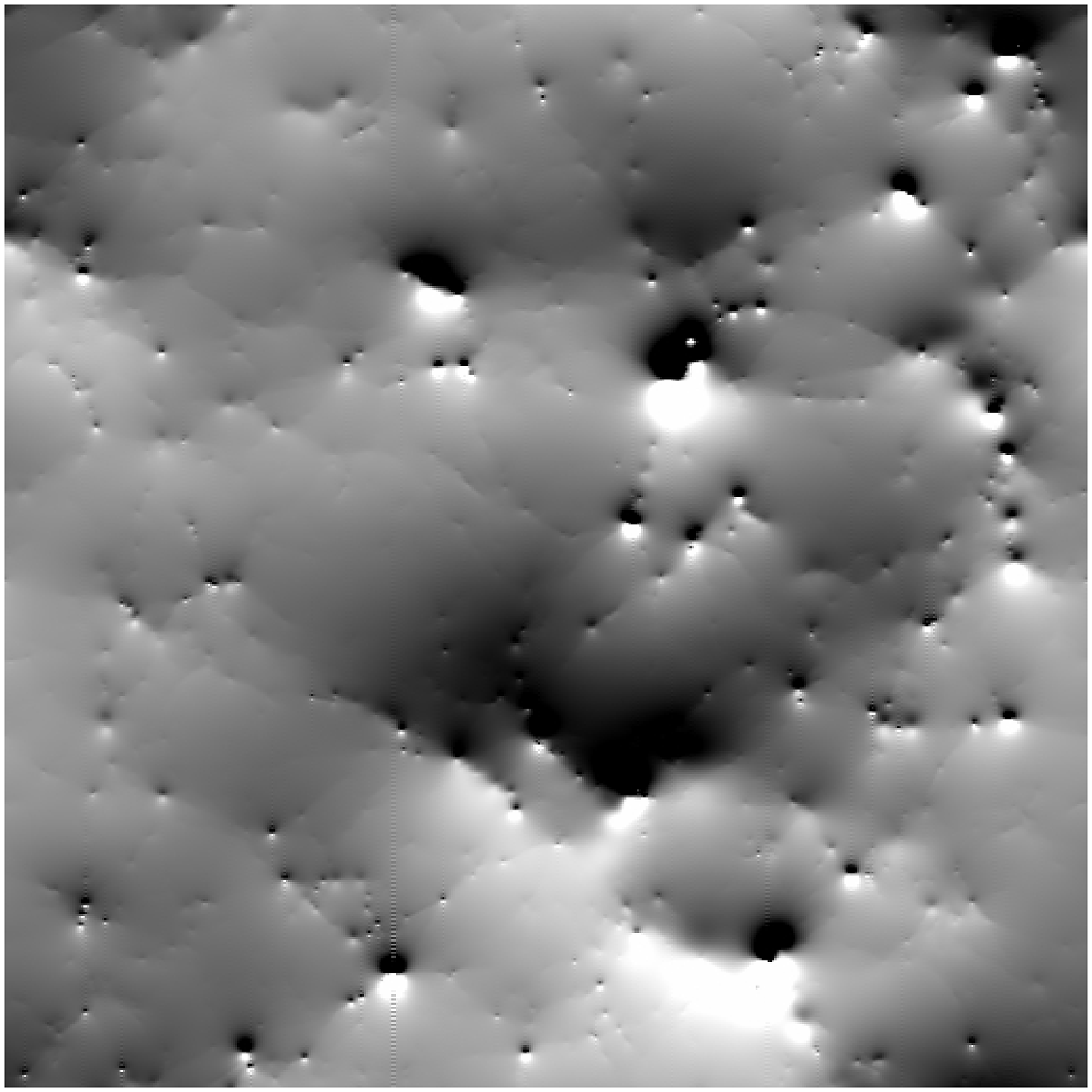,width=7cm}
\hskip 0.2cm
\psfig{file=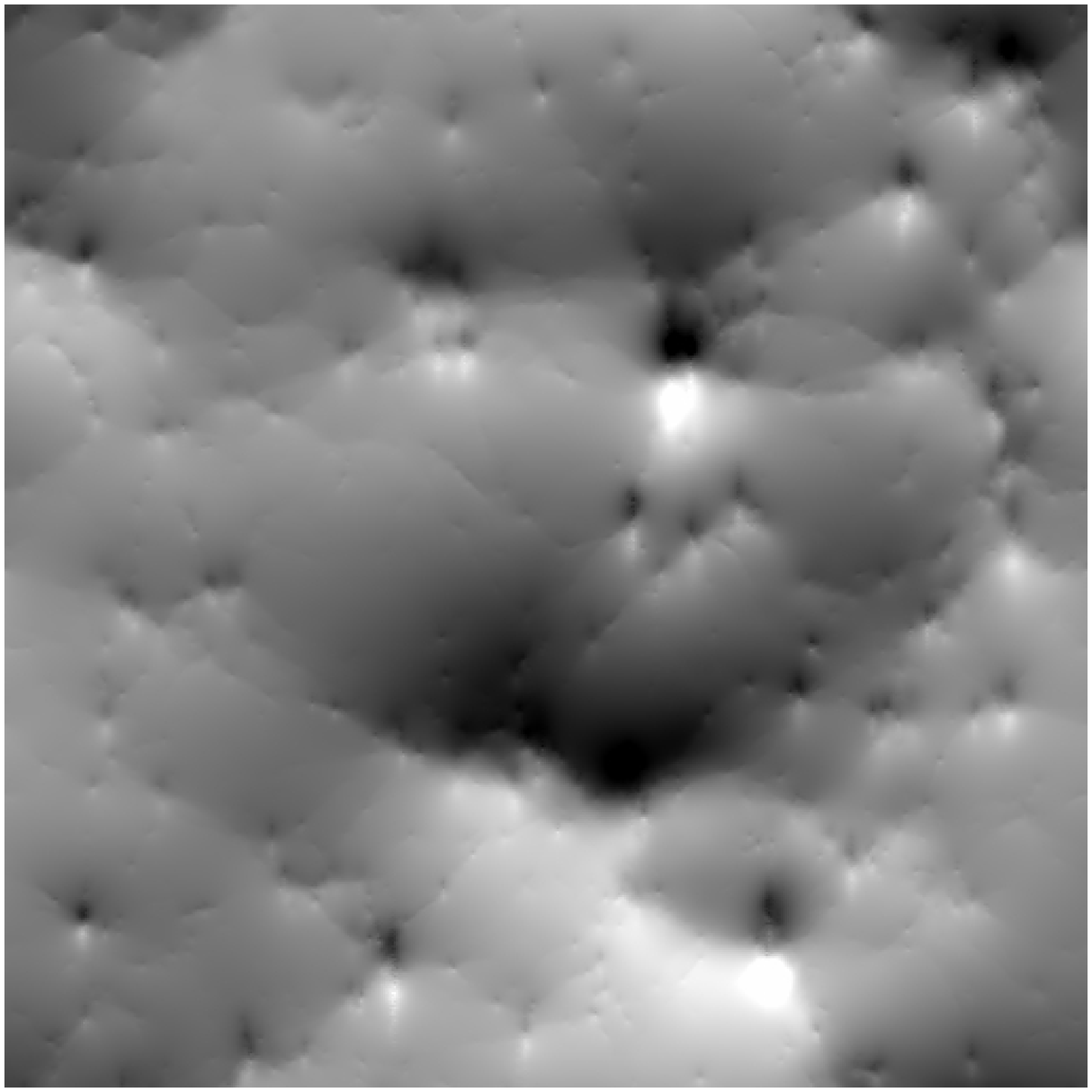,width=7cm}
}}
\vskip 0.2cm
\centerline{\hbox{
\psfig{file=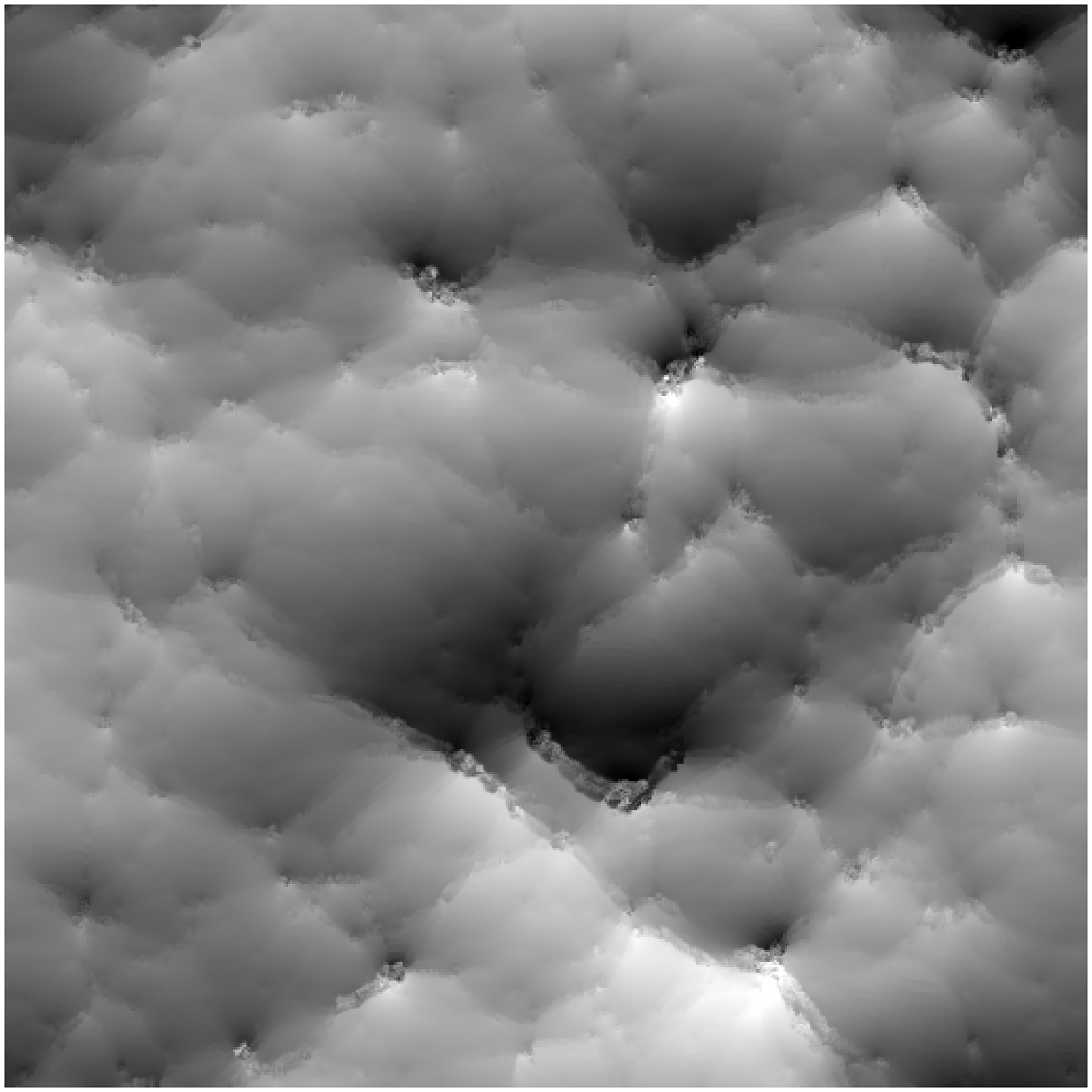,width=7cm}
\hskip 0.2cm
\psfig{file=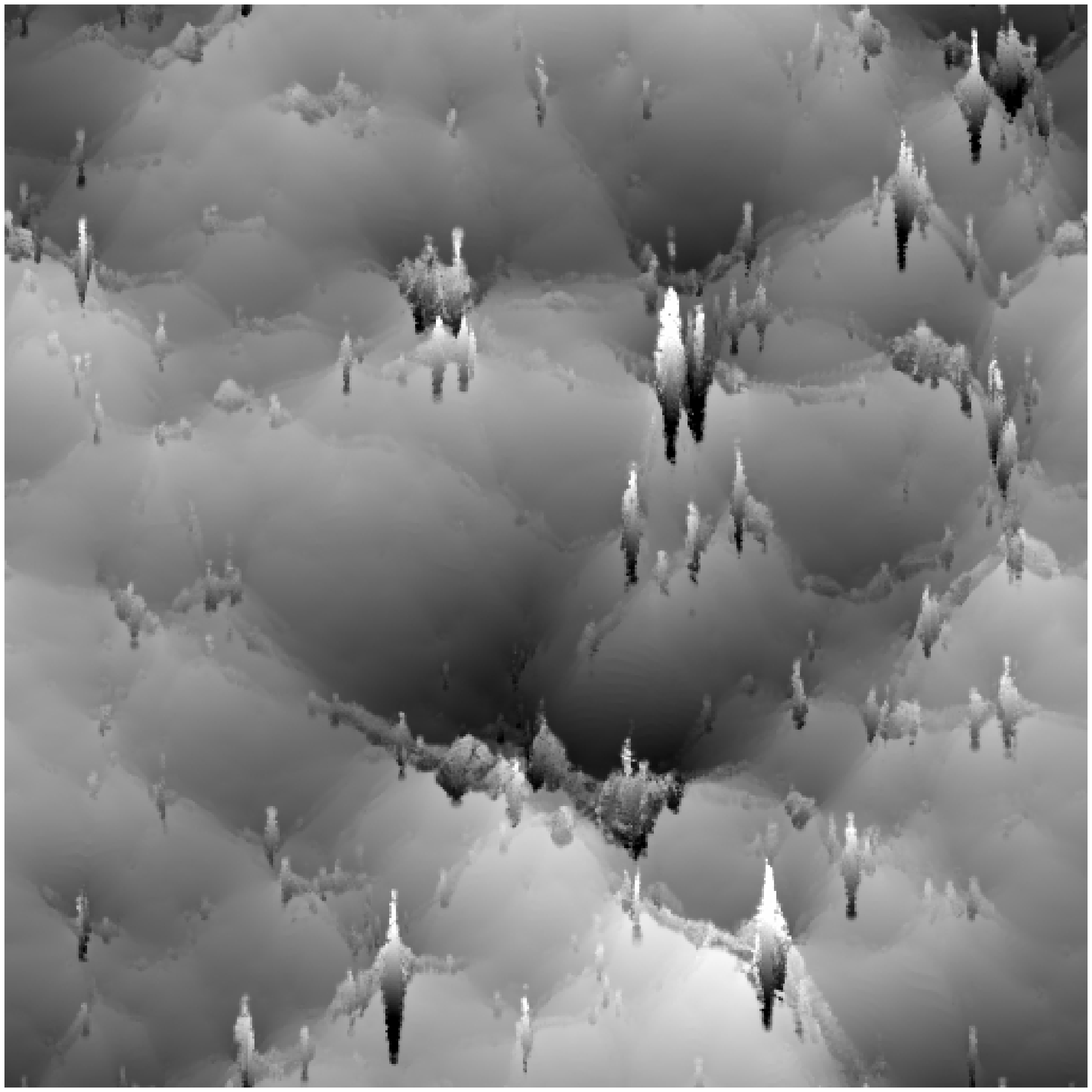,width=7cm}
}}
\caption[]{The density (upper panels), gravity (middle panels) and
velocity field (lower panels) in a one-mesh-element-thick slice
extracted from the samples (i) and (ii), which correspond to a
grid resolution of $512^3$. The left and right columns of panels
correspond respectively to real space and distant observer redshift
space. In the top panels, the color scale is logarithmic, from white
to black. In the four bottom panels, the $z$ coordinate of each field
is displayed. The color scale is linear and normalized in such a way
that direct visual comparison between the four panels can be
performed.  One can notice clearly `fingers of God' on the upper right
and lower right panels. In such `fingers of God', the velocity field
tends to have a different signature compared to that in the nearby
environment, as explained in the text.  Finger-of-God effects also
tend to soften the gravitational potential at small scales (cf.~
middle panels).}
\label{fig:images1}
\end{figure*}

Figure~\ref{fig:images1} shows the density, the gravity field and the
\begin{figure*}
\centerline{\hbox{
\psfig{file=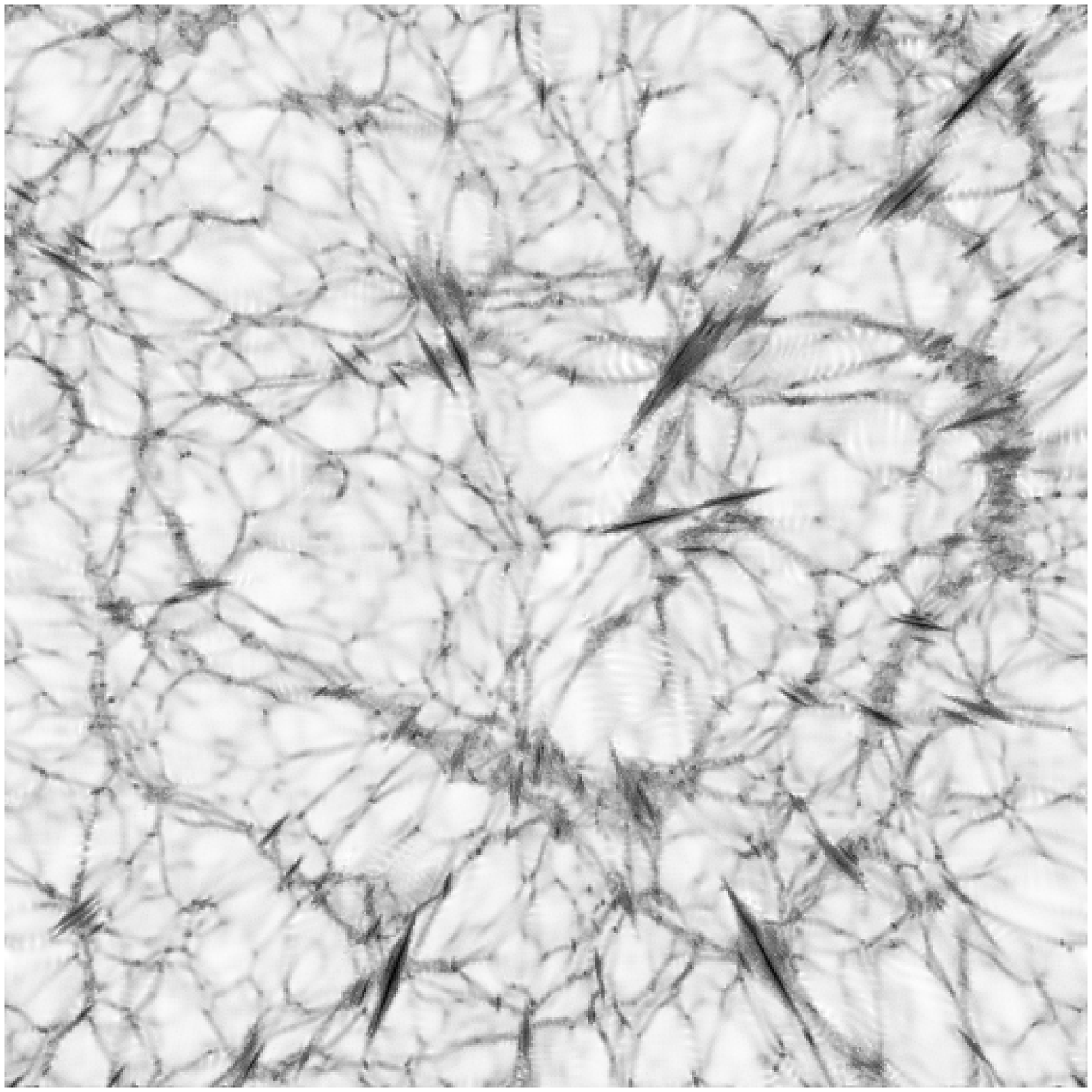,width=7cm}
}}
\vskip 0.2cm
\centerline{\hbox{
\psfig{file=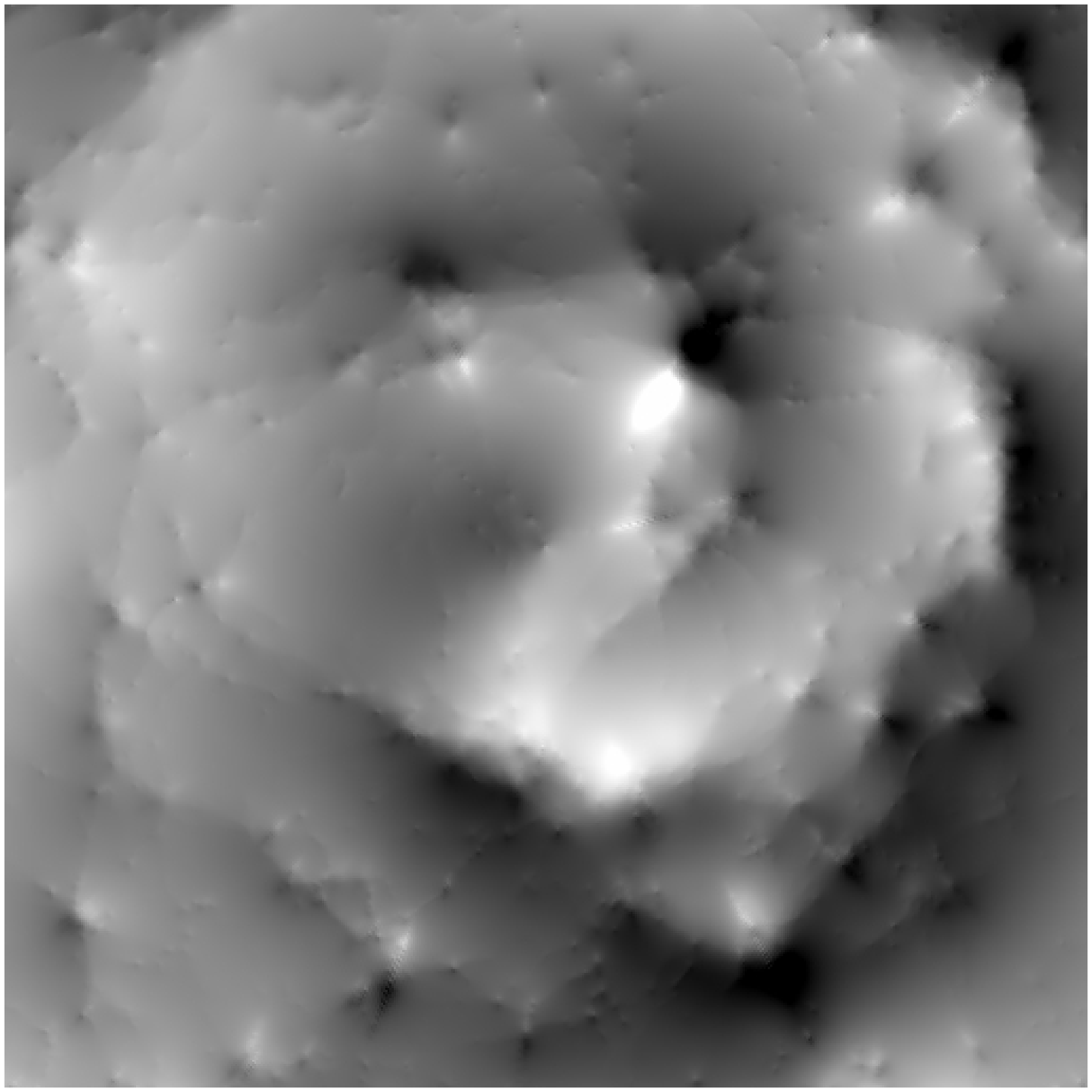,width=7cm}
}}
\vskip 0.2cm
\centerline{\hbox{
\psfig{file=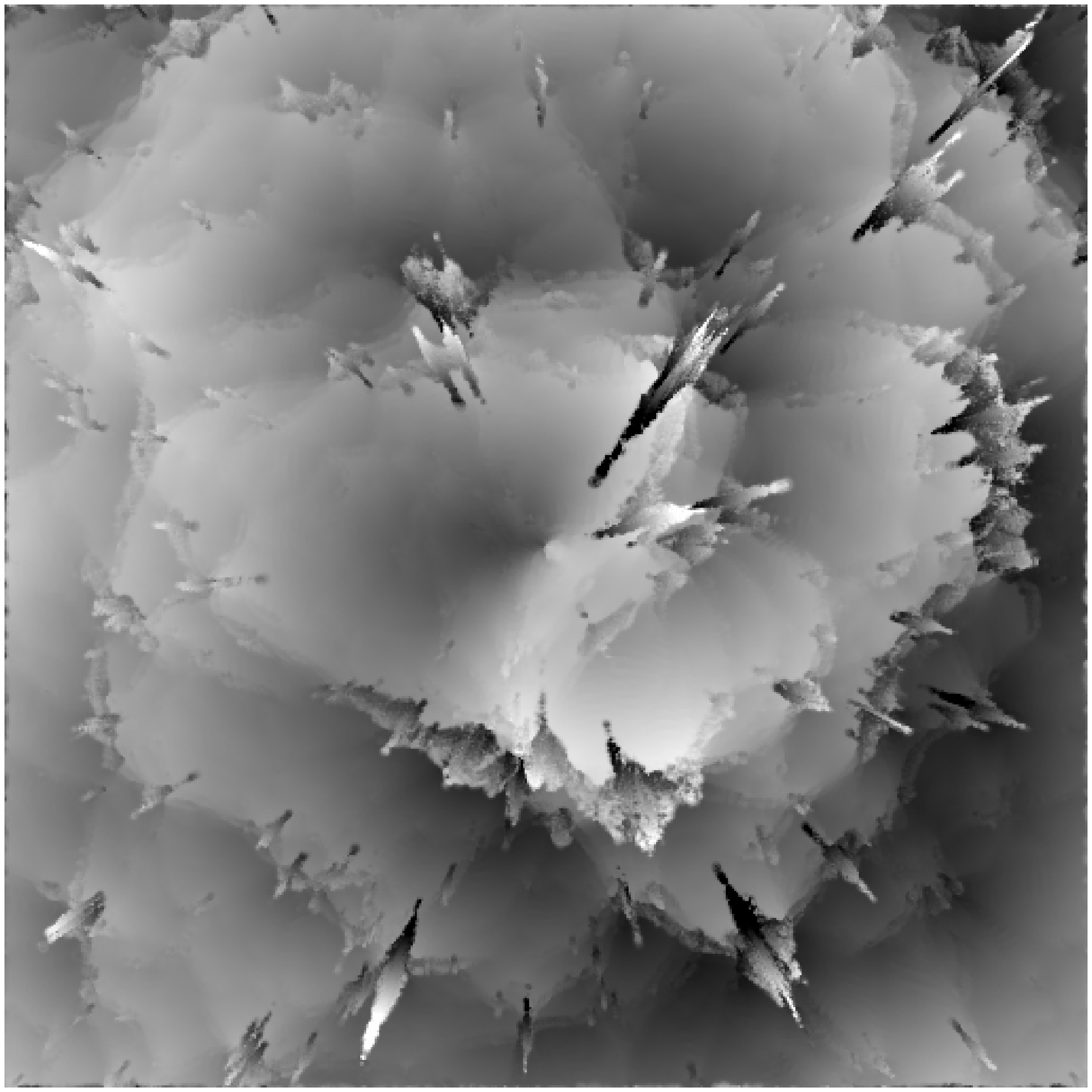,width=7cm}
}}
\caption[]{The same as in the right column of panels of
Fig.~\ref{fig:images1}, but for the sample (iii), where the assumption
of a distant observer is dropped. The two lower panels thus show the
radial component of the gravity and the velocity fields.  The observer
is at the center of the images, towards where `fingers of God' point.}
\label{fig:images2}
\end{figure*}
velocity field in a thin slice extracted from the simulation,
both in real space and in the DOL redshift space. Despite our
interpolation procedure, there are some minor discreteness artifacts
left, visible on the density field. One can notice for example a few
underdense regions where the initial grid pattern, distorted by large
scale dynamics, is still present. Such artifacts do not show up on the
smoother $z$ coordinate of the gravity field. On the other hand, this
latter seems to suffer from a few aliasing defects in real space, as
vertical spurious lines in the vicinity of deep potential wells. We
did not bother trying to understand such aliasing effects, because
they have negligible impact on the measurements.\footnote{These
aliasing effects might be related to some minor defect in the
particular Fourier transform algorithm we are using (Teuler 1999),
but this hypothesis seems to be contradicted by some accuracy tests
performed on this algorithm (Chergui 2000). A more sensible
explanation is that these effects are induced by the way we
interpolate the density field, in particular by the transition between
the $N_{\rm p} \geq N_{\rm SPH}$ and the $N_{\rm p} < N_{\rm SPH}$
regimes, combined with the fact that our Fourier Green function for
computing the gravitational acceleration is simply proportional to
${\bf k}/k^2$, without additional filtering.}  Interesting to
notice here is the nice agreement between the $z$ coordinate of the
velocity field and the $z$ coordinate of the gravity field in
underdense regions. These regions dominate the velocity--gravity
statistics, since we have a volume weighted approach.  Note the
particular features in the velocity field associated to the filaments
in the density distribution, as well as the `fingers of God' (FOG) in
redshift space.  These FOG, which are mainly associated to
the dark matter halos, are expected to induce some particular
properties on redshift space statistics:
\begin{itemize}
\item First, and this is a straightforward consequence of passing from
real space to redshift space, there is, on the velocity field, an
`inversion' effect inside FOG. In other words, inside a finger of God,
the variations of the velocity field are opposite to what happens in
its nearby environment. This can be explained in the following
way. Inside a halo, which can be, in real space, considered as a
point-like structure in first approximation, there are particles with
positive velocities and particles with negative velocities. Assume, to
simplify, that this halo has itself a zero center of mass velocity. It
will, in redshift space, look like an elongated structure. Particles
in this structure that have positive $z$ velocity will be above the
center of mass, and particles with negative velocity will be below the
center of mass. As a result, one expects, for the $z$ coordinate of
the velocity, the finger of God corresponding to this halo to have
positive velocities (towards light color on bottom right panel of
Fig.~\ref{fig:images1}) above the center of the halo and to have
negative velocities (towards dark color) below the center of the halo.
This is indeed what we can observe. However, in the nearby environment
of the halo, the situation is somewhat opposite. Indeed, the halo is
expected to lie in a filament, which itself represents a local
potential well. If this well is not too strong to induce
shell-crossing while passing from real to redshift space, the general
trend of the velocity field is not modified, compared to real space:
negative sign (towards dark color) above the filament, positive sign
below it. If the filament corresponds to a sufficiently deep potential
well such that shell crossing occurs, then the same effect than for
FOG is expected, as can be noticed on the bottom right panel of
Fig.~\ref{fig:images1} for the largest filaments.

\item Second, due to the FOG stretching, halos are more elongated in
redshift space, and a natural consequence is that the corresponding
potential well is less deep. That is why the $z$ coordinate of the
gravitational force is less contrasted in the middle right panel of
Fig.~\ref{fig:images1} than in the middle left one.
\end{itemize}
Still, we can see that both in real and in redshift space, the
velocity--gravity relation is going to be in quite good agreement with
expectation from linear theory. Indeed and again, our measurements are
volume weighted, dominated by underdense regions. The color scale on
four bottom panels of Fig.~\ref{fig:images1} has been chosen such that
if linear theory applies, the same color pattern should be found for
the gravity and velocity field, which is the case at first
glimpse, except in the densest regions, corresponding to halos
or rich filaments. 

Finally, similarly as the right column of panels of Fig.~\ref{fig:images1},
Fig.~\ref{fig:images2} displays the density, the gravity field and the velocity
field in redshift space, but without using the DOL. If one now takes 
into account the radial nature of the projection, it is clear that the 
conclusions of the previous discussion remain unchanged, at least
at the qualitative level.

\subsubsection{Quantitative comparison in real space}
\label{sub:real}
\begin{figure*}
\centerline{\hbox{
\psfig{file=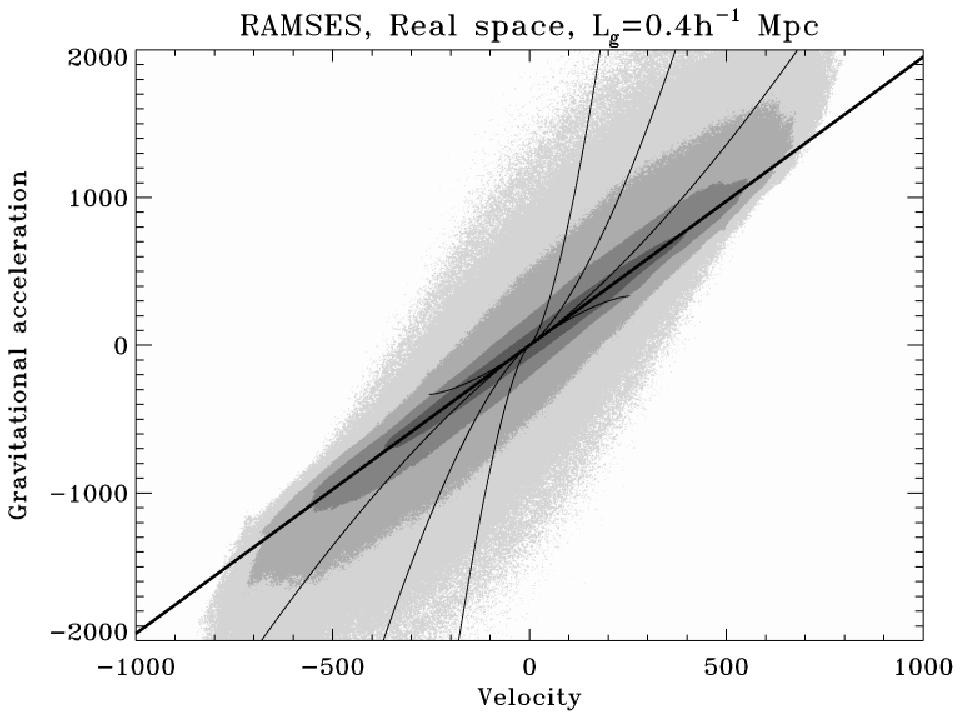,width=8cm}
}}
\centerline{\hbox{
\psfig{file=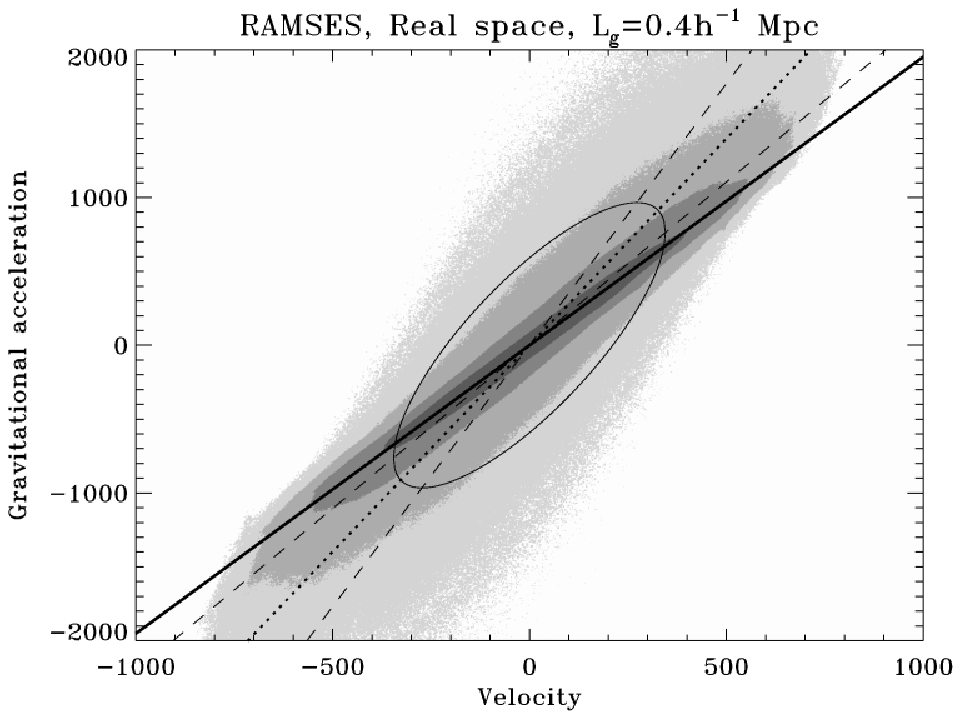,width=8cm}
}}
\centerline{\hbox{
\psfig{file=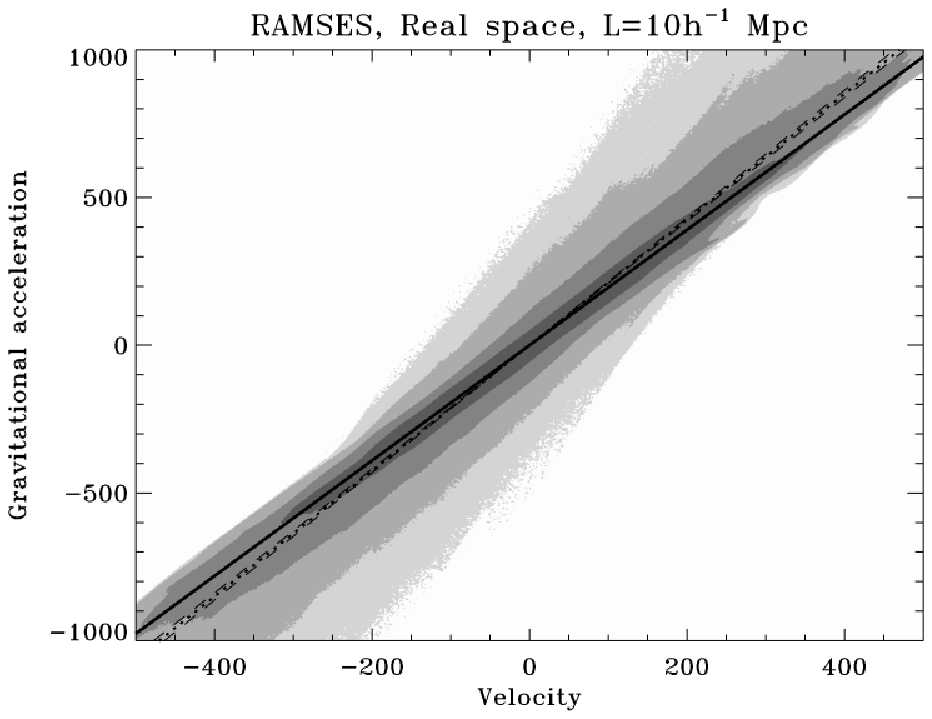,width=8cm}
}}
\caption[]{The velocity--gravity relation in real space as a scatter
plot, compared to theoretical prediction in the linear regime (thick
diagonal line), as measured from sample (i) extracted from the 
simulation using the $z$ coordinate of each field. The color scale
is defined as follows: the darkest area is encompassed by a 68 percent isocontour 
(eq. \ref{eq:con68}),
corresponding approximately to a 1.5$\sigma$ contour in the Gaussian case.
Then, lower isocontours, corresponding to lighter color, 
scale logarithmically, which emphasizes the tails of the PDF.

The two upper panels correspond to the raw data interpolated on the
grid with the SPH-like interpolation detailed in \S~\ref{sec:algomes},
while in the bottom panel, an additional Gaussian smoothing with
kernel of a fixed size $\ell=10 h^{-1}$ Mpc was performed. The
smoothing scale is indicated on three panels (in the two top panels,
it corresponds to the mesh element size). In the top panel,
predictions of velocity--gravity relation according to the spherical
top hat model are displayed as solid curves. As discussed in detail in
the text, these curves help to understand why the linear prediction
still represents the most prominent feature of the relation, even in
the highly nonlinear regime, while the tails due to nonlinear dynamics
induce a `propeller' shape of the scatter plot, as already noticed by
Cieciel\c{a}g et al. (2003). Such a propeller shape results in a bias
on the measurement of the slope by using directly second-order moments
of the bivariate PDF. This is demonstrated by plotting on middle (and
bottom) panel the dotted line which gives the estimated slope from the
ratio $\sqrt{\langle g_z^2 \rangle/\langle v_z^2 \rangle}$. This
dotted line should be compared to the thick solid line. The dashed
lines correspond to slopes obtained by the conditional averages
$\langle g_z^2 \rangle/\langle g_z v_z \rangle$ and $\langle g_z v_z
\rangle/\langle v_z^2 \rangle$. In the middle panel these slopes are
significantly different, consistent with significant amount of the
scatter in the relation, due to nonlinear effects. In the middle panel,
also the $1.5 \sigma$ contour expected in the Gaussian case is shown.}
\label{fig:gv}
\end{figure*}

Figure~\ref{fig:gv} shows in grey scale the measured joint probability
distribution function (PDF) of the $z$ coordinate of gravity and
velocity fields extracted from the real space sample (i).  The
striking result is that the regions of the best likelihood (larger
values of the PDF, darker places) match very well the prediction given
by linear theory (thick solid line), even in the highly non-linear
regime (upper panel).  This remarkable property is mainly related to
our volume-weighted approach: results are mainly influenced by
underdense regions, which are weakly evolved from the dynamical point
of view and are expected to match well linear theory
predictions. This, and the `propeller' shape of the bivariate PDF, can
be understood in more detail by using the spherical top-hat model as a
proxy of nonlinear dynamics.  In that case, up to shell crossing, the
velocity--gravity relation reads approximately 
\begin{equation}
  g \simeq \frac{R}{3}\left[ 1-\left( 1 - 2 \frac{v}{\beta R}
  \right)^{3/2} \right]
  \label{eq:sphe}
\end{equation}
(this can be easily derived from Bernardeau 1992; 1994), where $R$ is
the distance from the center of a top-hat fluctuation (in $\rm km
\cdot s^{-1}$). This equation is valid inside the fluctuation, which
can be overdense (negative $v$) or underdense (positive $v$).  One can
then imagine, to simplify, the density field as a patchwork of
spherical fluctuations, which correspond to a set of curves given by
Equation~(\ref{eq:sphe}), as shown in the top panel of
Fig.~\ref{fig:gv}. Note here that we should, for the picture to be
correct, take into account the fact that we are using only the $z$
coordinate of the fields, $g_z=g \cos\theta$ and $v_z=v \cos\theta$,
where $\theta$ is the angle between the $z$-axis and the radial vector
${\bf R}$.  In the top panel of Fig.~\ref{fig:gv}, we consider the
cases $R \cos\theta=\pm 10,\ \pm 100,\ \pm1000, \ \pm 3000$ km/s. 
Each curve has a stopping point corresponding to the
maximum possible value of $v$, $v_{\rm max}= \beta R/2$, which
reflects the fact that there is an expected upper bound for the
expansion speed of voids. This property excludes the upper left and
lower right quadrants of the velocity--gravity diagram to be populated
too far from linear theory prediction. On the other hand, overdense
fluctuations tend to populate the upper right and the lower left
quadrants, above and below linear theory prediction,
respectively. Furthermore, since the low-$v$ regime converges to
linear theory, all the curves corresponding to
Equation~(\ref{eq:sphe}) superpose in that regime, creating a
`caustic' of best likelihood nearby the maximum of the joint PDF,
explaining the very good agreement with linear expectation in that
region. As a result, we now understand, thanks to the spherical
top-hat model, both the `propeller' shape of the bivariate PDF, as
well as the remarkable agreement with linear theory prediction nearby
its maximum, even in the highly non-linear regime (see also
Cieciel\c{a}g et al., 2003).  The arguments developed here are
oversimplified, but capture the main features of the dynamics of the
large scale structures prior to shell crossing in real space. Beyond
shell crossing, there is a mixing effect that tends to decorrelate
velocity from gravity, implying a widening of the bivariate PDF and
even larger tails in the upper right and the lower left
quadrants. However such an effect does not affect significantly the
region of best likelihood.

A straightforward consequence of the above discussion is that the joint
PDF of gravity and velocity is substantially non Gaussian due to
non-linear contributions in the dynamics, as supported by the
examination on middle panel of Fig.~\ref{fig:gv} 
of the 1.5$\sigma$ elliptic isocontour of the
Gaussian distribution with the same second order moments as the
measured PDF.  As a result, the direct measurement of $\beta$
parameter from the moments of the joint PDF using linear theory
predictions is biased to lower values, due the propeller shape of the
PDF.  This is illustrated on middle panel of Fig.~\ref{fig:gv} by the dotted line,
which gives the velocity--gravity relation obtained from the second
order moments of the PDF, $\sqrt{ \langle g_z^2 \rangle/\langle v_z^2
\rangle }$, while the two dashed lines correspond to the conditional
averages $\langle g_z^2 \rangle/\langle g_z v_z \rangle$ and $\langle
g_z v_z \rangle/\langle v_z^2 \rangle$. In the linear regime, let us
remind from \S~\ref{sec:calcul} that these three quantities should be
equal.
Due to non-linear effects, the relation between gravity and velocity
is not deterministic anymore, $\langle g_z^2 \rangle/\langle g_z v_z
\rangle > \sqrt{ \langle g_z^2 \rangle/\langle v_z^2 \rangle } >
\langle g_z v_z \rangle/\langle v_z^2 \rangle $, and $\beta$ is
underestimated, leading to an effective bias or an underestimated
value of $\Omega$, as shown in Table~\ref{tab:table1}. Note, as
expected, that additional smoothing of the fields with a Gaussian
window of radius 10 $h^{-1}$ Mpc helps to reduce non Gaussian features as well as this effective
bias, and makes the overall relation between gravity and velocity
tighter.

However, Fig.~\ref{fig:gv} appeals for a more sophisticated way to
measure the velocity--gravity relation than simply using directly the
moments of the joint PDF. Since the region around the maximum seems to
agree rather well with the linear prediction, this suggests to
estimate the moments from the PDF only within that region. To perform
such an exercise, we selected the 68-percent PDF isocontour, ${\cal
P}_{68}$, such that
\begin{equation}
\int_{{\cal P}(g,v) > {\cal P}_{68}} dg\ dv\ {\cal P}(g,v)=0.68.
\label{eq:con68}
\end{equation}
This corresponds roughly to a $1.5 \sigma$ contour in the Gaussian
case. The PDF is then set to zero outside the contour of best
likelihood. From this truncated PDF, the moments are estimated again,
leading to a much better estimate of $\Omega$, agreeing at the few
percent level with the true value, as shown in Table~\ref{tab:table1},
regardless whether additional Gaussian smoothing with a $10\ h^{-1}$
Mpc size window is performed on the fields, or not. Furthermore, the
estimators $\sqrt{ \langle g_z^2 \rangle/\langle v_z^2 \rangle }$,
$\langle g_z^2 \rangle/\langle g_z v_z \rangle$ and $\langle g_z v_z
\rangle/\langle v_z^2 \rangle$ differ now only slightly from each
other, reflecting the narrowness of the region of best likelihood
around the linear expectation.

\subsubsection{Quantitative comparison in redshift space}
\label{sub:red}
\begin{figure*}
\centerline{\hbox{
\psfig{file=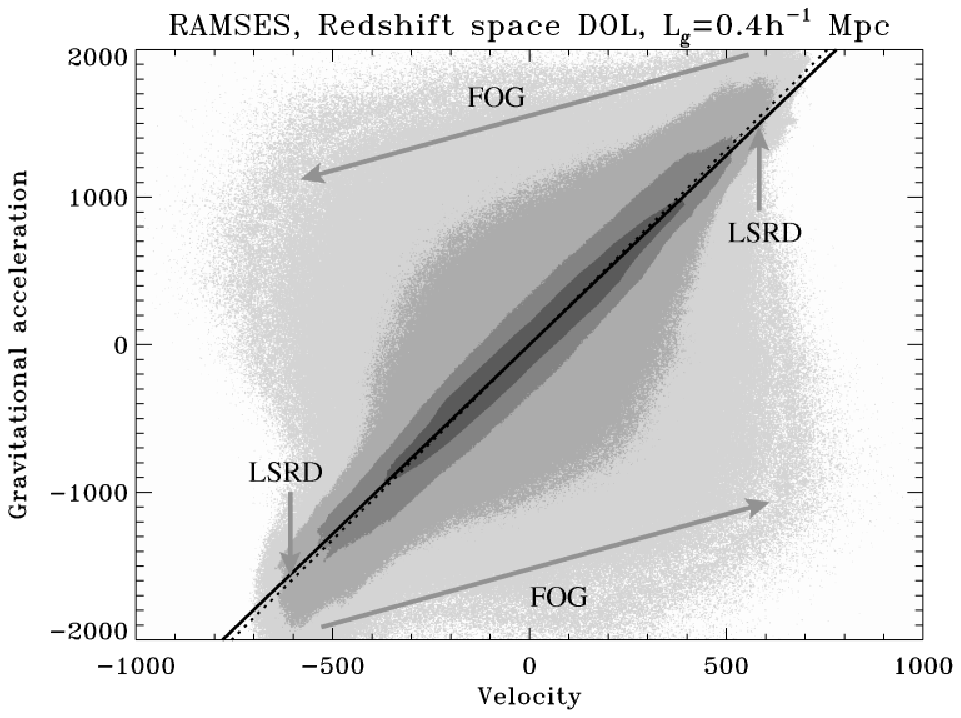,width=8cm}
\psfig{file=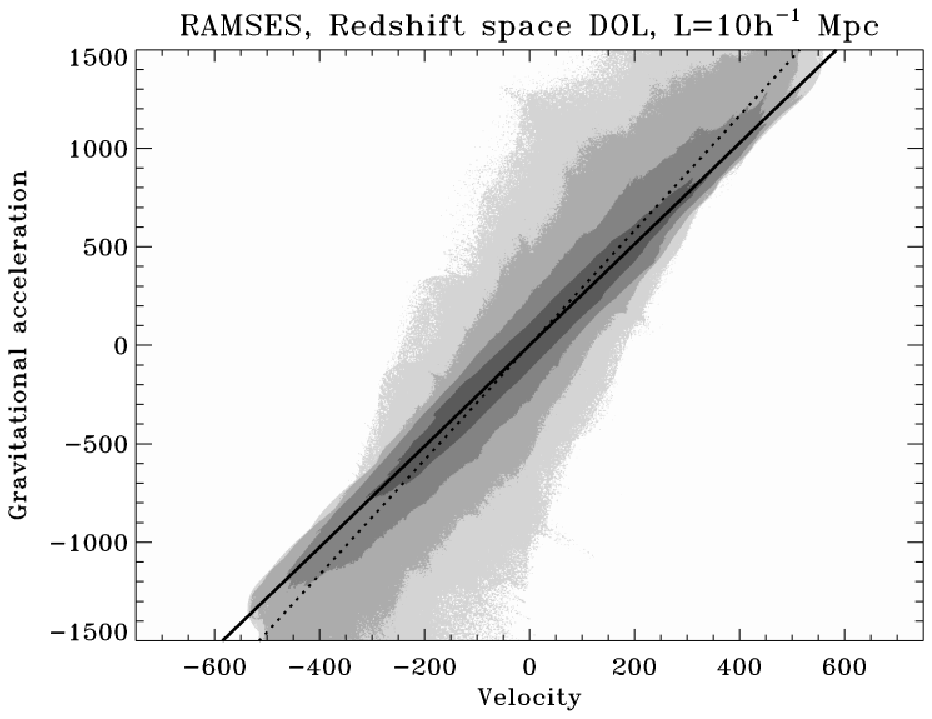,width=8cm}
}}
\caption[]{The velocity--gravity relation in distant observer redshift
space as measured from sample (ii) extracted from the
simulation, using the $z$ coordinate of the fields. The color
scaling is the same as in Fig.~\ref{fig:gv}. Again the
match between the most likely part of the bivariate PDF (darkest region) and the
prediction from linear theory (thick solid line) is close to perfect.
The dotted line gives the slope obtained directly from the ratio
$\sqrt{\langle g_z^2 \rangle/\langle v_z^2 \rangle}$.
  The left panel which corresponds to
minimum amount of smoothing can be compared to the two upper panels of
Fig.~\ref{fig:gv}. The differences with real space results are
explained by the arrows: large scale redshift distortion (LSRD)
increases the effective slope of the velocity--gravity relation, while
the finger of God (FOG) effect moves points from the upper right down
to the left, and points from the lower left up to the right, creating
two additional tails of the PDF. As discussed in the text, this is due
to the inversion of the signature of velocity happening in fingers of
God as seen in the bottom right panel of Fig.~\ref{fig:images1}, and
to the effective smoothing of gravitational potential due to the
extended nature of the FOG. When smoothing with a Gaussian window of
radius $\ell=10\; h^{-1} \rm Mpc$ is performed, these two tails move
back to the upper right and the lower left of the panel: one recovers
the propeller shape obtained in real space. However, its effect is
more pronounced, because there is still some remnant of the
anti-diagonal effect, depending on the level of smoothing.}
\label{fig:gvzinf}
\end{figure*}

Figure~\ref{fig:gvzinf} is similar to Fig.~\ref{fig:gv}, but shows the
results obtained from
the sample (ii), in the DOL redshift space. The same striking
result obtained as in real space holds: the region of maximum
likelihood agrees very well with the linear theory prediction, which
gives a velocity--gravity slope larger than in real space because of
the enhancement of large scale density contrasts due to projection in
redshift space (symbol LSRD -- Large Scale Redshift Distortion -- on
the left panel).

However, there is on Fig.~\ref{fig:gvzinf} a noticeable new feature on
the PDF visible on left panel, in addition to the propeller shape: the
joint PDF seems now to present tails in the directions orthogonal to
the maximum likelihood domain. Note that these tails tend to disappear
with smoothing which then makes the bivariate PDF look very much like
the real space one.

This new feature is due to the finger-of-God effects already discussed
at length in \S~\ref{sec:visu}.  These effects have mainly two
consequences: first, the gravitational potential is less contrasted at
small scales due to the elongated nature of fingers of God, which tend
to reduce the amplitude of the gravitational acceleration inside them
and in their neighborhood. Second, there is an inversion effect which
changes the sign of the velocity.  The consequence is that some points
that were, in real space, on the upper right quadrant are moved to the upper left
quadrant as indicated by the arrow on the left panel of
Fig.~\ref{fig:gvzinf}. These points create a tail on the upper left
quadrant; similarly some points that lied in the lower left quadrant
 contribute to the 
tail on the lower right quadrant.

In fact, in the same way as we did in real space, the results obtained
in redshift space can again be interpreted qualitatively using the top
hat spherical collapse model, as detailed in
Appendix~\ref{app:gen}. Note interestingly that in the linear regime,
the top hat model gives
\begin{equation}
\frac{g_z}{v_z}=\frac{1}{\beta}\left(1+\frac{\beta}{3} \right),
\label{eq:sphezlin}
\end{equation}
a value significantly smaller than what is obtained from
Eqs.~(\ref{eq:F}), (\ref{eq:I}) and (\ref{eq:U}):
equation (\ref{eq:sphezlin}) gives $g_z/v_z \simeq 2.3$ instead of $2.6$.
This disagreement with the statistical expectation from linear theory
is explained in Appendix~\ref{app:gen}. However, despite the
limitations of the spherical top hat model, the arguments developed
previously in the real space case to explain why the linear regime
dominates the most likely part of the joint PDF still hold.

Thanks to finger-of-God effects, if no additional smoothing is applied
to the interpolated fields, the measured joint PDF is now much more
symmetric about the linear prediction than in real space. As a
consequence, the slope obtained from the direct measurement of
$\sqrt{\langle g_z^2 \rangle/\langle v_z^2 \rangle}$ (dotted line on
left panel of Fig.~\ref{fig:gvzinf}) agrees now well with linear
theory, while we still have $\langle g_z^2 \rangle/\langle g_z v_z
\rangle > \sqrt{ \langle g_z^2 \rangle/\langle v_z^2 \rangle } >
\langle g_z v_z \rangle/\langle v_z^2 \rangle $.  However, when an
additional Gaussian smoothing is performed, the fingers of God tend to
be diluted. This implies on Fig.~\ref{fig:gvzinf} a clockwise rotation
of the upper left tail to the upper right and of the lower right tail
to the lower left. As a consequence, with our $10\ h^{-1}$ Mpc scale
Gaussian smoothing, one converges to similar behavior as obtained in
real space, with the propeller effect significantly biasing the
overall slope, implying again that $\Omega$ is biased low. This bias
is more pronounced in redshift space than in real space, because there
is still some remnant of the anti-diagonal effect, depending on the
level of smoothing. Of course, 
smoothing at larger scales would make
the agreement with linear theory prediction better again. 

To reduce the bias, one can play again the exercise of measuring the
slope of the velocity--gravity relation by selecting the region of
best likelihood, as in Equation~(\ref{eq:con68}). The agreement with
linear theory prediction is improved as expected, and the
corresponding measured value of $\Omega$ matches very well the real
value as illustrated by Table~\ref{tab:table1}.  However, additional
smoothing tends to mix the nonlinear finger of God effects with linear features,
contaminating the region of best likelihood. As a result, the measured
value of $\Omega$ is slightly biased to lower values ($0.28$ instead
of $0.3$) and changing the likelihood contour selection does not
improve significantly the results.  This mixing effect brought by
smoothing suggests that in fact it might be not so wise to perform
additional smoothing of the interpolated fields.  Even though
smoothing at sufficiently large scales brings better overall agreement
with linear theory, it makes the measurements much more sensitive to
finite volume and edge effects and it is furthermore not needed here
since linear theory regime always dominates the region of best
likelihood. Besides, we showed in \S~\ref{sec:valdol} that smoothing was
making the DOL approximation\footnote{Valid, however, by definition, for the
sample currently under consideration, (ii).} worse.

\begin{figure*}
\centerline{\hbox{
\psfig{file=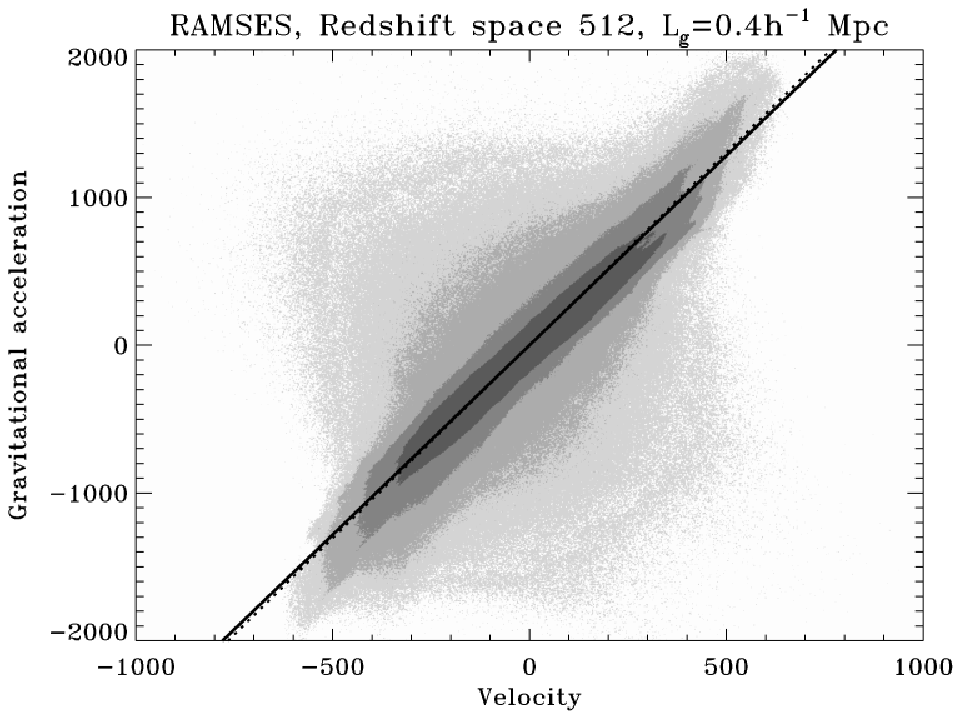,width=8cm}
\psfig{file=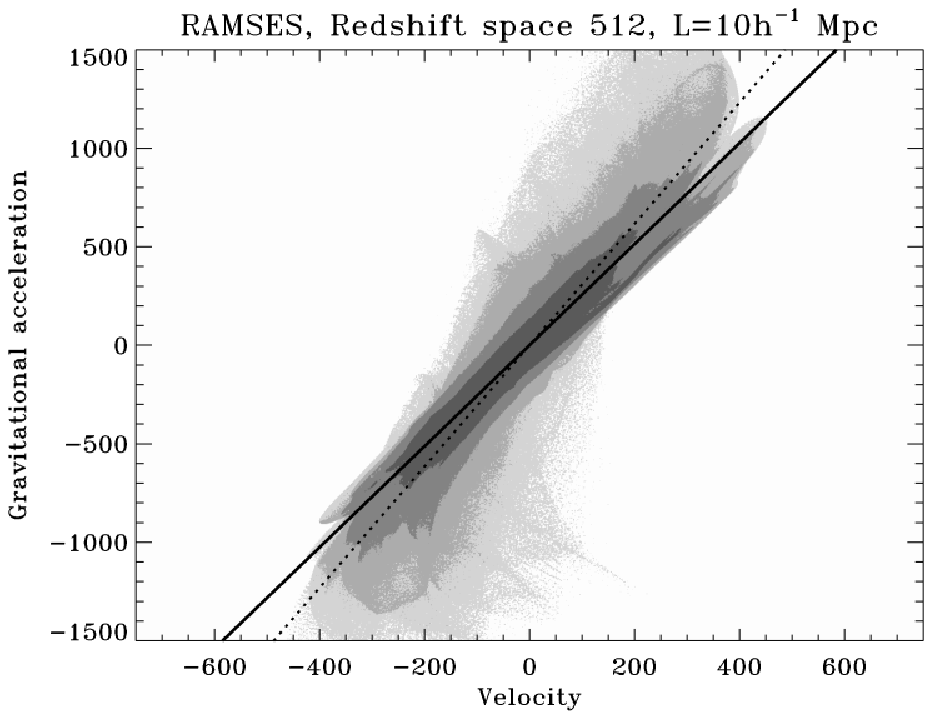,width=8cm}
}}
\caption[]{Same as in Fig.~\ref{fig:gvzinf}, but for the sample (iii)
which does not assume that the observer is (infinitely) distant.
Although the general behavior is exactly the same as in
Fig.~\ref{fig:gvzinf}, several explainable differences can be
noticed. The measurement is more noisy and the apparent width of the
bivariate distribution is smaller than on Fig.~\ref{fig:gvzinf}.  This
is mainly related to the fact that the sampled volume covers only 18
percent of the simulation box.  Still, the most likely part of the PDF
fits very well the slope predicted by linear theory in the infinite
remote observer approximation, lending credence to our simplified
approach.}
\label{fig:gvznotZinf}
\end{figure*}

\begin{figure*}
\centerline{\hbox{
\psfig{file=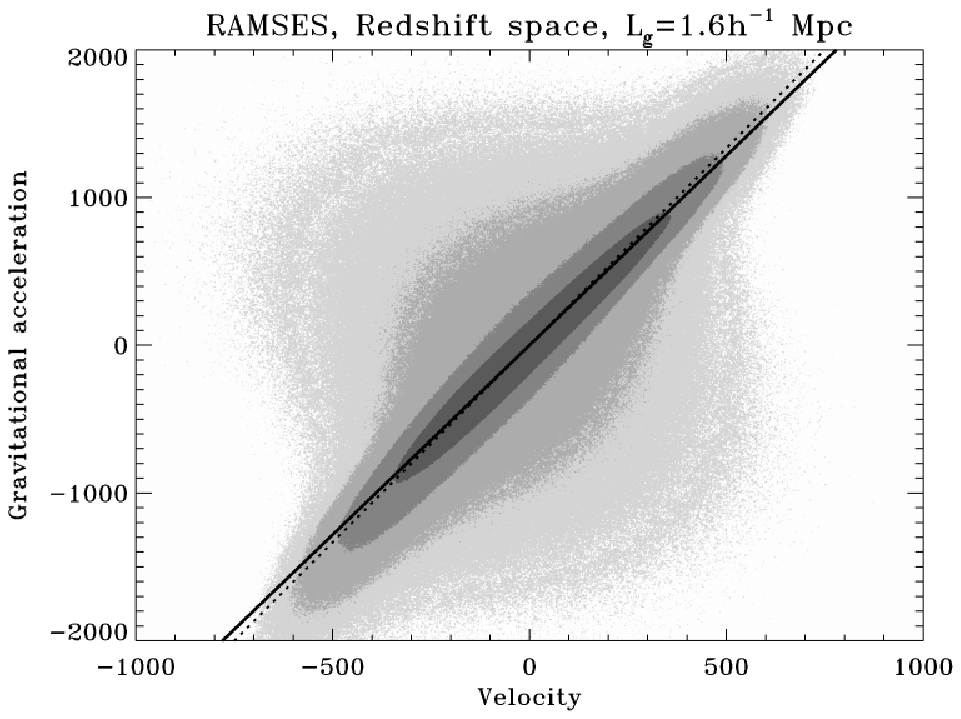,width=8cm}
\psfig{file=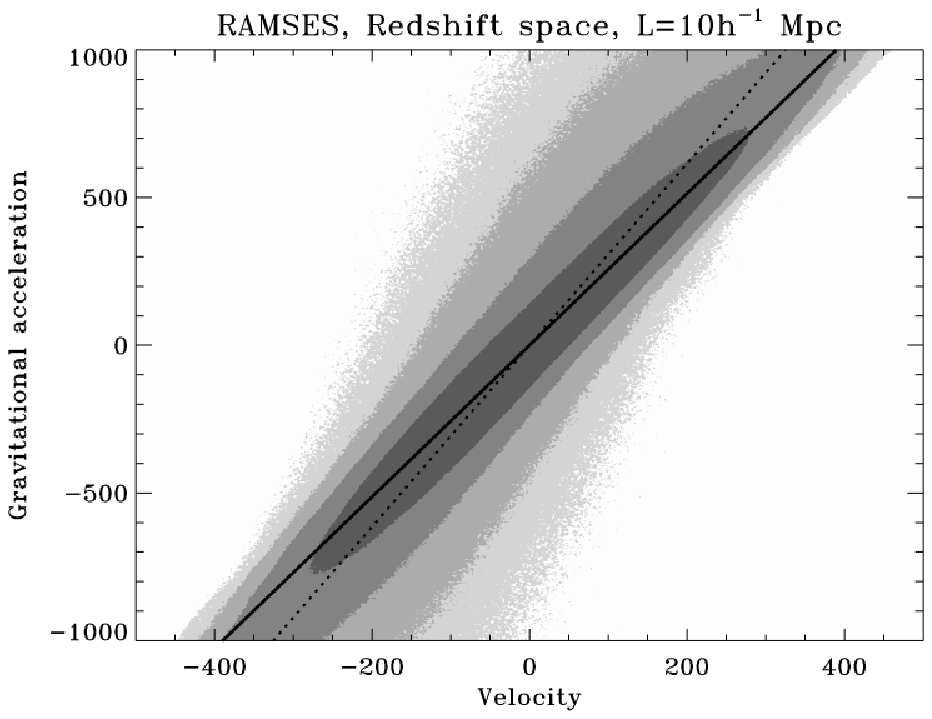,width=8cm} }}
\caption[]{Same as in Fig.~\ref{fig:gvznotZinf}, but we now consider
sample (iv), where 125 different observer positions are located on a
regular grid in the simulation box and where the interpolated
field is computed only on a $128^3$ grid. As expected, the resulting
bivariate PDF is smoother than in Fig.~\ref{fig:gvznotZinf}, since the
simulation volume is now fully sampled by all these observer
positions.

Note the significant asymmetry of the tails of the bivariate PDF
around the major axis of the maximum likelihood region.  This is an
expected consequence of edge effects.  Indeed, while performing
redshift space projection to construct a spherical catalog of finite
radius $R_{\rm max}$, the galaxies near the edge of the catalog with
positive radial velocity and radial position $r < R_{\rm max}$ tend to
get outside the sample. On the other hand, galaxies near the edge of
the catalog with negative radial velocity and radial position $r >
R_{\rm max}$ tend to get inside the catalog. As a result one expects
more galaxies with negative radial velocities than with positive ones,
hence the asymmetry of the PDF. This effect mainly affects fingers of
God, so the anti-diagonal tails of the bivariate PDF.  Importantly,
this effect does not affect too much the region of best likelihood.}
\label{fig:redzfig}
\end{figure*}

Figure~\ref{fig:gvznotZinf} is exactly the same as
Fig.~\ref{fig:gvzinf}, but the measurements
are now performed in sample (iii), without assuming that the observer
is (infinitely) remote. Qualitatively, all the conclusions derived
from the analysis of Fig.~\ref{fig:gvzinf} still hold. The main
difference is that the measurements are more noisy due to the size of
the sampled volume, now 5 times smaller.  Therefore, the recovered
value of $\Omega$ from the measurements of the moments in the region
of best likelihood of the joint PDF of the unsmoothed fields is good
only at the ten percent level ($\Omega=0.33$ instead of $0.3$, see
Table~\ref{tab:table1}).  At this level of accuracy, we find that the
linear prediction, which was derived in the approximation of DOL,
agrees well with the measurements. To improve the quality of this
comparison, Figure~\ref{fig:redzfig} gives the same scatter plots as
in Fig.~\ref{fig:gvznotZinf},
but for sample (iv), i.e. after co-adding the contributions of 125
different observers. The value of $\Omega$ derived from the best
likelihood region of left panel of Fig.~\ref{fig:redzfig} agrees now
at a few percent level with the true value of $\Omega$ (the estimated
value is $0.294$, see Table~\ref{tab:table1}), showing that the
distant observer limit is an approximation good enough for deriving
the linear prediction. Note again the bias to lower values ($0.262$) 
on the measured value of $\Omega$ brought by additional Gaussian
smoothing with a 10 \hmpc\ Gaussian window. This bias is more pronounced
than in the DOL sample (ii), and this is certainly at least partly due to the fact
that, as discussed in  \S~\ref{sec:valdol}, deviations from
the DOL limit are not negligible anymore for such a smoothing scale, given
the sample depth, and they add to the mixing between linear and nonlinear
features discussed above. It could also come partly from the edge effect
discussed in caption of Fig.~\ref{fig:redzfig}.

The dispersion among the 125 different observers leads to a typical
error on $\Omega$ of the order of $0.04$, suggesting that the errors
related to the choice of the position of the observer -- 
which probes the space of configurations for the ${\bf s}={\bf 0}$ singularity
discussed in \S~\ref{sec:valdol} -- are of the order
of 10 percent for our $R=70\ \hmpc$ radius catalog.  These errors
include cosmic variance effects, but these latter are probably
underestimated because our spherical redshift samples represent a
rather significant fraction of the total simulation volume. 

Nevertheless these measurements illustrate the relative robustness in that respect of
our velocity--gravity estimator to derive $\Omega$ from a large-scale
galaxy survey. However, we reiterate that, as already mentioned in
beginning of \S~\ref{sec:algomes}, the volume used to compute gravity
field should be significantly larger than the actual volume used to
perform the velocity--gravity comparison. The present analyses suppose
it is the case as indeed achieved by most recent three dimensional
galaxy catalogs such as the 2MRS, for
which the median redshift corresponds to a half-depth of $\simeq 60\,h^{-1}$ Mpc
(Erdo$\breve{\rm g}$du et al. 2006).

%
\subsubsection{Effects of dilution}
\label{sec:diluef}
%
%
Up to now, we have measured the velocity--gravity relation in very
rich catalogs, where the number of objects was so large that
discreteness effects could be considered as negligible. In real galaxy
catalogs, the number of objects is much smaller, particularly when
tracers of the velocity field are taken at concern. Before addressing
the issue of biasing between the galaxy distribution and the dark
matter distribution, we therefore examine pure discreteness effects,
by diluting our dark matter samples. This dilution will not only bring
a shot noise contribution, it will also increase the overall level of
smoothing which is performed by our interpolation procedure.

To be able to quantify as accurately as possible the biases induced by
discreteness in the mock ``galaxy'' catalogs considered in next
section, we dilute randomly our $N$-body sample in two kind of
subsamples (see Table~\ref{tab:table3}):

\begin{table*}
\caption[]{Same as Table~\ref{tab:table1}, but for the dilute dark matter sub-samples, D1 and D2.}
\label{tab:table3}
\begin{tabular}{llclcccccl}
\hline
Samp. & Content & $N_{\rm rea}$      & Size & $N_{\rm g}$ & $N_{\rm obj}$ &  \multicolumn{3}{c}{Measured $\Omega$ from:} & Comment \\
       &         &                    &      &             &               &  $\frac{\langle g^2 \rangle}{\langle v g \rangle}$ &
                                                                              $\sqrt{\frac{\langle g^2 \rangle}{\langle v^2 \rangle}}$ &
                                                                              $\frac{\langle v g \rangle}{\langle v^2 \rangle}$ & \\ 
\hline \hline
(D1)   & 50000, real space                              & 125      & $L=200$ & 128 & 50000                    & 0.163 & 0.213 & 0.278 & $\langle$ all $\rangle$, no smoothing \\
       &                                                         &          &         &     &                          & 0.004 & 0.006 & 0.008 & error from dispersion \\
       &                                                         &          &         &     &                          & 0.237 & 0.252 & 0.268 & $\langle$ all $\rangle$, smoothed\\
       &                                                         &          &         &     &                          & 0.007 & 0.008 & 0.009 & error from dispersion \\
       &                                                         &          &         &     &                          & 0.283 & 0.300 & 0.319 & $\langle$ 1.5$\sigma$ 
                                                                                                                                                 isocontour $\rangle$,
                                                                                                                                                 no smoothing \\
       &                                                         &          &         &     &                          & 0.011 & 0.012 & 0.013 & error from dispersion \\
       &                                                         &          &         &     &                          & 0.277 & 0.288 & 0.299 & $\langle$ 1.5$\sigma$ 
                                                                                                                                                 isocontour $\rangle$,
                                                                                                                                                 smoothed \\
       &                                                         &          &         &     &                          & 0.012 & 0.013 & 0.013 & error from dispersion \\
       &  50000, redshift space                              & 125      & $R=70$  & 128 & $8980^{*}$               & 0.062 & 0.248 & 3.841 & $\langle$ all $\rangle$, no smoothing \\
       &                                                         &          &         &     &                          & 0.025 & 0.027 & 16.11 & error from dispersion \\
       &                                                         &          &         &     &                          & 0.100 & 0.184 & 0.378 & $\langle$ all $\rangle$, smoothed\\
       &                                                         &          &         &     &                          & 0.027 & 0.022 & 0.118 & error from dispersion\\
       &                                                         &          &         &     &                          & 0.227 & 0.264 & 0.309 & $\langle$ 1.5$\sigma$ 
                                                                                                                                                 isocontours $\rangle$,
                                                                                                                                                 no smoothing\\
       &                                                         &          &         &     &                          & 0.030 & 0.034 & 0.050 & error from dispersion\\
       &                                                         &          &         &     &                          & 0.209 & 0.237 & 0.269 & $\langle$ 1.5$\sigma$ 
                                                                                                                                                 isocontours $\rangle$,
                                                                                                                                                 smoothed\\
       &                                                         &          &         &     &                          & 0.026 & 0.028 & 0.039 & error from dispersion\\
(D2)   & 11000, real space                              & 125      & $L=200$ & 128 & 11000                    & 0.192 & 0.250 & 0.326 & $\langle$ all $\rangle$, no smoothing \\
       &                                                         &          &         &     &                          & 0.011 & 0.017 & 0.026 & error from dispersion\\
       &                                                         &          &         &     &                          & 0.216 & 0.255 & 0.300 & $\langle$ all $\rangle$, smoothed\\
       &                                                         &          &         &     &                          & 0.015 & 0.020 & 0.026 & error from dispersion\\
       &                                                         &          &         &     &                          & 0.251 & 0.292 & 0.340 & $\langle$ 1.5$\sigma$ 
                                                                                                                                                 isocontour $\rangle$,
                                                                                                                                                 no smoothing \\
       &                                                         &          &         &     &                          & 0.023 & 0.027 & 0.033 & error from dispersion\\
       &                                                         &          &         &     &                          & 0.249 & 0.281 & 0.318 & $\langle$ 1.5$\sigma$ 
                                                                                                                                                 isocontour $\rangle$,
                                                                                                                                                 smoothed \\
       &                                                         &          &         &     &                          & 0.024 & 0.028 & 0.033 & error from dispersion\\
       & 11000, redshift space                              & 125      & $R=70$  & 128 & $1976^{*}$               & 0.074 & 0.226 & 1.323 & $\langle$ all $\rangle$, no smoothing \\
       &                                                         &          &         &     &                          & 0.028 & 0.035 & 1.687 & error from dispersion \\
       &                                                         &          &         &     &                          & 0.096 & 0.197 & 0.492 & $\langle$ all $\rangle$, smoothed\\
       &                                                         &          &         &     &                          & 0.031 & 0.034 & 0.238 & error from dispersion\\
       &                                                         &          &         &     &                          & 0.190 & 0.253 & 0.347 & $\langle$ 1.5$\sigma$ 
                                                                                                                                                 isocontours $\rangle$,
                                                                                                                                                 no smoothing\\
       &                                                         &          &         &     &                          & 0.040 & 0.048 & 0.099 & error from dispersion\\
       &                                                         &          &         &     &                          & 0.182 & 0.233 & 0.308 & $\langle$ 1.5$\sigma$ 
                                                                                                                                                 isocontours $\rangle$,
                                                                                                                                                 smoothed\\
       &                                                         &          &         &     &                          & 0.040 & 0.044 & 0.089 & error from dispersion\\
\hline
\end{tabular}
\end{table*}
\begin{description}
\item[(D1)] 125 realizations of dark matter catalogs involving 50000
points, similarly as for the mock with mass thresholding $M_{\rm
th}=5.10^{11} \ M_{\odot}$ used in \S~\ref{sec:mockmes};
\item[(D2)] 125 realizations of dark matter catalog involving 11000
points, similarly as for the mock
catalogs with mass thresholding $M_{\rm
th}=4.10^{12} \ M_{\odot}$ used in \S~\ref{sec:mockmes}.
\end{description}
We perform exactly the same measurements in these dilute catalogs than
in sample (i) (in real space, except that we use $128^3$ grids for
interpolating the fields and we have 125 different realizations) and
in samples (iv) (in redshift space). In the latter case, note that the
observer position in the simulation cube is different for each
realization, chosen exactly to be on a regular patter covering the
full simulation volume as in samples (iv).

\begin{figure*}
\centerline{\hbox{
\psfig{file=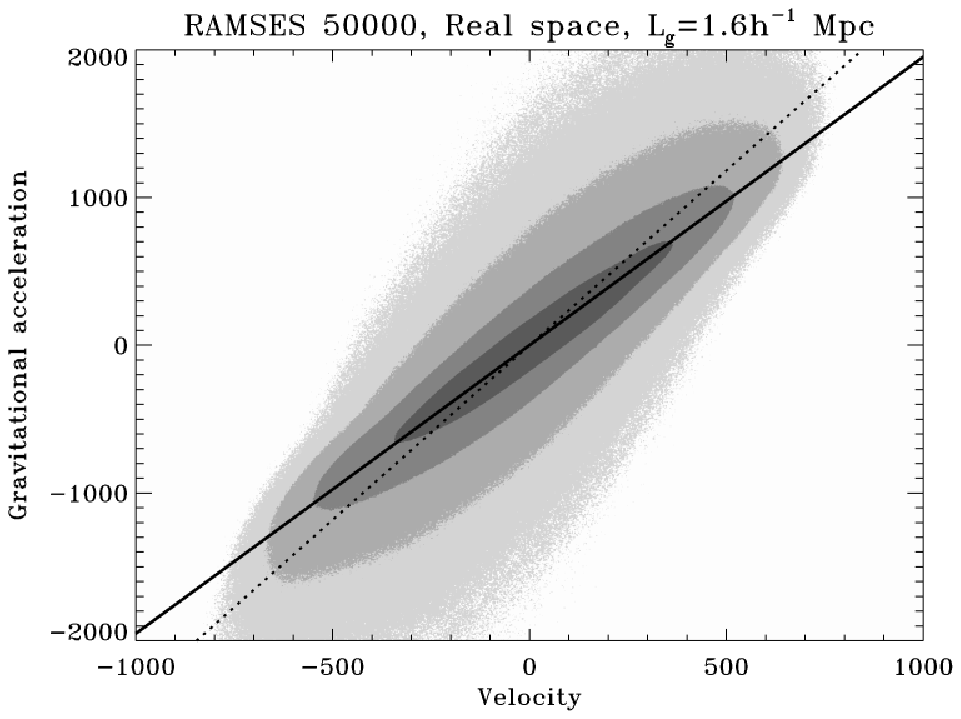,width=8cm}
\psfig{file=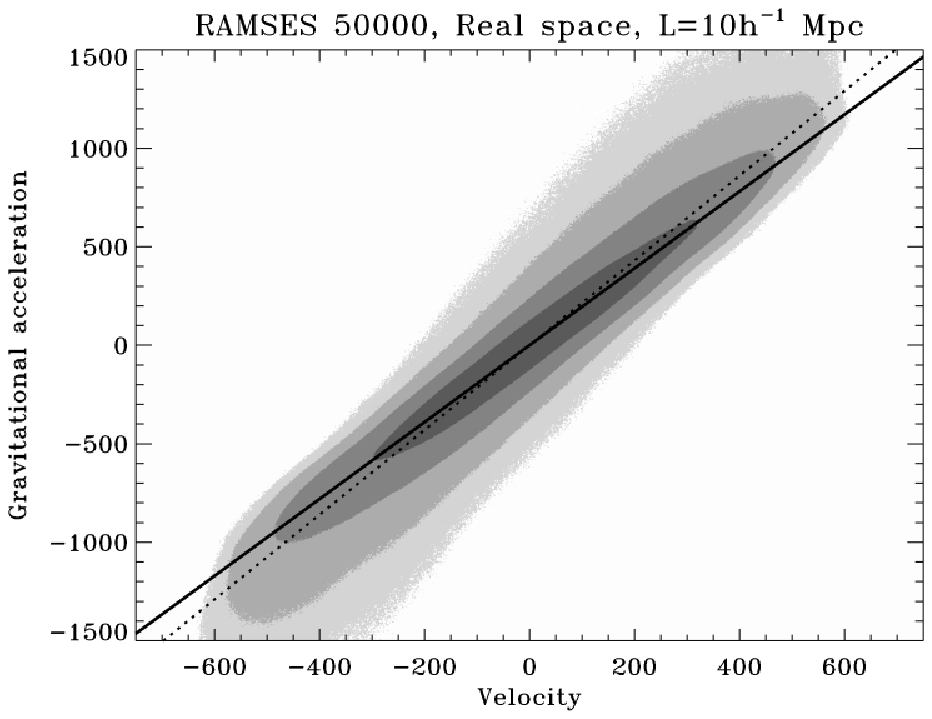,width=8cm}
}}
\centerline{\hbox{
\psfig{file=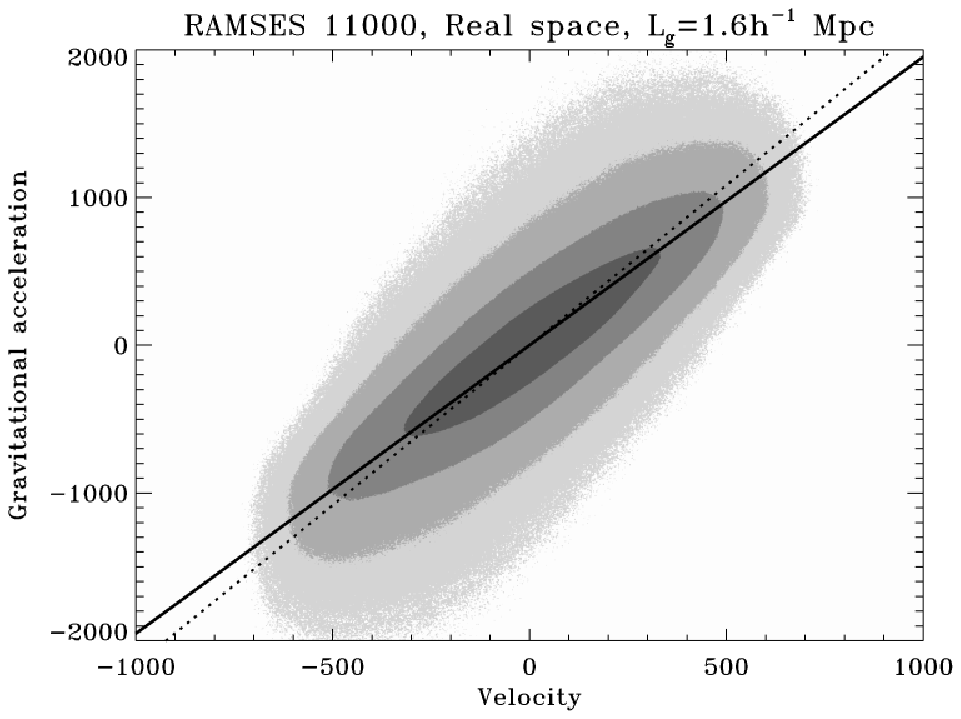,width=8cm}
\psfig{file=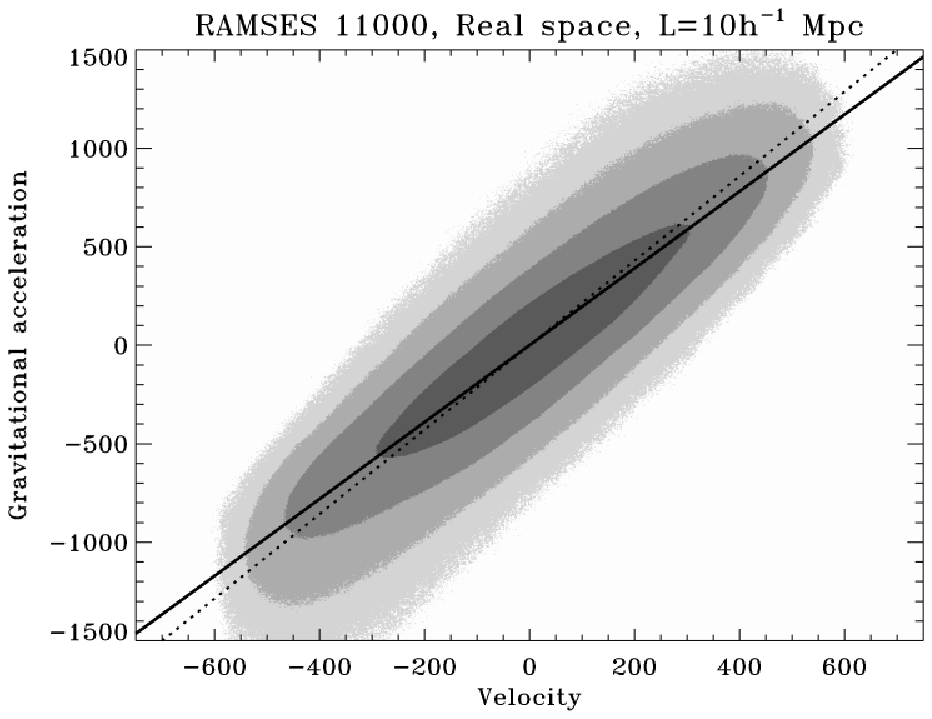,width=8cm}
}}
\caption[]{Similarly as in Fig.~\ref{fig:gv}, but for the 125 realizations of
 diluted samples D1 (upper panels) and D2 (lower panels).
the left panels correspond to the raw interpolated fields on a $128^3$ grid, while for the right panels, a smoothing
with a Gaussian window of size $\ell=10\ h^{-1}$ Mpc was performed prior to the measurements. Since the samples are now
more diluted, the adaptive interpolating kernel is of much larger size, which decreases considerably nonlinear tails:
the propeller shape of the PDF is less apparent but it still affects the measurements: the slope of the dotted line is larger
than the linear expectation given by the thick solid line.}
\label{fig:realdil}
\end{figure*}
\begin{figure*}
\centerline{\hbox{
\psfig{file=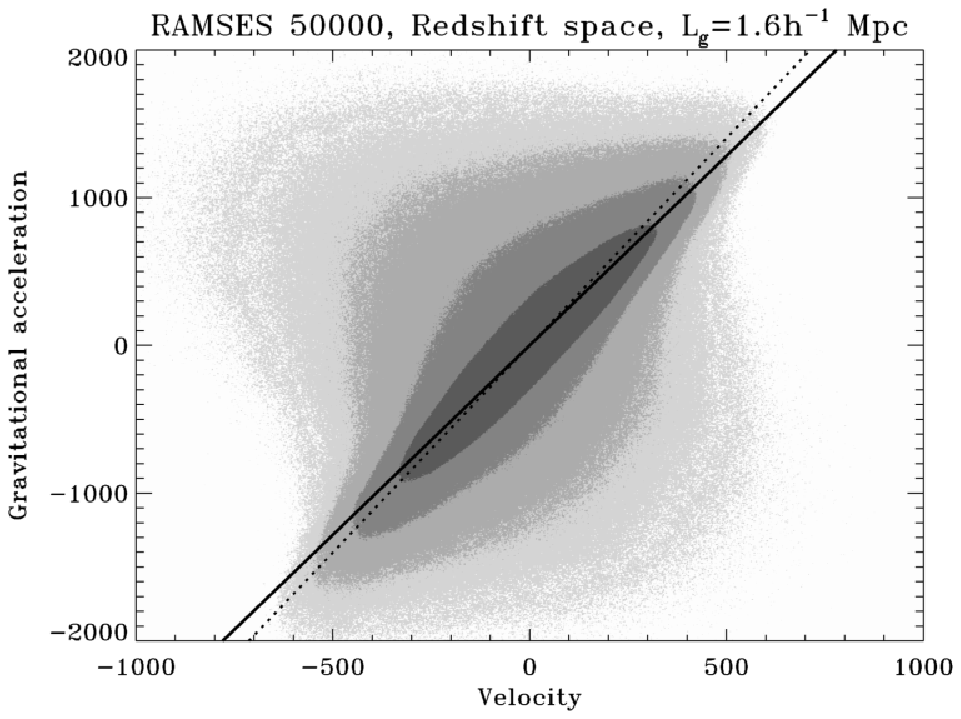,width=8cm}
\psfig{file=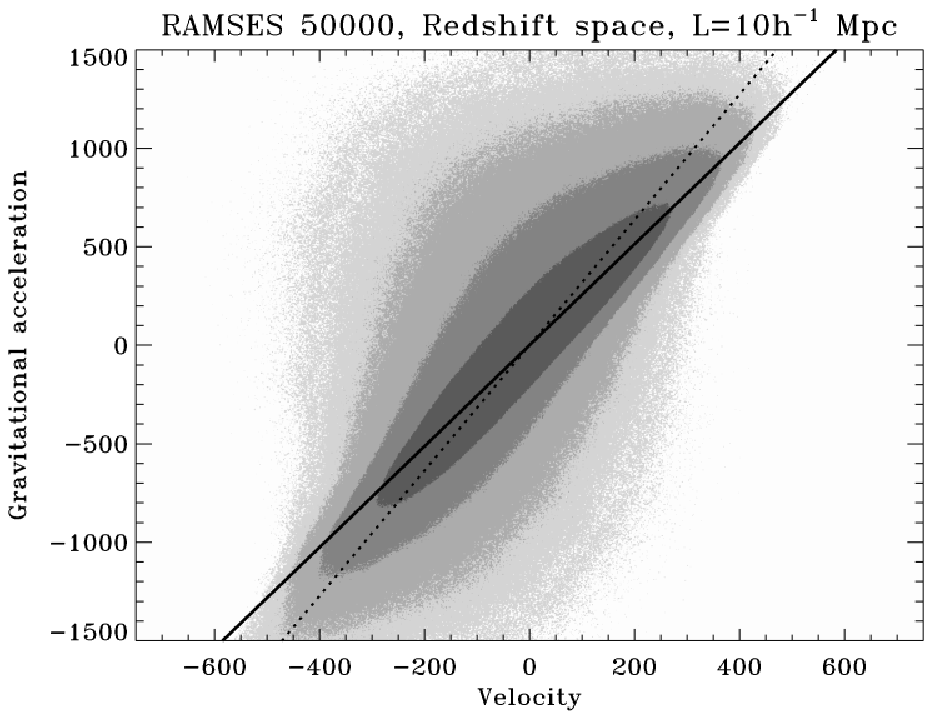,width=8cm}
}}
\centerline{\hbox{
\psfig{file=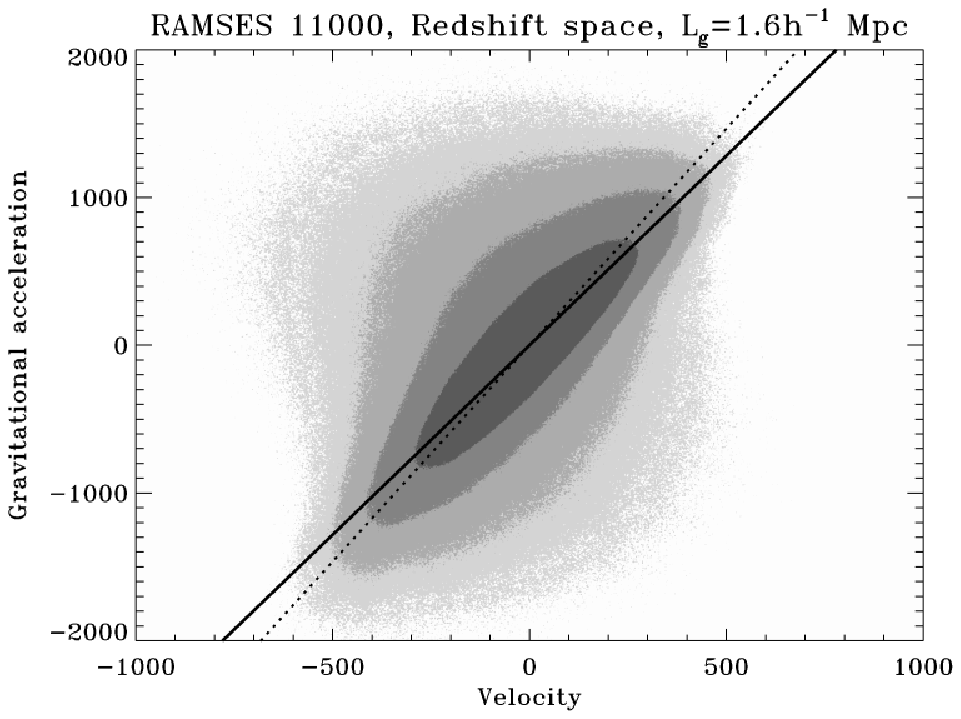,width=8cm}
\psfig{file=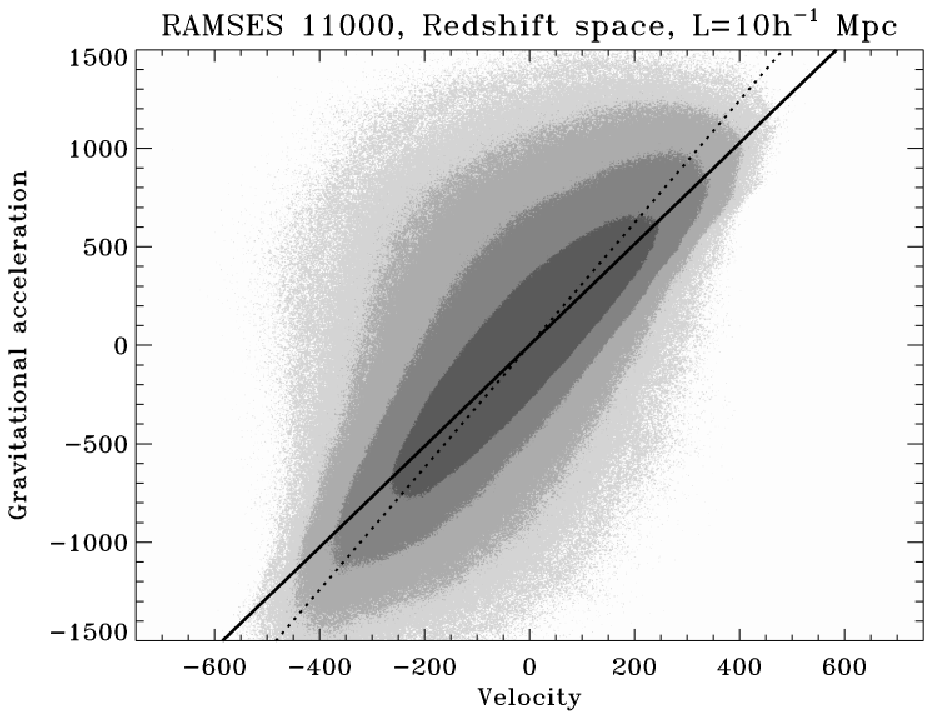,width=8cm}
}}
\caption[]{Similarly as in Fig.~\ref{fig:redzfig}, but for the diluted samples D1 (upper panels) and D2 (lower panels).
}
\label{fig:zdil}
\end{figure*}

The results are summarized in Table~\ref{tab:table3} and illustrated by Figs.~\ref{fig:realdil} and \ref{fig:zdil}. 
We first discuss real space measurements using the second order moments of the full PDF and compare
the values obtained for $\Omega$ to those from the undiluted sample (i), given in Table~\ref{tab:table1}.
Note that this sample uses a $512^3$ grid for the interpolation instead of the $128^3$ one for samples
D1 and D2. Using a $128^3$ resolution grid increases the measured value of $\Omega$ in first line
of Table~\ref{tab:table1} from $0.157$ to $0.178$.
Taking that fact into account, we notice that, due
to the dilute nature of the samples, our interpolation procedure uses an adaptive kernel of larger size: 
as shown by Fig.~\ref{fig:realdil}, this
makes the fields more linear, closer to the Gaussian limit, and decreases 
the level of effective bias brought by the ``propeller'' shape of the bivariate PDF. 
As a result, prior to additional smoothing with a Gaussian window of size $\ell=10\ h^{-1}$ Mpc, 
the measured $\Omega$ is larger for the dilute samples than for the full one, and the convergence
with linear theory prediction improves with level of dilution.
The mean inter-particle distance in the sparser sample is of the order of $\lambda=9\ h^{-1}$ Mpc, slightly
lower than the size of the post-processing Gaussian smoothing kernel. One thus expects rough agreement
between full sample and samples D1 and D2 after smoothing with such a window, which is approximately
the case. Furthermore, the measured value of $\Omega$ in D2
is not very sensitive to whether additional smoothing is performed or not, 
since $\lambda \simeq \ell$. 

Still examining real space measurements, we now consider the values of $\Omega$ measured from second
order moments of the PDF, but using only the region of best likelihood, Eq.~(\ref{eq:con68}), 
similarly as in previous
paragraph.
As expected, the effective bias due to nonlinear contributions (or propeller effects) is tremendously reduced, 
and one recovers a value of $\Omega$ compatible
with the true value, given the errorbars. These latter, which estimate pure shot noise,
are of the order of 4 and 10\% for D1 and D2 respectively. As for the full
sample, this result stands also when additional smoothing is applied to the data, although
one can notice a slight bias to lower values of $\Omega$.

We now turn to redshift space measurements and first consider the measured value of $\Omega$ using the second
order moments of the full PDF. The results obtained in previous section still hold (compare Table~\ref{tab:table1}
to Table~\ref{tab:table3}), except that the measurements
performed before additional Gaussian smoothing give a lower value of $\Omega$, $0.248$ for D1 and
$0.226$ for D2: the low bias effect on $\Omega$ gets worse with dilution,
since the larger size of the adaptive kernel tends to reduce the effect of fingers of God:
 as noticed previously, finger-of-God effects 
help to reduce the asymmetry brought by the ``propeller'', the main source of the low bias on $\Omega$. 
Additional Gaussian smoothing with a window
of size $\ell=10\ h^{-1}$ Mpc improves the convergence
between D1, D2 and the full sample, as expected, but induces a highly underestimated value of $\Omega \sim 0.2$, 
because of the propeller effects which are then prominent. These arguments are supported by examination
of Fig.~\ref{fig:zdil}. Note that the asymmetry of the tails of the 
bivariate PDF around the major axis of the maximum likelihood region,
already observed and explained in Fig.~\ref{fig:redzfig}, is now more pronounced, at least
from the visual point of view. 

Measurements are improved while selecting the region of best likelihood, but not as well as in the real space case
or in the redshift space case with full sampling:
at best, $\Omega$ is underestimated by about 12 percent. Additional smoothing or passing from
D1 to D2 expectingly increases the bias.
This underestimate comes again from the fact that the adaptive
interpolating kernel is now of much larger extension than for the full particles sample, which
induces biases comparable to what was observed for the full samples  with additional smoothing (\S~\ref{sub:red}). 

Note that it is possible to reduce the effective bias observed on $\Omega$ by narrowing
the region of best likelihood at the price of an increase of the errors. 
For instance, taking a 38 percent confidence region enclosed 
by a $\simeq 1 \sigma$ isocontour gives $\Omega=0.285$ and $\Omega=0.273$ 
for D1 and D2, respectively. Our procedure for measuring $\Omega$ 
can thus certainly be improved with a more sophisticated treatment of the region of best likelihood.
For instance, a way to extract in an unbiased way the parameter $\Omega$ from the
data could consist in measuring the local slope of the skeleton of the surface representing the bivariate 
PDF  (see Novikov, Colombi \& Dor\'e, 2006) after appropriate (adaptive) smoothing of the velocity--gravity
scatter plot.
This is left for future work; in what follows, we shall still use for
the sake of simplicity our $1.5\sigma$ likelihood contour technique,
while staying aware of the bias brought by dilution.
%
%
%
%
%
\subsection{Measurements on simple mock ``galaxy'' catalogs}
\label{sec:mockmes}
%
%
%
To estimate in a sufficiently realistic way how biasing affects
the results on the velocity--gravity relation, we extracted from
the simulation four ``galaxy'' catalogs, corresponding
to two methods of treating dark matter halos.
In the first method, we consider each dark matter halo as a galaxy.
In the second method, we consider each substructure present in dark
matter halos as a galaxy. The details are
given in \S~\ref{sec:mockdesc}. For each kind of ``galaxy catalogs'',
two level of biasing are considered, corresponding to small and large
threshold on the dark matter (sub)-halos masses, $M > 5.10^{11} \ M_{\odot}$
and $M > 4.10^{12} \ M_{\odot}$ respectively. 
In all cases, each galaxy is given {\em the same
weight} while density and velocity are interpolated as explained in
\S~\ref{sec:algomes}. This rather extreme way of treating the bias is
far from being optimal: appropriate weighting could be given to each object
in order to correct at least partly for the effects of the bias, as further
discussed in the conclusions of this paper. 

The quantitative analysis
of the velocity-gravity relation is performed in \S~\ref{sec:mockmes2}. The measurements
are interpreted in the light of the previous paragraphs analyses, using as a guideline
the large scale bias obtained from the measurement of the power-spectrum of the ``galaxy'' 
distribution.
%
%
\subsubsection{The mock catalogs}
\label{sec:mockdesc}
%
%

To extract halos and substructures from the simulation, we use the
publically available software {\sl adaptaHOP} (Aubert, Pichon \&
Colombi, 2004).\footnote{The parameters used in {\sl adaptaHOP} are
the same as in Aubert et al. (2004, Appendix B), namely $N_{\rm
SPH}=64$, $N_{\rm HOP}=16$, $\rho_{\rm TH}=81$ and $f_{\rm
Poisson}=4$.}  {\rm AdaptaHOP} builds an ensemble of trees. Each tree
corresponds to a halo which is a connected ensemble of particles with
SPH density $\rho > 81$.  The branches of the trees are composite
structures of which the connectivity is controlled by the saddle
points in the particle distribution.  The leaves of the trees
correspond to the smallest possible substructures one can find in the
simulation.  From this ensemble of trees, we extract two kinds of
catalogs, one where each galaxy is identified to a tree, the other one
where each galaxy is identified to the leaves of the trees (if a tree
has only one leaf, it means that the halo is its own single
substructure). Note that velocities of these galaxies are computed
as the average velocity of all the particles belonging to the
corresponding (sub-)structure.

Additional mass thresholding is used to control the number of ``galaxies'' in the catalogs,.
In final, 4 mock catalogs are obtained (see also Table~\ref{tab:table2}): 
\begin{table*}
\caption[]{Same as Table~\ref{tab:table1}, but for the ``galaxy'' catalogs, (v)--(viii).
In the second column, it is specified if they correspond to a halo (``HAL.'') or a substructure catalog (``SUBS.'') 
and  what mass threshold was used to select the dark matter (sub-)structures. The values of $\Omega$ displayed
in columns seven to nine are obtained from the measured $\beta$ assuming linear theory predictions with no bias.
As a guidance to understand the results, we show in the second column of the table
the measured value of the bias $b$ at the simulation box size scale (using the power-spectrum
of the interpolated density), as well as $\Omega_{\rm eff}=\Omega/b^{9/5}$, the expected value of $\Omega$ from the
measured $\beta$ if unity bias is assumed.}
\label{tab:table2}
\begin{tabular}{llclcccccl}
\hline
Samp. & Content & $N_{\rm rea}$      & Size & $N_{\rm g}$ & $N_{\rm obj}$ &  \multicolumn{3}{c}{Measured $\Omega$ from:} & Comment \\
       &         &                    &      &             &               &  $\frac{\langle g^2 \rangle}{\langle v g \rangle}$ &
                                                                              $\sqrt{\frac{\langle g^2 \rangle}{\langle v^2 \rangle}}$ &
                                                                              $\frac{\langle v g \rangle}{\langle v^2 \rangle}$ & \\ 
\hline \hline
(v)   & HAL. $> 5{\rm e}11 M_{\odot}$, real space      & 1        & $L=200$ & 128 & 43482                    & 0.304 & 0.375 & 0.464 & all, no smoothing \\
       & $b=0.80$, $\Omega_{\rm eff}=0.45$                       &          &         &     &                          & 0.330 & 0.379 & 0.435 & all, smoothed \\
       &                                                         &          &         &     &                          & 0.318 & 0.368 & 0.426 & 1.5$\sigma$ 
                                                                                                                                                 isocontour,
                                                                                                                                                 no smoothing \\ 
       &                                                         &          &         &     &                          & 0.329 & 0.370 & 0.415 & 1.5$\sigma$ 
                                                                                                                                                 isocontour,
                                                                                                                                                 smoothed \\
       & HAL. $> 5{\rm e}11 M_{\odot}$, red. space  & 125      & $R=70$  & 128 & $7874^{*}$               & 0.208 & 0.313 & 0.487 & $\langle$ all $\rangle$, no smoothing \\
       & $b=0.80$, $\Omega_{\rm eff}=0.45$                       &          &         &     &                          & 0.033 & 0.035 & 0.089 & error from dispersion \\
       &                                                         &          &         &     &                          & 0.259 & 0.312 & 0.378 & $\langle$ all $\rangle$, smoothed \\
       &                                                         &          &         &     &                          & 0.035 & 0.039 & 0.055 & error from dispersion \\ 
       &                                                         &          &         &     &                          & 0.311 & 0.360 & 0.419 & $\langle$ 1.5$\sigma$ 
                                                                                                                                                 isocontours $\rangle$,
                                                                                                                                                 no smoothing\\
       &                                                         &          &         &     &                          & 0.054 & 0.063 & 0.082 & error from dispersion\\
       &                                                         &          &         &     &                          & 0.310 & 0.345 & 0.385 & $\langle$ 1.5$\sigma$ 
                                                                                                                                                 isocontours $\rangle$,
                                                                                                                                                 smoothed\\
       &                                                         &          &         &     &                          & 0.058 & 0.064 & 0.078 & error from dispersion\\
(vi)  & HAL. $> 4{\rm e}12 M_{\odot}$, real space      & 1                 & $L=200$ & 128 & 11934                    & 0.206 & 0.254 & 0.313 & all, no smoothing \\
       & $b=0.95$, $\Omega_{\rm eff}=0.33$                       &          &         &     &                          & 0.218 & 0.254 & 0.296 & all, smoothed \\
       &                                                         &          &         &     &                          & 0.225 & 0.259 & 0.297 & 1.5$\sigma$ 
                                                                                                                                                 isocontour,
                                                                                                                                                 no smoothing \\
       &                                                         &          &         &     &                          & 0.231 & 0.258 & 0.288 & 1.5$\sigma$ 
                                                                                                                                                 isocontour,
                                                                                                                                                 smoothed \\
       & HAL. $> 4{\rm e}12 M_{\odot}$, red. space  & 125        & $R=70$  & 128 & $2143^{*}$               & 0.155 & 0.220 & 0.320 & $\langle$ all $\rangle$, no smoothing \\
       & $b=0.95$, $\Omega_{\rm eff}=0.33$                       &          &         &     &                          & 0.029 & 0.030 & 0.055 & error from dispersion \\
       &                                                         &          &         &     &                          & 0.178 & 0.221 & 0.277 & $\langle$ all $\rangle$, smoothed \\
       &                                                         &          &         &     &                          & 0.030 & 0.032 & 0.045 & error from dispersion \\
       &                                                         &          &         &     &                          & 0.215 & 0.255 & 0.305 & $\langle$ 1.5$\sigma$ 
                                                                                                                                                 isocontours $\rangle$,
                                                                                                                                                 no smoothing\\
       &                                                         &          &         &     &                          & 0.044 & 0.049 & 0.065 & error from dispersion\\
       &                                                         &          &         &     &                          & 0.215 & 0.245 & 0.282 & $\langle$ 1.5$\sigma$ 
                                                                                                                                                 isocontours $\rangle$,
                                                                                                                                                 smoothed\\
       &                                                         &          &         &     &                          & 0.045 & 0.049 & 0.061 & error from dispersion\\
(vii) & SUBS.   $> 5{\rm e}11 M_{\odot}$, real space      & 1        & $L=200$ & 128 & 55044                    & 0.215 & 0.253 & 0.298 & all, no smoothing \\
       & $b=0.97$, $\Omega_{\rm eff}=0.32$                       &          &         &     &                          & 0.248 & 0.265 & 0.283 & all, smoothed\\
       &                                                         &          &         &     &                          & 0.260 & 0.281 & 0.303 & 1.5$\sigma$ 
                                                                                                                                                 isocontour,
                                                                                                                                                 no smoothing \\
       &                                                         &          &         &     &                          & 0.264 & 0.277 & 0.292 & 1.5$\sigma$ 
                                                                                                                                                 isocontour,
                                                                                                                                                 smoothed \\
       & SUBS.   $> 5{\rm e}11 M_{\odot}$, red. space  & 125     & $R=70$  & 128 & $9886^{*}$               & 0.101 & 0.228 & 0.607 & $\langle$ all $\rangle$, no smoothing \\
       & $b=0.97$, $\Omega_{\rm eff}=0.32$                       &          &         &     &                          & 0.024 & 0.020 & 0.223 & error from dispersion \\
       &                                                         &          &         &     &                          & 0.144 & 0.198 & 0.277 & $\langle$ all $\rangle$, smoothed \\ 
       &                                                         &          &         &     &                          & 0.023 & 0.019 & 0.041 & error from dispersion \\
       &                                                         &          &         &     &                          & 0.234 & 0.264 & 0.299 & $\langle$ 1.5$\sigma$ 
                                                                                                                                                 isocontours $\rangle$,
                                                                                                                                                 no smoothing\\
       &                                                         &          &         &     &                          & 0.030 & 0.035 & 0.045 & error from dispersion \\
       &                                                         &          &         &     &                          & 0.224 & 0.244 & 0.267 & $\langle$ 1.5$\sigma$ 
                                                                                                                                                 isocontours $\rangle$,
                                                                                                                                                 smoothed\\
       &                                                         &          &         &     &                          & 0.028 & 0.031 & 0.038 & error from dispersion\\
(viii)   & SUBS.   $> 4{\rm e}12 M_{\odot}$, real space      & 1        & $L=200$ & 128 & 11221                    & 0.189 & 0.226 & 0.269 & all, no smoothing \\
       &      $b=1.035$, $\Omega_{\rm eff}=0.28$                 &          &         &     &                          & 0.204 & 0.228 & 0.254 & all, smoothed\\
       &                                                         &          &         &     &                          & 0.215 & 0.239 & 0.266 & 1.5$\sigma$ 
                                                                                                                                                 isocontour,
                                                                                                                                                 no smoothing \\
       &                                                         &          &         &     &                          & 0.218 & 0.236 & 0.255 & 1.5$\sigma$ 
                                                                                                                                                 isocontour,
                                                                                                                                                 smoothed \\
       & SUBS.   $> 4{\rm e}12 M_{\odot}$, red. space  & 125     & $R=70$  & 128 & $2015^{*}$               & 0.102 & 0.176 & 0.327 & $\langle$ all $\rangle$, no smoothing \\
       & $b=1.035$, $\Omega_{\rm eff}=0.28$                      &          &         &     &                          & 0.024 & 0.020 & 0.082 & error from dispersion \\
       &                                                         &          &         &     &                          & 0.123 & 0.170 & 0.241 & $\langle$ all $\rangle$, smoothed\\
       &                                                         &          &         &     &                          & 0.023 & 0.021 & 0.040 & error from dispersion\\
       &                                                         &          &         &     &                          & 0.178 & 0.211 & 0.251 & $\langle$ 1.5$\sigma$ 
                                                                                                                                                 isocontours $\rangle$,
                                                                                                                                                 no smoothing\\
       &                                                         &          &         &     &                          & 0.030 & 0.033 & 0.046 & error from dispersion\\
       &                                                         &          &         &     &                          & 0.177 & 0.201 & 0.230 & $\langle$ 1.5$\sigma$ 
                                                                                                                                                 isocontours $\rangle$,
                                                                                                                                                 smoothed\\
       &                                                         &          &         &     &                          & 0.030 & 0.032 & 0.041 & error from dispersion\\
\hline
\end{tabular}
\end{table*}
\begin{description}
\item[(v)] A ``halos'' catalog, involving 43482
dark matter halos with masses larger than $M_{\rm th}=5.10^{11} \ M_{\odot}$; 
\item[(vi)] A ``massive halos'' catalog, involving 11934
dark matter halos with masses larger than $M_{\rm th}=4.10^{12} \ M_{\odot}$;
\item[(vii)] A ``sub-structures'' catalog, involving 55044
dark matter sub-halos with masses larger than  $M_{\rm th}=5.10^{11} \ M_{\odot}$;
\item[(viii)] A ``massive sub-structures'' catalog, 
involving 11221
dark matter sub-halos with masses larger than $M_{\rm th}=4.10^{12} \ M_{\odot}$.
\end{description} 
From each of these catalogs, we compute the velocity and gravity field
on a $128^3$ grid as explained in \S~\ref{sec:algomes}, in real and
redshift space. In the latter case, we use 125 different observers and
perform the measurements in a sphere of radius $70\, \hmpc$ Mpc
centered on the observer exactly as was done to generate the pure dark
matter realizations (iv) in \S~\ref{sec:dmsamples}.

The philosophy underlying to the making of these catalogs relies on the facts that: 
\begin{description}
\item[(a)] dark matter substructures are reasonable tracers of galaxies
(e.g. Springel et al. 2001; Weinberg et al. 2006). In that
sense, one can consider the sub-structure catalogs as the most realistic. 
This is known to be true only to a limited extent. Indeed, sub-halos tend to be tidally disrupted
while they spiral in their host halo, therefore one expects less
sub-structures than galaxies in the core of dark matter halos
(e.g.  Diemand et al. 2004; Nagai \& Kravtsov 2005). This is all the more
true since we apply a mass thresholding to control the number of objects in our catalogs.
However, the effect of such a depletion should be noticeable only at the smallest scales and
should not affect significantly the results of the analyses at the level of accuracy 
reached in this paper. 
\item[(b)] dark matter halos are representative of the galaxy distribution in terms
of an ensemble of structures composed of 
clusters, groups of galaxies and field galaxies. Therefore, if one considers the
galaxy distribution from a slightly different perspective, these structures 
can be used as well to
study the velocity--gravity relation, with the appropriate weighting.
The potential advantage of such an approach is to reduce considerably 
the fingers of God effects discussed previously, since all the galaxies belonging
to one cluster or group of galaxies are collapsed to a single point. 
Note that we oversimplify the analyses here 
by purposely giving the same weight to all the dark matter halos: this 
rather extreme procedure is expected to introduce
quite significant (anti-)biasing effects on the gravity field determination. 
\end{description}

%
%
\subsubsection{Quantitative measurements and the effect of the bias}
\label{sec:mockmes2}
\begin{figure*}
\centerline{\hbox{
\psfig{file=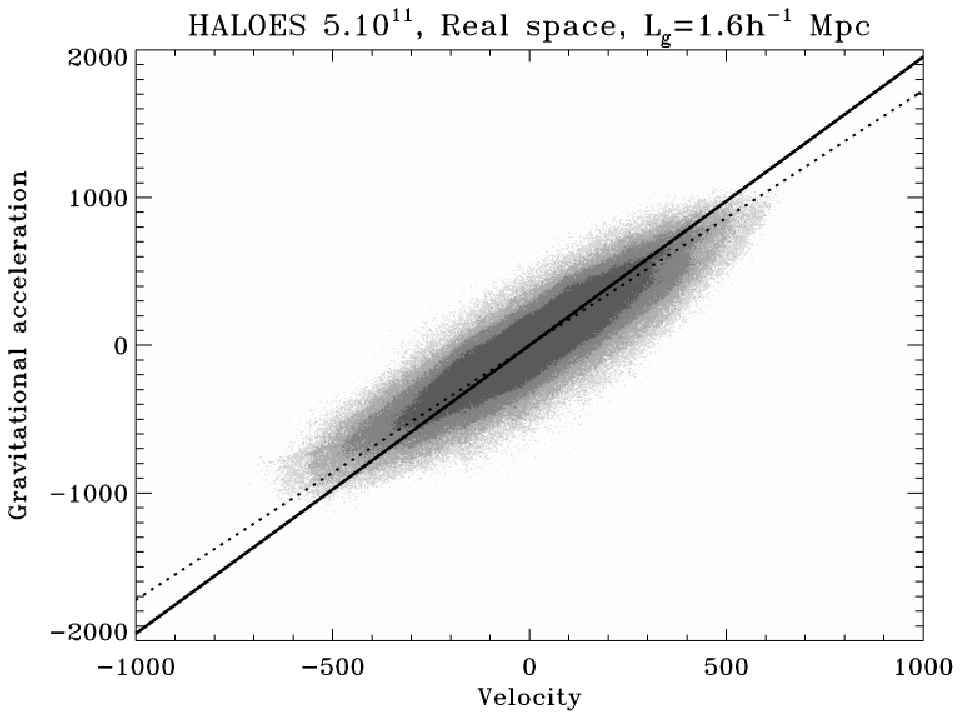,width=6cm}
\psfig{file=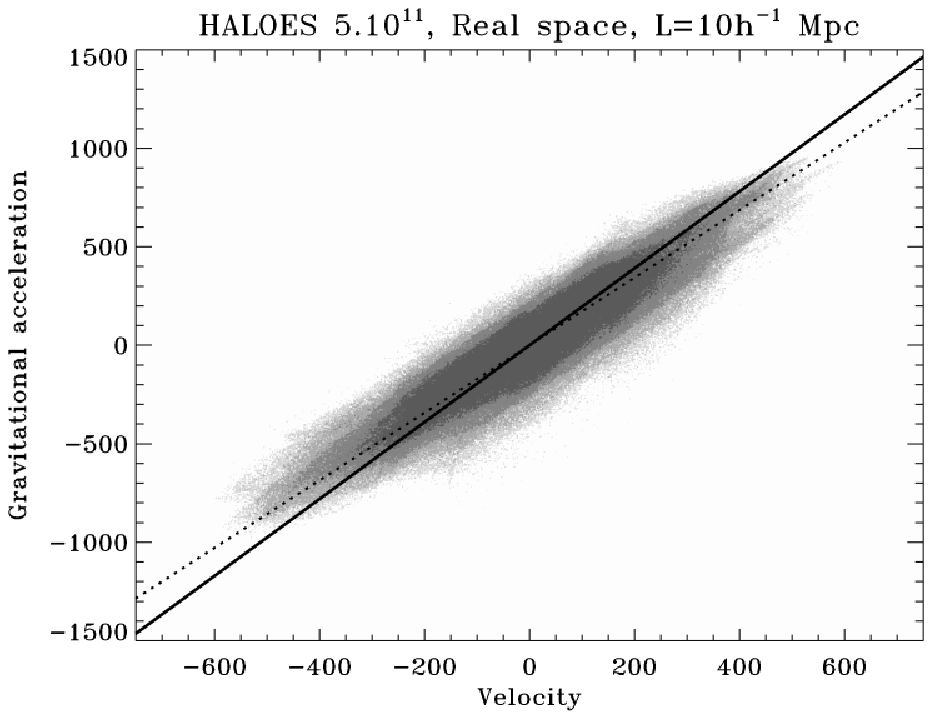,width=6cm}
}}
\centerline{\hbox{
\psfig{file=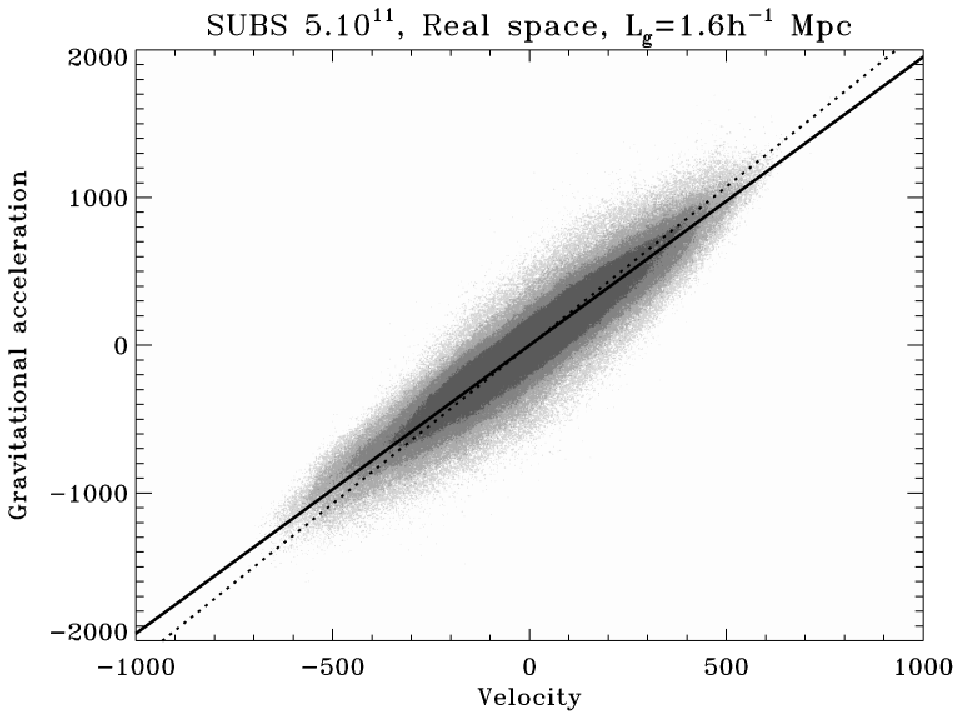,width=6cm}
\psfig{file=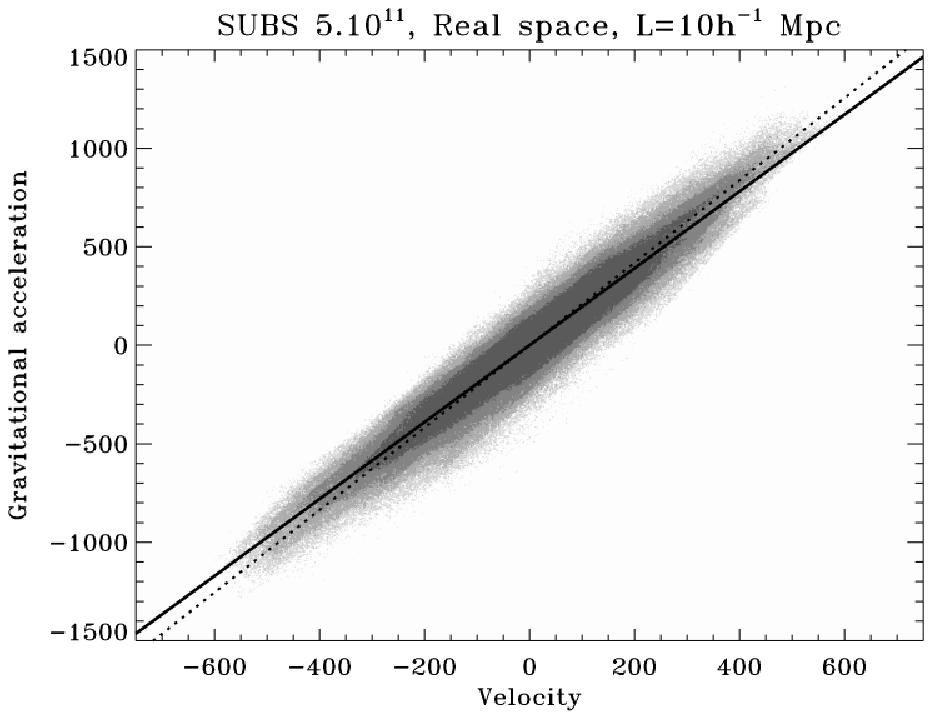,width=6cm}
}}
\centerline{\hbox{
\psfig{file=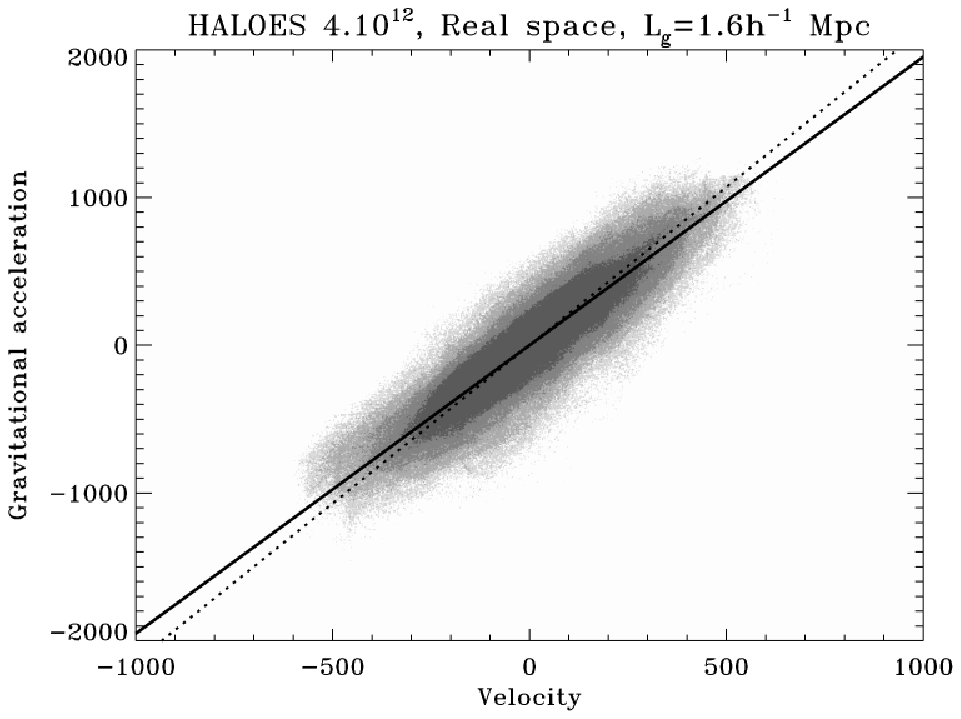,width=6cm}
\psfig{file=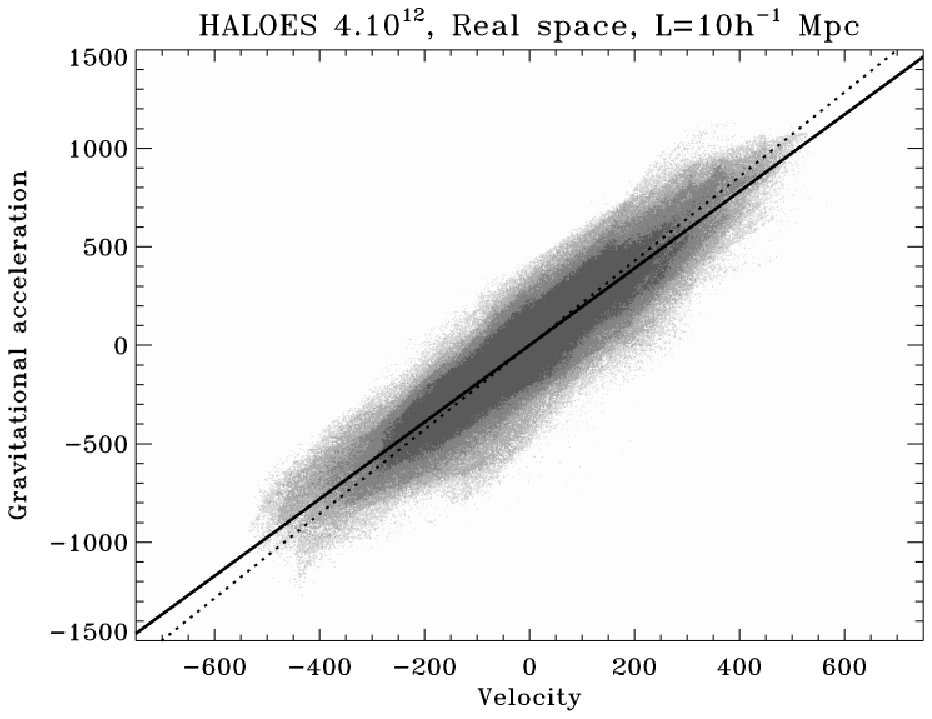,width=6cm}
}}
\centerline{\hbox{
\psfig{file=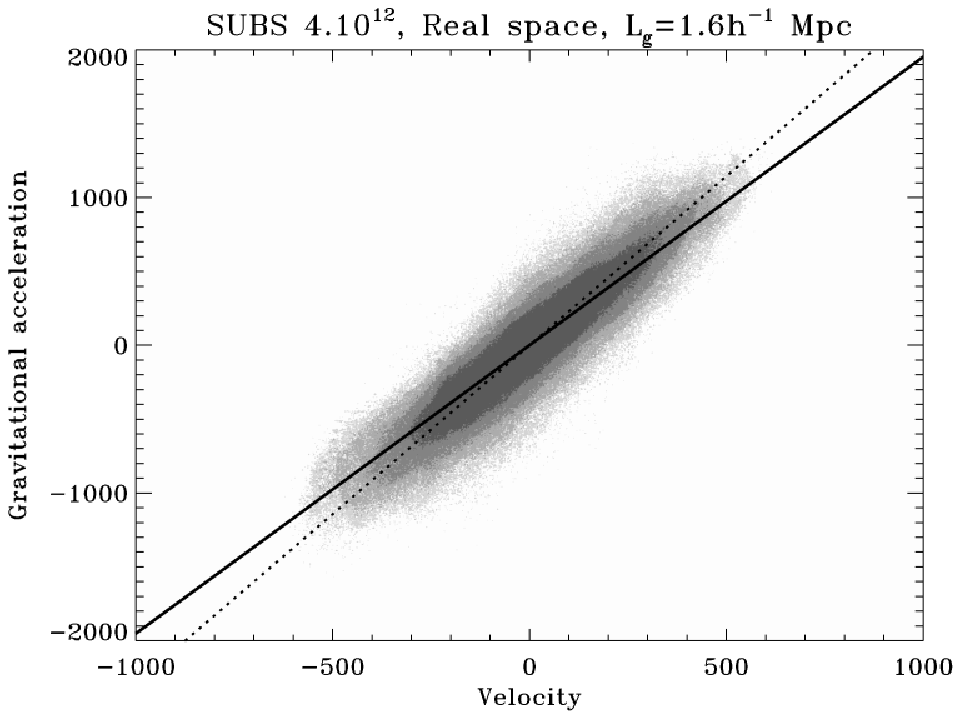,width=6cm}
\psfig{file=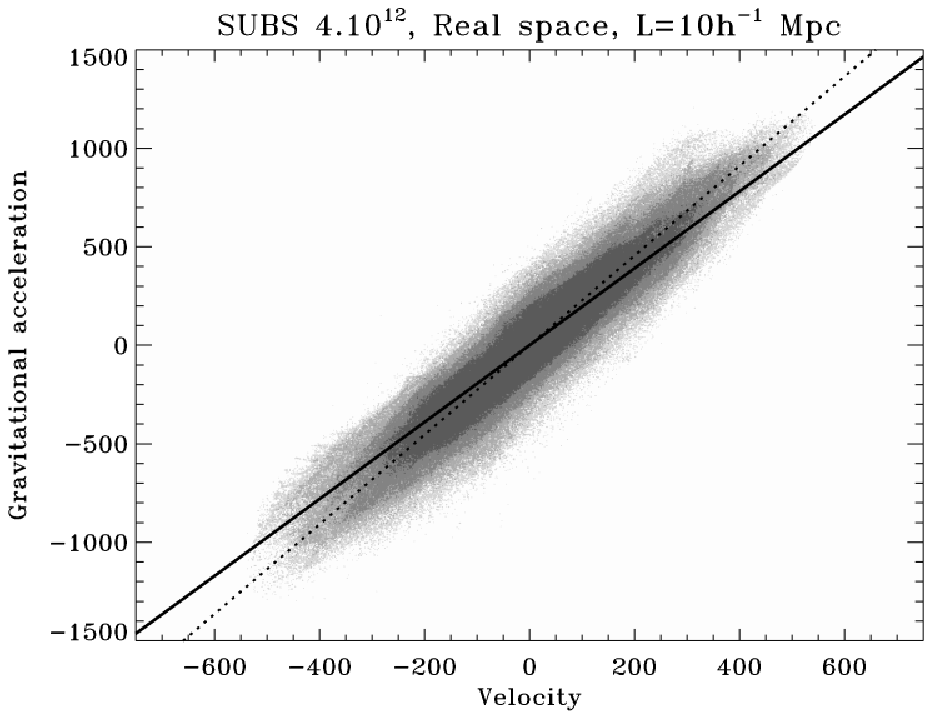,width=6cm}
}}
\caption[]{The velocity--gravity relation in real space as measured in our mock ``galaxy'' catalogs
(v)--(viii). The left column of panels corresponds to the direct SPH alike interpolation 
on a $128^3$ grid, while an additional Gaussian smoothing with a window of size $10\ h^{-1}$ Mpc
was performed in the right column of panels. The informations on the catalogs 
are given as titles on each panel, namely, from top to bottom, halo catalog with mass thresholding
$M_{\rm th}=5.10^{11} M_{\odot}$, sub-structure catalog with $M_{\rm th}=5.10^{11} M_{\odot}$,
halo catalog with $M_{\rm th}=4.10^{12} M_{\odot}$ and 
sub-structure catalog with $M_{\rm th}=4.10^{12} M_{\odot}$. 
The prediction from linear theory in absence of bias is given as the 
solid line while the slope obtained directly from the ratio
$\sqrt{\langle g_z^2 \rangle/\langle v_z^2 \rangle}$ is represented as
a dotted line.}
\label{fig:catreal}
\end{figure*}

\begin{figure*}
\centerline{\hbox{
\psfig{file=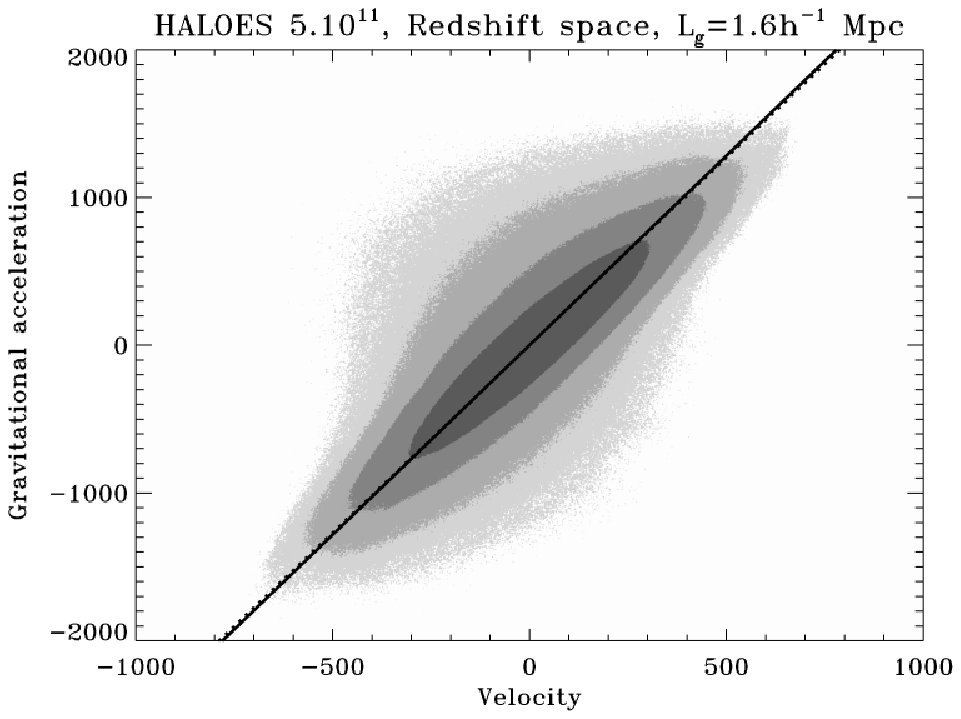,width=6cm}
\psfig{file=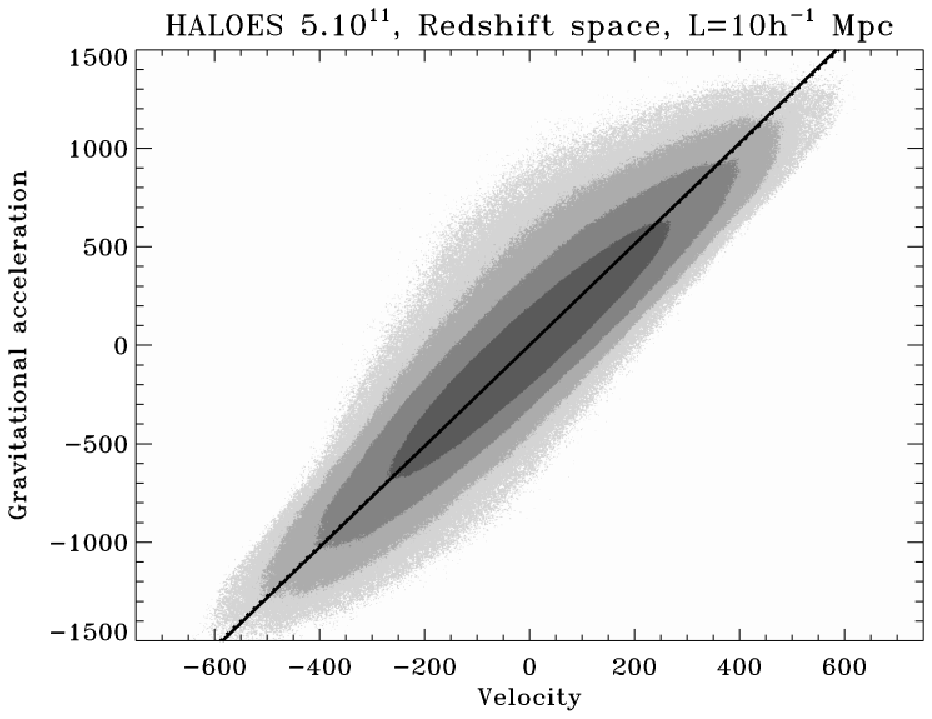,width=6cm}
}}
\centerline{\hbox{
\psfig{file=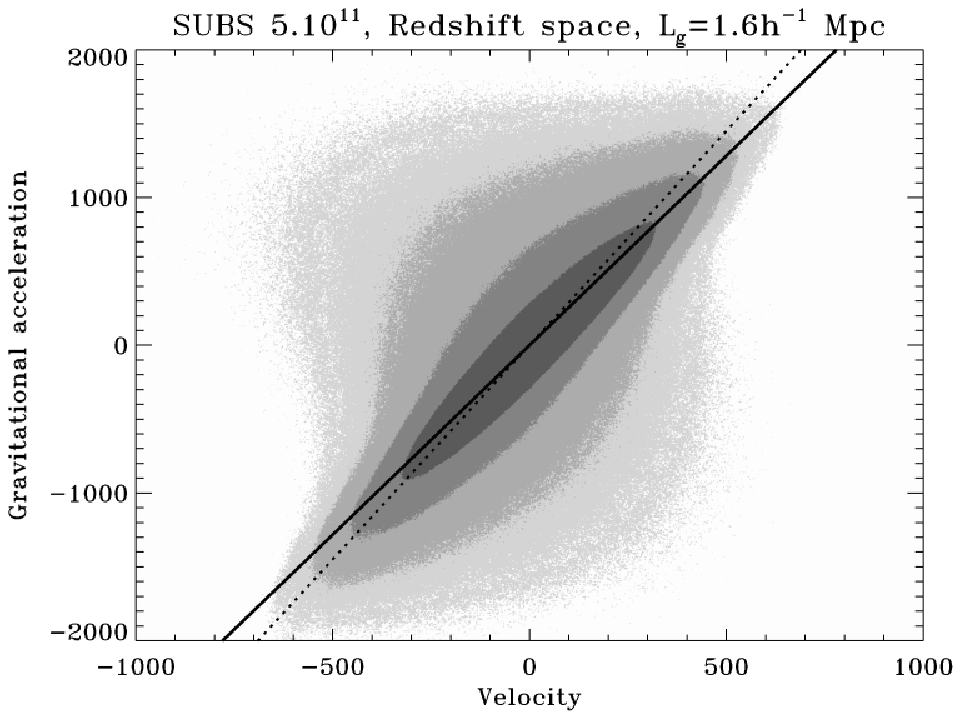,width=6cm}
\psfig{file=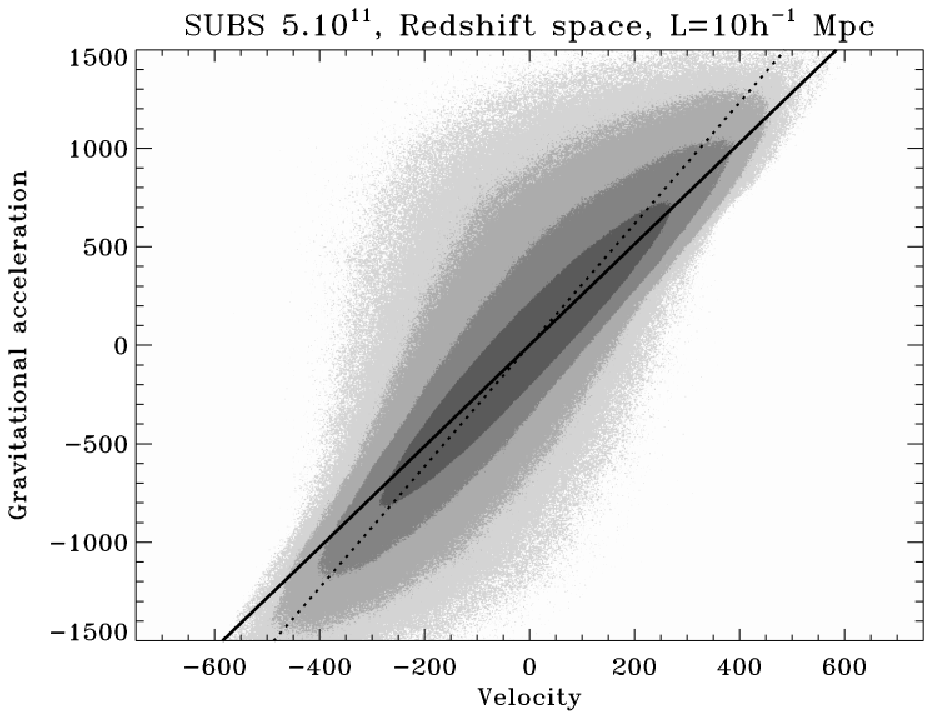,width=6cm}
}}
\centerline{\hbox{
\psfig{file=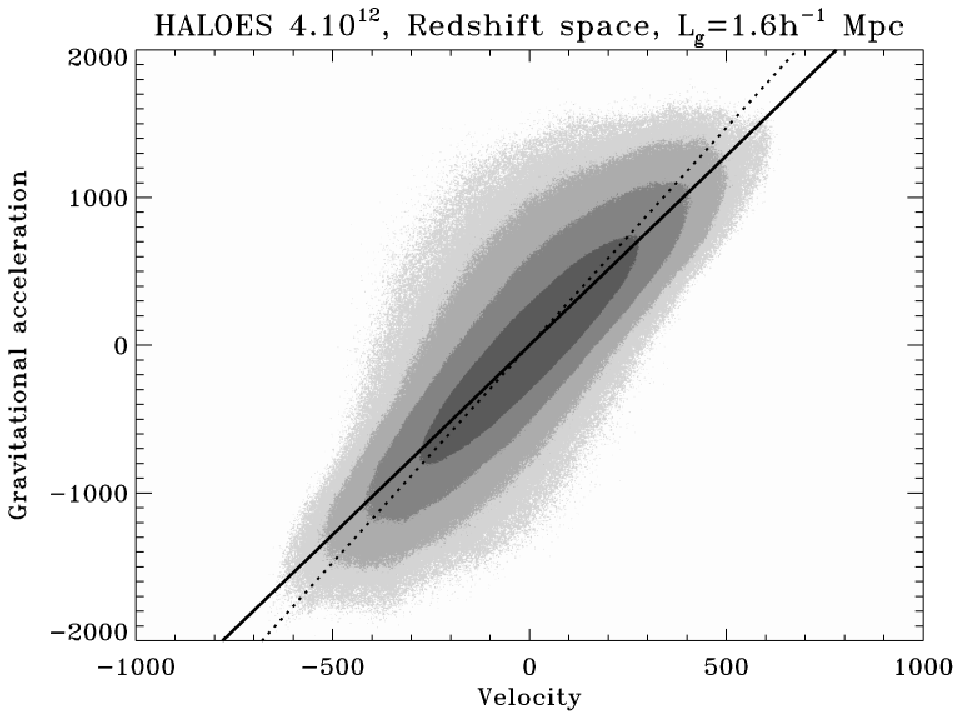,width=6cm}
\psfig{file=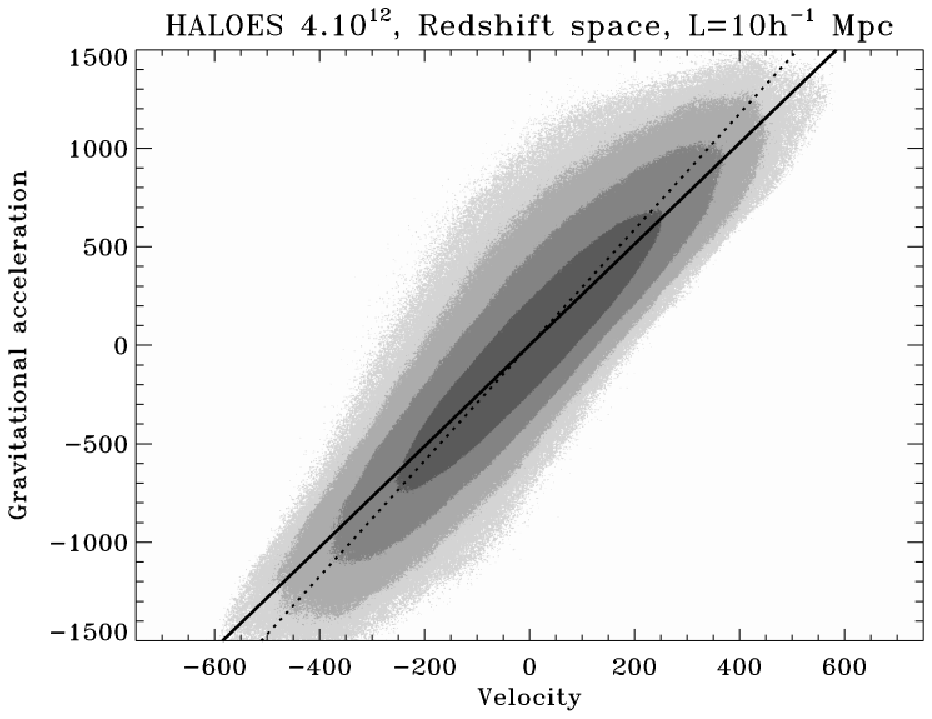,width=6cm}
}}
\centerline{\hbox{
\psfig{file=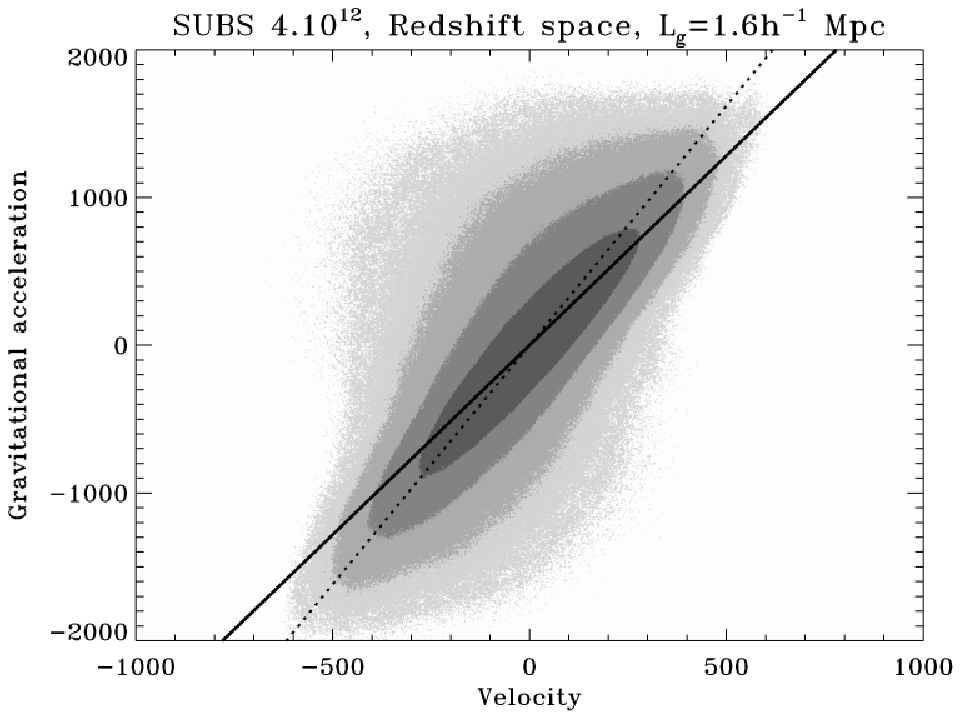,width=6cm}
\psfig{file=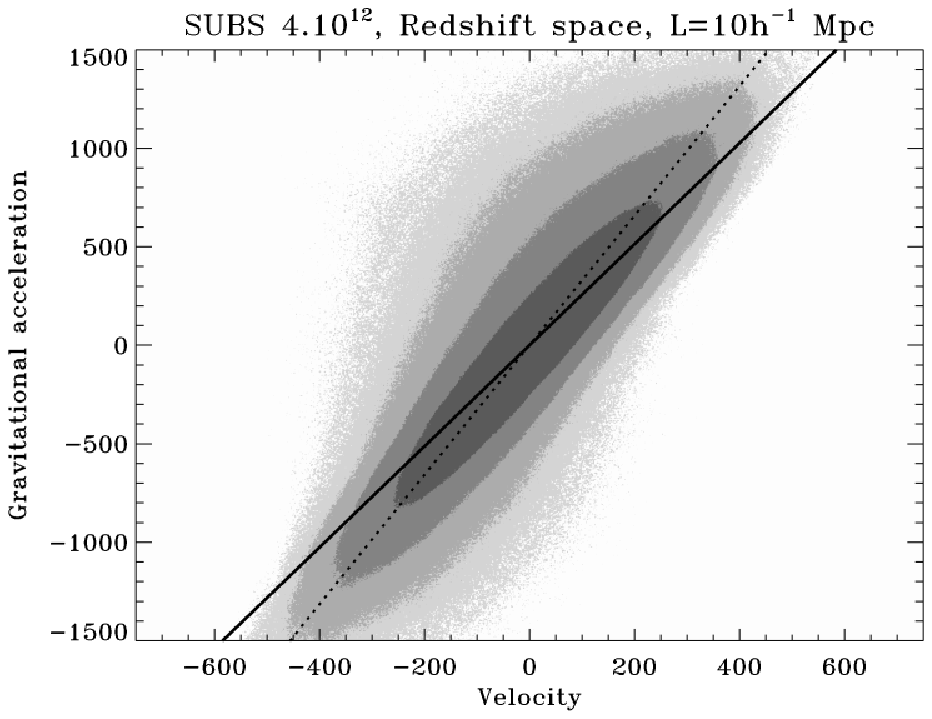,width=6cm}
}}
\caption[]{Same as in Fig.~\ref{fig:catreal}, but in redshift space.
The scatter plots are obtained from the average of 125 observer
positions disposed on a regular pattern spanning the simulation
volume. As expected, finger of God effects are more prominent on the
sub-structures catalogs than on the halos catalogs.}
\label{fig:catred}
\end{figure*}

%
The results of our analyses are summarized in Table~\ref{tab:table2} and illustrated
by Figs.~\ref{fig:catreal} and \ref{fig:catred}. 
As shown in \S~\ref{sec:dmsamples}, due to nonlinear effects, measurements
based on the direct estimation of the moments of the joint probability of velocity and gravity
tend to bias the estimated linear $\beta$ to lower values, both in real and redshift space. 
To reduce this systematic effect, we propose to consider only the region of best likelihood.
Strictly speaking, this approach makes sense only if the bias between
the galaxy distribution and the dark matter distribution is either inexistent or linear.  
The concept of a non local, scale dependent bias as experienced here 
complicates considerably the analyses. However, the aim is to capture the linear 
contribution of the dynamics, which
in terms of our volume-weighted measurements is dominated by underdense regions.
In these regions, the size of the adaptive kernel is large: inside the contour of best likelihood
of the joint PDF, the measurements are expected to be dominated
by the large scale bias, $b(k)$, with small wavenumber $k$. In that regime, the bias is in general
roughly linear or close to linear. The need to actually
consider the region of best likelihood of the PDF to perform the measurements is therefore 
furthermore justified. As a result, 
in what follows, only results of Table~\ref{tab:table2} obtained from selecting the
region inside the $1.5\sigma$ contour of the PDF are considered. Since additional smoothing tends
to introduce additional bias in the redshift space measurements, it will not be discussed either, although
the results are shown in Table~\ref{tab:table2} for completeness.

We first discuss real space measurements
and try to understand the effects of the bias.
Sample (v) is the halo catalog with low mass threshold. This catalog is
expected to present significant antibias. Indeed, if small halos, which mainly lie
in moderately dense and underdense regions, trace rather well the underlying dark matter
distribution, the largest ones, which are collapsed to a single point, induce a
significant underestimate of the overall strength of the dark matter gravity field, resulting
in the antibias effect. As shown in Table~\ref{tab:table2}, the value of the bias measured
at the simulation box size (using the Fourier modes at the largest scales) gives $b(k=2\pi/L)=0.8$.
If one uses, as argued above, this value of $b$ as a reference, 
the effective value of $\Omega$ obtained from linear theory, ignoring the bias,
is larger than the true one, $\Omega_{\rm eff}=\Omega/b^{9/5}=0.45$. This is to be compared
to the value of $0.37$ obtained from $\sqrt{\langle g^2 \rangle/\langle v^2 \rangle}$. Although
this latter result is significantly lower than the ``expected'' value, two facts have to be taken into account:
\begin{enumerate}
\item[(a)] As an effect of non local, scale dependent biasing, the correlation
between gravity and velocity field is not as tight as for the dark matter 
diluted sample, D1, albeit this latter
contains the same number of objects. 
Indeed, one measures $\Omega=0.31$ and $0.42$ from 
$\langle g^2 \rangle/\langle v g \rangle$ and $\langle v g \rangle/\langle v^2 \rangle$ respectively in
sample (v), while $\Omega$ was ranging between $0.28$ and $0.32$ for D1. This loss
of tightness in the velocity-gravity relation obviously affects in a non trivial way the region of best likelihood.
\item[(b)] Our approach
for estimating the effects of the bias is rather crude and does not give account of the 
additional subtleties related to our adaptive smoothing procedure. 
\end{enumerate}
Taking these two points into account, 
the measurements are rather consistent with the expectations, at least at the qualitative level.

In the halo catalog (vi), small mass
``galaxies'' are removed, so the underdense regions are not populated anymore, leaving us with a set
of more clustered objects than in sample (v), since the higher is their mass, the more 
significantly the halos are
clustered. The consequence is that the antibias found in catalog (v) is reduced to $b(k=2\pi/L)=0.95$,
resulting in an effective expected value of $\Omega_{\rm eff}=0.33$, to be compared with the
value of $0.26$ obtained from the measurement of $\sqrt{\langle g^2 \rangle/\langle v^2 \rangle}$.
Again the arguments (a) and (b) developed just above apply, leading us to conclude that this
result is consistent with expectations, and similarly for the substructures catalogs:
sample (vii) is the closest to the underlying dark matter distribution, with a very slight 
antibias of $b(k=2\pi/L)=0.97$ and $\Omega_{\rm eff}=0.32$ to be compared to $0.28$ from 
$\sqrt{\langle g^2 \rangle/\langle v^2 \rangle}$; sample (viii) presents a small positive
large scale bias, with $b(k=2\pi/L)=1.035$ and $\Omega_{\rm eff}=0.28$, to be compared to $0.24$ from 
$\sqrt{\langle g^2 \rangle/\langle v^2 \rangle}$.

We can conclude here that we understand what are the various effects
influencing the measurements in the mock catalogs. The general trend
is that the ``true'' value of parameter $\beta$ (i.e. the very large
scale one) seems to be underestimated by about 10 to 20 percent. We
give here quotes, because such a value is not well defined. Indeed,
the interpretation of the measurements is complicated by adaptive
filtering, combined with a non local, scale dependent bias. But as
discussed above, we really want to capture the value of $\beta$ in the
linear regime limit, where the bias is expected to be roughly linear,
scale independent and in fact rather close to unity, as observed in
our ``galaxy'' catalogs.  Note interestingly that the conditional
moment $\langle v g \rangle/\langle v^2 \rangle$ gives a rather good
estimate of the expected value of $\beta$. Without going too far
exploiting such a property, which may be specific to our catalogs, it
is clear that the slopes given by the two conditional moments could be
used as proxies to estimate a range of possible values for $\beta$, or
in other words, errorbars. Indeed, the difference between these slopes
is related to the width of the likelihood region, i.e. it estimates
the tightness of the correlation between gravity and velocity as
measured in the catalog. However, note that this difference does not
properly give account of systematic effects discussed at
length in this paper, as well as cosmic variance effects, although it is
indirectly related to them. 

We now turn to redshift space measurements. What matters is that they should be self-consistent
with the real space ones. This is nearly the case, except again that they tend to
underestimate real space measurements by up to ten percent, an effect of 
sparse sampling discussed in \S~\ref{sec:diluef}, with no additional surprise. 
Overall, the difference between the measured value of $\Omega$ in redshift space 
and the ``expected', effective one, $\Omega_{\rm eff}$, 
is about 25 percent, with a systematic bias to lower values, as always. 
Obviously our mock catalogs are very peculiar in a sense that they represent quite
an extreme case in the range of possibilities. And still, without any assumption on
the bias, or in other words, assuming that $b$ is of the order of unity, we can determine
$\Omega$ with an accuracy of the order of 30 percents without significantly strong prior.
Furthermore, the quality of this estimate can certainly be improved with appropriate
weighting of the galaxies. However, as discussed in
the conclusion that follows, the sources of systematic errors related to the instrumental
noise or to the method used to provide velocities were not considered here. 
%
\section{Conclusions}
\label{sec:concl}
%
In cosmic density--velocity comparisons, since the density field is
given solely in redshift space, redshift-space effects have to be
accounted for. One method is to perform the analysis directly in
redshift space, as proposed by ND (Nusser \& Davis 1994). 
Another one is to reconstruct the
real-space density field, but this method is complicated by the fact
that redshift-space corrections require assuming a value for
$\beta$. As $\beta$ is not known a priori, one is forced to
reconstruct the density field for a range of values of $\beta$, which
makes the real-space comparison involved, despite its apparent
simplicity. In the present paper we have proposed a method that works
in redshift space, but is simpler than the formalism of ND, which is
based on a spherical harmonics expansion.  It is essentially as simple
as in real-space; the only difference is that here a directly
estimated parameter is not just $\beta$, but its simple function.

This method relies on a tight correlation between cosmic gravity and
cosmic velocity in redshift space. We have derived the
velocity--gravity relation in redshift space and its scatter
analytically, under the assumptions of the linear regime,
distant observer limit (DOL) and CMB rest frame for redshift 
measurements.\footnote{Using CMB rest frame is {\em crucial}
to avoid the so-called {\em rocket effect} on the redshift
space gravitational acceleration, as discussed in \S~\ref{sec:valdol}.}
This relation has turned out to be a
simple modification of the corresponding one in real space and in the
linear regime. Then we have performed a dark matter $N$-body experiment
to test the 
velocity--gravity relation both in real space and in
redshift space. 
We also extracted mock galaxy catalogs from this data set,
with rather extreme prescription for selecting ``galaxies'' from the dark 
matter distribution in order to analyze
the effects of non-trivial biasing on the velocity--gravity relation.
The main results are the following:
\begin{itemize}
\item To perform the measurements, we propose to interpolate the
gravity and the velocity fields on a regular grid using a new adaptive
smoothing procedure described in \S~\ref{sec:algomes}. This method is
close to SPH interpolation, but without the defect of having regions
of space where the velocity field is undefined. Although our algorithm
is expected to be less optimal than those based on Voronoi or Delaunay
interpolation (e.g., Bernardeau et al. 1997; Pelupessy, Schaap \& van
de Weygaert 2003), it has the advantage of being simpler to implement
and much faster, and can be applied to very large sets such as the
$512^3$ sample analyzed in this paper.

\item In real space, simple measurements using moments of the full
joint probability distribution function (PDF) 
bias $\Omega$ to low values. This is due to the fact that
these moments are contaminated by the tails of the PDF. These latter
are influenced by nonlinear dynamics, which induces a global
`propeller' shape for the PDF. 

\item In redshift space, fingers of God (FOG) induce additional 
``anti-diagonal'' tails to the PDF. This effect seems to somewhat
compensate the bias due to the propeller shape. However this result
could be coincidental and is too sensitive to the level of smoothing
applied to the data. Note that in actual peculiar velocity catalogs, clusters
of galaxies are in general collapsed into one point to improve the
signal to noise ratio of the distance estimates. Although FOG 
are thus less of an issue in real data, rich filaments can
still contribute to an effect similar to FOG as discussed in \S~\ref{sec:visu}.

\item Selecting about a $1.5\sigma$ (68\% confidence) region around
the maximum of the joint PDF of gravity and velocity fields makes the
data to match the linear theory predictions very well, both in real
and in redshift space. This result
stands even in the highly nonlinear regime (no additional smoothing
except the adaptive interpolation on the grid). It is easily explained
by the fact that our measurements are volume weighted, giving most of
the statistical weight to underdense regions. Therefore, it is
sufficient to exclude only the tails of the distribution to obtain
unbiased estimate of $\Omega$.

\item Additional smoothing does not improve the agreement with linear
theory predictions: first, it is expected to increase
deviations from the DOL limit; second, it makes the situation worse in the
region of best likelihood, where it mixes linear contributions with
nonlinear ones and biases $\beta$ to
lower values, an effect which is minimized as much as possible
with our adaptive smoothing procedure. 

\item The measurements obtained in our mock catalogs provide a value
of $\beta$ in agreement with intuition, namely that large scale bias
dominates if the region of best likelihood is selected. This is
inherent to our volume weighted approach which gives more weight to
underdense regions (see also Berlind, Narayanan \& Weinberg 2001).
However, due to the dilute nature of real catalogs, we expect an
effective bias on $\beta$ to lower values.  This bias was found to be
of the order of 10--20 $\%$ for the mock samples considered in this
paper. In addition, the velocity--gravity relation looses some
tightness due to effects of non-local biasing between luminous and
dark matter distributions.  Nevertheless, our analyses show that
direct measurement of $\Omega$ from real catalogs assuming no bias
should give an answer accurate to about $30$ $\%$ at worse. Moreover,
this error can probably be significantly improved to values as low as
10 $\%$, with the appropriate weighting of the data (purposely not
performed here), as discussed below.
\end{itemize}

There are several points we did not address in our numerical analyses,
and we leave them for future work:
\begin{description} 
\item[{\em - Noise:}] through the analysis of
dilute samples and mock catalogs, we addressed to some extent the effects of shot
noise for realistic modern catalogs. However, we used the same
catalog for velocity and gravity fields, implying that 
the same adaptive kernel can be used for both fields, which is not realistic.\footnote{If the
catalogs are different, a common smoothing kernel must be determine to find a compromise between
the velocity and the gravity samples.}
Furthermore we did not take into
account the fact that there is in practice a significant relative error on galaxy distance 
estimates, which can be as large as 20 percent. 
Thus the velocity--gravity relation is certainly not expected to
be as tight in real catalogs as in the samples considered in this paper.
Note interestingly, though, that since the measurements are performed in redshift space,
issues related to Malmquist bias (e.g., Strauss \&
Willick 1995) should be irrelevant, as long as the distance
estimates are unbiased (just noisy). 

\item[{\em - Incompleteness:}] Another closely related issue is 
that the catalogs used to estimate the gravity field are incomplete. 
First, edge effects are expected to be significant because gravity
force is of long range. The galaxy catalog used to estimate the gravity
field has to be significantly deeper than the region used
to perform the velocity--gravity comparison. We did not estimate in this
paper how deep it has to be, not to mention the problem of obscuring
by our own Galaxy. Second, the apparent density of
galaxies decreases with increasing distance from the observer. In
order to estimate the interpolated density field, one can, under the
assumption of no bias between the galaxy distribution and the mass
distribution, weight the galaxies by the inverse of the selection
function, $\phi$. In the linear regime, in the absence of bias and
for redshifts measured relative to the CMB rest frame, the redshift
space density contrast measured that way reads (Kaiser 1987)
\begin{equation}
\delta_{\bf s}=\delta-\left(2 + \frac{d\ln\phi}{d\ln r} \right) \frac{v}{r}
-\frac{\partial v}{\partial r}.
\label{eq:kaiser}
\end{equation}
Under the DOL, the second term on the right hand side
of this equation drops, so the effect of the selection function
disappears.  However, since remote parts of the catalogs are sparser,
they might be given lesser statistical weight, hence augmenting the relative
contribution of this deviation from the distant observer limit. 

\item[{\em - Better handling of the bias:}] our analyses in the simple
mock catalogs show that biasing affects the velocity--gravity relation
in a non trivial way.  However, we used the most naive prescription to
compute the gravity field, giving equal weight to all galaxies.
Clearly, a better description would be to assign to each galaxy a
weight proportional to its supposed host halo mass, for example
assuming constant mass to light ratio, $M/L$, or a more sophisticated
weighting using a function $M/L=f(L)$ derived from observations or
obtained from theoretical models of galaxy formation. That would allow
one to correct to a large extent for the effects of biasing, at the
cost of additional priors. Still, unless the galaxy catalog used to
compute the gravity field is very deep, i.e. includes faint galaxies,
even giving the proper mass to light ratio to each galaxy does not
correct for the fact that underdense regions can be artificially
underpopulated, especially far away from the observer.  To tackle with
that, one could add a background population of low mass objects that
would give account of the missing mass in the catalog and that would
act as a shielding effect on the gravity field (see, e.g., Phelps et
al. 2006).  Again, some strong priors are required to deal with such a
background, in particular on its clustering properties and how it
correlates with the population already present in the real catalog.
\end{description}

Note finally that while our approach presents the advantage of
simplicity compared to ND (and most other methods), we do not expect
it to be as accurate as of ND, since it relies on the distant observer
approximation and uses in its current form adaptive smoothing, which
complicates the interpretation of the results. (On the other hand, we
clearly showed that the region of best likelihood of the joint
velocity--gravity distribution is dominated by the linear regime
prediction, even in the highly nonlinear regime, which presents a
noticeable advantage.) Similarly, our method is not expected to
perform as well as sophisticated reconstructions of Lagrangian nature,
which try to minimize the Euler-Lagrange action (e.g., Peebles 1989;
Shaya, Peebles \& Tully 1995; Nusser \& Branchini 2000; Phelps 2002;
Phelps et al. 2006) or to solve optimal assignment problem (e.g.,
Croft \& Gazta\~naga 1997; Frisch et al. 2002; Mohayaee et al. 2003;
Mohayaee \& Tully 2005), although this remains to be verified.

\section*{acknowledgments}
This work was carried out within the framework of the European
Associated Laboratory ``Astronomy Poland--France'' and was performed
within the Numerical Investigations in Cosmology (NIC) group as a task
of the HORIZON project. This research has been also supported in part
by the Polish State Committee for Scientific Research grant No.~1 P03D
012 26, allocated for the period 2004--2007. The computational
resources (HP cluster)  for the present numerical
simulation were made available to us by  
Centre de Calcul en Recherche et Technologie (CCRT, CEA).

\appendix
\section{Spherical top-hat in redshift space}
\label{app:gen}

The general result in real space is

\be
\frac{g_3}{v_3} = \beta^{-1} \frac{\delta}{\theta} \,,
\label{eq:real_space}
\ee 
where $\delta$ and $\theta$ are the real-space density contrast
and velocity divergence, respectively. In redshift space, in the
distant observer limit, the general result is

\be 
\frac{g_3}{v_3} = \frac{3}{2} C\, \beta^{-1}
\left(\frac{\delta}{\theta} + \frac{\beta}{3}\right).
\label{eq:redshift_space}
\ee
Here, 

\be
C = \frac{2(1-e^2)}{e^2} \left[(1-e^2)^{-1/2} - \frac{\sin^{-1}{\!e}}{e}
 \right],
\label{eq:C}
\ee
eccentricity

\be
e = (1 - \varrho^2)^{1/2}, 
\label{eq:eccent}
\ee
and

\be
\varrho = 1 - \beta \theta/3 \,.
\label{eq:ratio}
\ee

In the linear regime, $C = 2/3$ and $\theta = \delta$. Hence, 

\be 
\frac{g_3}{v_3} = \beta^{-1} \left(1 + \frac{\beta}{3}\right) .
\label{eq:red_lin}
\ee 
We see thus that the spherical top-hat model does not give the
statistical result obtained from linear theory. The reason for this is
the following. In the spherical model, the velocity--gravity relation
is fully deterministic, but we could also derive it calculating the
relevant statistical averages. While in the linear regime, in general 

\be
\tilde\de^{(g)}_\bfs(\bfk) = b (1+ \beta \mu^2) \tilde\de(\bfk) \,
\label{eq:dg_s2}
\ee
[where $\mu = \bfk\cdot\bfs/(k s)$], in the spherical top-hat we have 

\be
\de^{(g)}_{\bfs,\rm TH} = b (1+ \beta/3)\de \,,
\label{eq:dg_s_TH}
\ee that immediately yields 
\be 
\tilde\de^{(g)}_{\bfs,\rm TH}(\bfk) = b (1+ \beta\lan\mu^2\ran)
\tilde\de(\bfk) \,,
\label{eq:dg_s_TH_k}
\ee 
since $\lan\mu^2\ran = 1/3$. In other words, the expression for the
redshift space density contrast in the top-hat model is an average of
generally valid expression~(\ref{eq:dg_s2}) over possible orientations
of the vector $\bfk$. In the case of the estimator $\lan v_\bfs
g_\bfs \ran/\lan v_\bfs^2 \ran$, the average over angles yields an
expression proportional to

\be
\frac{\lan \mu^2 (1 + \beta \mu^2)\ran}{\lan \mu^2 \ran} = 1 + \beta
\frac{\lan \mu^4 \ran}{\lan \mu^2 \ran} = 1 + \frac{3}{5}\beta \,.
\label{eq:average}
\ee
In the top-hat model, this is modified to

\be 
\frac{\left\lan \mu^2 \left(1 + \beta \lan\mu^2\ran\right)
\right\ran}{\left\lan \mu^2 \right\ran} = 1 + \beta \frac{\left\lan \mu^2
\right\ran^2}{\left\lan \mu^2 \right\ran} = 1 + \frac{1}{3}\beta \,.
\label{eq:average_TH}
\ee 
Loosely speaking, the top-hat model performs a part of this average
`too early'. So the top-hat model yields quantitatively different
results from the results of a rigorous statistical
calculation. However, it is sufficient for qualitative purposes,
needed here. In particular, it correctly predicts that the slope of
the velocity--gravity relation in redshift space gets steeper. 

Turn-around in real space corresponds to extreme flatness of the
pancake in redshift space. When $\theta \to \theta_t \equiv 3
\beta^{-1}$, then $\varrho \to 0$ and $C$ tends to its maximal value,
$2$. Hence,

\be 
\frac{g_3}{v_3} \to 3 \beta^{-1}
\left(\frac{\delta_t}{\theta_t} + \frac{\beta}{3}\right) \simeq 11.6 \,.
\label{eq:red_pancake}
\ee 
Here, $\delta_t$ is related to $\theta_t$ by the formula of
Bernardeau (1992). In fact, Bernardeau \etal (1999) invented its more
accurate modification:

\be
1 + \delta \simeq (1 + \theta/\alpha)^\alpha ,
\label{eq:de-te}
\ee 
where $\alpha$ is slightly greater than $3/2$. However, for our
purposes $\alpha = 3/2$ should be accurate enough; moreover, it better
describes the evolution of voids [for this value of $\alpha$,
$\delta(\theta = -1.5) = -1$]. This formula works for mildly
non-linear densities. 

For $\theta > 3 \beta^{-1}$, the structure in redshift space is
inverted: velocity, and so the ratio of $g_3$ to $v_3$, changes
sign. Then 

\be
\varrho = \beta \theta/3 - 1\,,
\label{eq:ratio_invert}
\ee
and

\be 
\frac{g_3}{v_3} = - \frac{3}{2} C\, \beta^{-1}
\left(\frac{\delta}{\theta} + \frac{\beta}{3}\right).
\label{eq:red_invert}
\ee 
For $\theta = \theta_i \equiv 6 \beta^{-1}$ we have $\varrho = 1$, so the
structure in redshift space is momentarily an (inverted) sphere, hence
without further effort, $C = 2/3$. Therefore, then

\be 
\frac{g_3}{v_3} = - \beta^{-1} \left(\frac{\delta_i}{\theta_i} +
\frac{\beta}{3}\right) \simeq - 5.42.
\label{eq:red_inv_sphere}
\ee 
For $\theta \gg 3 \beta^{-1}$ (highly nonlinear infall), the
(inverted) structure in redshift space becomes very elongated (finger
of God). Then $\varrho = \beta \theta/3 - 1\simeq \beta \theta/3 \gg
1$ and $e \simeq (1/3)\beta \theta \rm i$. After some algebra, one can
show that then

\be
C \to 2 \frac{\ln{(\beta\theta/3)}}{(\beta\theta/3)^2} \,.
\label{eq:C_asympt}
\ee 
Moreover, then one cannot use formula~(\ref{eq:de-te}), but $\delta
\to (\theta/3)^2$ (Bilicki \& Chodorowski, in preparation), hence
$\delta/\theta \propto \theta$. This yields

\be
- \frac{g_3}{v_3} \propto \theta^{-1} \ln{\theta} .
\label{eq:slope_asympt}
\ee
In other words, $g_3$ becomes very small compared to $|v_3|$. 


\begin{thebibliography}{}
\bibitem{} Aubert D., Pichon C., Colombi S., 2004, MNRAS 352, 376
\bibitem{} Bardeen J. M., Bond J. R., Kaiser N., Szalay A. S., 1986, ApJ 304, 15
\bibitem{} Barnes J., Hut P., 1986, Nature 324, 446
\bibitem{} Barnes J., Hut P., 1989, ApJS 70, 389
\bibitem{} Berlind A. A., Narayanan V. K., Weinberg D., 2001, ApJ 549, 688 
\bibitem{r3} Bernardeau F., 1992, ApJ 390, L61
\bibitem{r4} Bernardeau F., 1994,  ApJ 427, 51
\bibitem{} Bernardeau F., Chodorowski M. J., {\L}okas E. L., Stompor R., Kudlicki A., 1999, MNRAS 349, 543
\bibitem{r5} Bernardeau F., van de Weygaert R., Hivon E., Bouchet F. R., 1997, MNRAS 290, 566
\bibitem{r6} Bertschinger E., 1995, astro-ph/9506070
\bibitem{} Bouchet F. R., Colombi S., Hivon E., Juszkiewicz R., 1995, A\&A 296, 575 
\bibitem{} Branchini E., et al., 1999, MNRAS 308, 1
\bibitem{r8} Chergui J., 2000, Transform\'ees de Fourier Rapides Monoprocesseur
sur NEC SX-5, publications de l'IDRIS (CNRS/IDRIS 2000),
http://www.idris.fr/data/publications/fft-SX5.pdf
\bibitem{r9} Cieciel\c{a}g P., Chodorowski M. J., Kiraga M., Strauss M. A., 
Kudlicki A., Bouchet F. R., 2003, MNRAS 339, 641
\bibitem{} da Costa L. N., Bernardi M., Alonso M. V., Wegner G., Willmar C. N. A., Pellegrini P. S.,
Rit\'e C., Maia M. A. G., 2000, AJ 120, 95
\bibitem{} Croft R. A., Gazta\~naga E., 1997, MNRAS 285, 793
\bibitem{} Davis M., Nusser A.,  Willick J. A., 1996, ApJ 473, 22
\bibitem{} Dekel A., Bertschinger E., Yahil A., Strauss M. A., Davis M., Huchra J. P.,
1993, ApJ 412, 1
\bibitem{} Diemand J., Moore B., Stadel J., 2004, MNRAS 352, 535
\bibitem{r11} Eisenstein D. J., Hu W., 1998, ApJ 496, 605 
\bibitem{} Erdo$\breve{\rm g}$du P., et al., 2006, MNRAS 368, 1515 
\bibitem{} Frisch U., Matarrese S., Mohayaee R., Sobolevski A., 2002, Nature 417, 260
\bibitem{} Kaiser N., 1987, MNRAS 227, 1
\bibitem{} Kaiser N., Efstathiou G., Ellis R., Frenk C., Lawrence A., Rowan-Robinson M., 
Saunders W., 1991, MNRAS 252, 1
\bibitem{} Kaiser N., Lahav O., 1988, In Large Scale Motions in the Universe,
Proceedings of the Vatican Study Week, G. Coyne \& V. C. Rubin, Eds. (Princeton
University Press 1988), p. 339 
\bibitem{} Kaiser N., Stebbins A., 1991, in
Large-Scale Structure and Peculiar Motions in the Universe, ASP Conferences
Series, Vol. 15, D. W. Latham \& L. N. da Costa, Eds., p. 111
\bibitem{r18} Kudlicki A., Plewa T.,  R\'o\.{z}yczka M., 1996, Acta A. 46, 297
\bibitem{r19} Kudlicki A., Chodorowski M. J., Plewa T., R\'o\.{z}yczka M., 2000, MNRAS 316, 464
\bibitem{} Masters K. L., Springob C. M., Haynes M. P., Giovanelli R., Huchra J. P., 2005,
American Astronomical Society Meeting 207, 170.01
\bibitem{} Mohayaee R., Frisch U., Matarese S., Sobolevski A., 2003, A\&A 406, 393
\bibitem{} Mohayaee R., Tully R. B., 2005, ApJ 635, L113
\bibitem{r20} Monaghan J. J., 1992, ARA\&A 30, 543
\bibitem{r21} Moutarde F., Alimi J.-M., Bouchet F.R., Pellat R., Ramani
           A., 1991, ApJ 382, 377 
\bibitem{} Nagai D., Kravtsov A. V., 2005, ApJ 618, 557
\bibitem{} Novikov D., Colombi S., Dor\'e O, 2006, MNRAS 366, 1201
\bibitem{} Nusser A., Branchini E., 2000, MNRAS 313, 587
\bibitem{} Nusser A., Davis M., 1994, ApJ 421, L1 (ND)
\bibitem{} Peebles P. J. E., 1980,  The Large-Scale
Structure of the Universe (Princeton University Press, 1980)
\bibitem{} Peebles P. J. E., 1989, ApJ 344, L53
\bibitem{r24} Pelupessy F. I., Schaap W. E., van de Weygaert R., 2003, A\&A 403, 389
\bibitem{} Phelps S. D., 2002, ApJ 575, 1 
\bibitem{} Phelps S. D., Desjacques V., Nusser A., Shaya E. J., 2006, MNRAS 370, 1361
\bibitem{} Saunders W., et al., 2000, MNRAS 317, 55
\bibitem{} Shaya E. J., Peebles P. J. E., Tully R. B., 1995, ApJ 454, 15
\bibitem{} Sigad Y., Eldar A., Dekel A., Strauss M. A., Yahil A., 1998, ApJ 495, 516
\bibitem{} Springel V., White S. D. M., Tormen G., Kauffmann G., 2001, MNRAS 328, 726
\bibitem{} Strauss M. A., Willick J. A., 1995, PhR 261, 271
\bibitem{r28} Teuler J.-M., 1999, JMFFT: Emulation des Transformees de Fourier CRAY
(CNRS/IDRIS 1999), http://www.idris.fr/data/publications/JMFFT/
\bibitem{r29} Teyssier, R., 2002, A\&A 385, 337
\bibitem{} Weinberg D. H., Colombi S., Dav\'e R., Katz N., 2006, submitted to ApJ (astro-ph//0604393)
\bibitem{} Willick J. A., Courteau S., Faber S. M.,  Burstein D., Dekel A., Strauss M. A.,
1997, ApJS 109, 333
\bibitem{} Willick J. A., Strauss M. A., 1998, ApJ 507, 64
\bibitem{} Yahil A., Strauss M. A., Huchra J. P., 1991, ApJ 372, 380
\bibitem{r32} Zel'dovich Ya. B., 1970, A\&A 5, 84
\end{thebibliography}
\end{document}